\pdfoutput=1
\documentclass[journal,onecolumn]{IEEEtran}

\newcommand{\descr}[1]{\smallskip \noindent \textbf{#1}}
\newcommand{\descrit}[1]{\smallskip \noindent \textbf{\textit{#1}}}
\usepackage{color}
\usepackage[hyphens]{url}
\usepackage{hyperref}
\usepackage[moderate]{savetrees}
\usepackage{fancyhdr}

\usepackage{xspace}
\usepackage{enumerate}
\usepackage{lipsum}
\usepackage{multirow}
\usepackage[shortlabels]{enumitem}
\usepackage{amssymb}
\usepackage{makecell}
\usepackage{booktabs} 
\usepackage{amsmath}
\usepackage{float}
\usepackage{microtype}
\usepackage[left=0.6in,right=0.6in,top=0.5in,bottom=0.5in,includehead,includefoot]{geometry}
\usepackage{graphicx}
\usepackage{subcaption}
\usepackage{mathtools}
\DeclarePairedDelimiter\ceil{\lceil}{\rceil}

\allowdisplaybreaks 
\usepackage{amsthm} 
\usepackage{graphicx}
\usepackage{comment}
\usepackage{xcolor}
\usepackage{alltt}
\usepackage{arydshln}
\usepackage{pifont}% http://ctan.org/pkg/pifont
\makeatletter
\newenvironment{protocolalg}[1][H]{%
    \renewcommand{\ALG@name}{Protocol}% Update algorithm name
    \begin{algorithm}[#1]%
    \small
    }{\end{algorithm}
}
\usepackage{tikz}
\usetikzlibrary{decorations.pathreplacing,calc}
\newcommand{\tikzmark}[1]{\tikz[overlay,remember picture] \node (#1) {};}
\usepackage{algorithm,algpseudocode}
\usepackage{setspace}
\usepackage{caption}

\usepackage{subcaption}
\newcommand{\sys}{\textsc{Poseidon}\xspace}
\usepackage{tabularx}
\usepackage[T1]{fontenc}
\usepackage[utf8]{inputenc}
\usepackage{bm}
\definecolor{mygreen}{rgb}{0.0, 0.5, 0.0}
\DeclareCaptionType{copyrightbox}

\newtheorem{proposition}{Proposition}
\newtheorem{lemma}{Lemma}

\newif\ifcomment
\commenttrue 
\ifcomment
	\newcommand{\ap}[1]{\textbf{\em\color{blue}[AP: #1]}}
	
	\newcommand{\sinem}[1]{\textbf{\em\color{magenta}[SS: #1]}}
		\newcommand{\added}[1]{{\color{black} #1}}
	\newcommand{\changed}[1]{{\color{black} #1}}
	
\else  
    \newcommand\ap[1]{}
\fi

\begin{document}

\title{\sys: Privacy-Preserving Federated Neural Network Learning}

\author{\IEEEauthorblockN{Sinem Sav, Apostolos Pyrgelis, Juan Ramón Troncoso-Pastoriza,}\hfill\and\\

\IEEEauthorblockN{David Froelicher, Jean-Philippe Bossuat, Joao Sa Sousa, and Jean-Pierre Hubaux}
%\and\\

%\hfill\IEEEauthorblockN{\textsc{Paper under submission - Please do not disseminate}}
%\thanks{This work was partially supported by the grant \#2017-201 of the Strategic Focal Area “Personalized Health and Related Technologies (PHRT)” of the ETH Domain.}
%\thanks{D. Froelicher is with the Laboratory for Data Security and DeDiS Laboratory, École Polytechnique Fédérale de Lausanne, 1015 Lausanne, Switzerland, e-mail: david.froelicher@epfl.ch. % <-this % stops a space
%	Sinem Sav, Juan R. Troncoso-Pastoriza, Apostolos Pyrgelis, Jean-Philippe Bossuat, Joao Sa Sousa and Jean-Pierre Hubaux are with the Laboratory for Data Security, École %Polytechnique Fédérale de Lausanne, 1015 Lausanne, Switzerland, e-mail: name.surname@epfl.ch.}
}

\maketitle
\thispagestyle{plain}
\pagestyle{plain}

\begin{abstract}
In this paper, we address the problem of privacy-preserving training and evaluation of neural networks in an $N$-party, federated learning setting. We propose a novel system, \sys, the first of its kind in the regime of privacy-preserving neural network training. It employs multiparty lattice-based cryptography to preserve the confidentiality of the training data, the model, and the evaluation data, under a passive-adversary model and collusions between up to $N-1$ parties. To efficiently execute the secure backpropagation algorithm for training neural networks, we provide a generic packing approach that enables Single Instruction, Multiple Data (SIMD) operations on encrypted data. We also introduce arbitrary linear transformations within the cryptographic bootstrapping operation, optimizing the costly cryptographic computations over the parties, and we define a constrained optimization problem for choosing the cryptographic parameters. Our experimental results show that \sys achieves accuracy similar to centralized or decentralized non-private approaches and that its computation and communication overhead scales linearly with the number of parties. \sys trains a 3-layer neural network on the MNIST dataset with 784 features and 60K samples distributed among 10 parties in less than 2 hours. %Our implementation is available for review.
\end{abstract}

%  and that scales linearly with the number of parties
% by performing homomorphic operations on the participants' encrypted data. 
% To provide privacy and security guarantees, \sys extends the MapReduce abstraction  to preserve the confidentiality of the data and model parameters in a semi-honest model with dishonest majority. 

\section{Introduction}\label{intro}
%NEW INTRO
%1st parag intro on distributed NN and its importance
In the era of big data and machine learning (ML), neural networks (NNs) are the state-of-the-art models, as they achieve remarkable predictive performance in various domains such as healthcare, finance, and image recognition~\cite{ABIODUN2018,deep-learning-apps,deep-learning-apps2}. However, training an accurate and robust deep learning model requires a large amount of diverse and heterogeneous data~\cite{Zhu2016}. This phenomenon raises the need for data sharing among multiple data owners who wish to collectively train a deep learning model and to extract valuable and generalizable insights from their joint data. Nonetheless, data sharing among entities, such as medical institutions, companies, and organizations, is often not feasible due to the sensitive nature of the data~\cite{Zhang_BigDataSecurity}, strict privacy regulations~\cite{HIPAA,GDPR}, or the business competition between them~\cite{Stoica2017}. Therefore, solutions that enable privacy-preserving training of NNs on the data of multiple parties are highly desirable in many domains.

% collecting large amounts of data from multiple parties remains a challenge due to the sensitive nature of the data and business competition between parties. 
% In a medical setting, for example, hospitals need decision-support machine-learning systems employing neural networks (NNs) that are trained on large amounts of data~\cite{Lisboa2002,Delen2008}. One approach is to aggregate or pool all the data from different hospitals and train the model in a centralized server. However, such a solution is not feasible in most of the real-life applications due to strict privacy regulations~\cite{HIPAA,GDPR} or business competition~\cite{Stoica2017}. Thus, privacy-preserving machine-learning systems are highly desirable in many domains. From those machine-learning models, enabling collective training neural networks in a privacy-preserving manner is particularly crucial as it is a well-known fact that neural networks achieve better performance than other machine-learning methods, e.g., even a simple multilayer perceptron perform better than a logistic regression model with 95\% confidence level \cite{Adeodato2004}.

%2nd parag. centralized solutions (DP+HE) -> do I miss a citation here? 
A simple solution for collective training is to outsource the data of multiple parties to a \textit{trusted party} that is able to train the NN model on their behalf and to retain the data and model's confidentiality, based on established stringent non-disclosure agreements. These confidentiality agreements, however, require a significant amount of time to be prepared by legal and technical teams~\cite{ndas1} and are very costly~\cite{ndas2}. Furthermore, the trusted party becomes a single point of failure, thus both data and model privacy could be compromised by data breaches, hacking, leaks, etc. Hence, solutions originating from the cryptographic community replace and emulate the trusted party with a group of computing servers. In particular, to enable privacy-preserving training of NNs, several studies employ multiparty computation (MPC) techniques and operate on the two~\cite{SecureML,Chen}, three~\cite{mohassel2018aby,wagh2019securenn,falcon}, or four~\cite{flash,trident} server models. Such approaches, however, limit the number of parties among which the trust is split, often assume an honest majority among the computing servers, and require parties to communicate (i.e., secret share) their data outside their premises. This might not be acceptable due to the privacy and confidentiality requirements and the strict data protection regulations. Furthermore, the computing servers do not operate on their own data or benefit from the model training; hence, their only incentive is the reputation harm if they are caught, which increases the possibility of malicious behavior.

%4th parag. promising FL approach!
A recently proposed alternative for privacy-preserving training of NNs -- without data outsourcing -- is \textit{federated learning}. Instead of bringing the data to the model, the model is brought (via a coordinating server) to the clients, who perform model updates on their local data. The updated models from the parties are averaged to obtain the global NN model~\cite{federatedLearning1,Konency2016fed}. Although federated learning retains the sensitive input data locally and eliminates the need for data outsourcing, the model, that might also be sensitive, e.g., due to proprietary reasons, becomes available to the coordinating server, thus placing the latter in a position of power with respect to the remaining parties. Recent research demonstrates that sharing intermediate model updates among the parties or with the server might lead to various privacy attacks, such as extracting parties' inputs~\cite{Hitaj2017,Wang2019,NIPS2019_9617} or membership inference~\cite{Melis2019,Nasr2019}. Consequently, several works employ differential privacy to enable privacy-preserving exchanges of intermediate values and to obtain models that are free from adversarial inferences in federated learning settings~\cite{Nvidia_Fed,shokri2015privacy,McMahan2018}. Although differentially private techniques partially limit attacks to federated learning, they decrease the utility of the data and the resulting ML model. Furthermore, training robust and accurate models requires high privacy budgets, and as such, the level of privacy achieved in practice remains unclear~\cite{jayaraman2019evaluating}. Therefore, a distributed privacy-preserving deep learning approach requires strong cryptographic protection of the intermediate model updates during the training, as well as of the final model weights.

Recent cryptographic approaches for private distributed learning, e.g.,~\cite{zheng2019helen,Drynx}, not only have limited ML functionalities, i.e., regularized or generalized linear models, but also employ traditional encryption schemes that make them vulnerable to post-quantum attacks. This should be cautiously considered, as recent advances in quantum computing~\cite{Quantum1,Quantum4,Quantum3,Quantum2_GoogleAI}, increase the need for deploying quantum-resilient cryptographic schemes that eliminate potential risks for applications with long-term sensitive data. Froelicher et al. recently proposed \textsc{SPINDLE}~\cite{spindle}, a generic approach for the privacy-preserving training of ML models in an $N$-party setting that employs multiparty lattice-based cryptography, thus achieving post-quantum security guarantees. However, the authors~\cite{spindle} demonstrate the applicability of their approach only for generalized linear models, and their solution lacks the necessary protocols and functions that can support the training of complex ML models, such as NNs.

% To this end, our work extends the \textsc{SPINDLE}~\cite{spindle} framework, employing distributed lattice-based cryptography for the training of NNs. While enabling privacy-preserving training of generalized linear models with $N$ parties, \textsc{SPINDLE} lacks the protocols for  enabling  several  functions necessary  for  the  neural  networks,  e.g.,  transpose  or  pooling. Besides, we show that the proposed solution in \textsc{SPINDLE}, when used for training NNs, introduces a significant computation and communication overhead due to the underlying packing scheme of the plaintexts. This overhead makes the neural network executions under \textsc{SPINDLE} \cite{spindle} infeasible. 

%6th parag
In this work, we extend the approach of \textsc{SPINDLE}~\cite{spindle} and build \sys, a novel system that enables the training and evaluation of NNs in a distributed setting and provides end-to-end protection of the parties' training data, the resulting model, and the evaluation data. Using multiparty lattice-based homomorphic encryption~\cite{mouchet2019distributedbfv}, \sys enables NN executions with different types of layers, such as fully connected, convolution, and pooling, on a dataset that is distributed among $N$ parties, e.g., a consortium of tens of hospitals, that trust only themselves for the confidentiality of their data and of the resulting model. \sys relies on mini-batch gradient descent and protects, from any party, the intermediate updates of the NN model by maintaining the weights and gradients encrypted throughout the training phase. \sys also enables the resulting encrypted model to be used for privacy-preserving inference on encrypted evaluation data.

We evaluate \sys on several real-world datasets and various network architectures such as fully connected and convolutional neural network structures and observe that it achieves training accuracy levels on par with centralized or decentralized non-private approaches. Regarding its execution time, we find that \sys trains a 2-layer NN model on a dataset with 23 features and 30,000 samples distributed among 10 parties, in 8.7 minutes. Moreover, \sys trains a 3-layer NN with 64 neurons per hidden-layer on the MNIST dataset~\cite{MNIST} with 784 features and 60K samples shared between 10 parties, in 1.4 hours, and a NN with convolutional and pooling layers on the CIFAR-10~\cite{cifarPaper} dataset (60K samples and 3,072 features) distributed among 50 parties, in 175 hours. Finally, our scalability analysis shows that \sys's computation and communication overhead scales linearly with the number of parties and logarithmically with the number of features and the number of neurons in each layer.

% It does not require collection of data in encrypted form, as several works do \cite{cryptoDL,Karthik2019}, thus reduces the number computational complexity.

% Each party of collaborative training of NN holds their data and trust only themselves for the confidentiality of their data. \sys is based on distributed homomorphic encryption scheme and enables the NN executions with different types of layers, i.e., fully-connected, convolutional, and pooling, on $N$ parties for an unbounded $N$. 
% We rely on mini-batch gradient descent and build on the well-known MapReduce framework~\cite{parallel_SGD,MapReduce_ML} to enable the distributed NN training. \sys protects the intermediate updates of the NN model from any party, keeping the weights and gradients encrypted throughout the training and the inference phases in the federated learning setting. 
% It provides privacy-preserving inference on the encrypted model using the encrypted data, by leveraging on the key-switching functionality of the underlying crypto scheme.

%7th parag
In this work, we make the following contributions:
\begin{itemize}
    \item We present \sys, a novel system for privacy-preserving, quantum-resistant, federated learning-based training of and inference on NNs with $N$ parties with unbounded $N$, that relies on multiparty homomorphic encryption and respects the confidentiality of the training data, the model, and the evaluation data.
   % \item We provide a novel transpose operation on diagonalized matrices under encryption and show its' performance in our system \sinem{We don't use this anymore, should we still give and compare with our approach? }.
    \item We propose an alternating packing approach for the efficient use of single instruction, multiple data (SIMD) operations on encrypted data, and we provide a generic protocol for executing NNs under encryption, depending on the size of the dataset and the structure of the network.
    \item We improve the distributed bootstrapping protocol of~\cite{mouchet2019distributedbfv} by introducing arbitrary linear transformations for optimizing computationally heavy operations, such as pooling or a large number of consecutive rotations on ciphertexts.
    \item We formulate a constrained optimization problem for choosing the cryptographic parameters and for balancing the number of costly cryptographic operations required for training and evaluating NNs in a distributed setting.
  \item \sys advances the state-of-the-art privacy-preserving solutions for NNs based on MPC~\cite{wagh2019securenn,SecureML,mohassel2018aby,quotient,trident,falcon}, by achieving better flexibility, security, and scalability:\\
  \textbf{Flexibility.} \sys relies on a federated learning approach, eliminating the need for communicating the parties' confidential data outside their premises which might not be always feasible due to privacy regulations~\cite{HIPAA,GDPR}. This is in contrast to MPC-based solutions which require parties to distribute their data among several servers, and thus, fall under the cloud outsourcing model.\\
  \textbf{Security.} \sys splits the trust among multiple parties, and guarantees its data and model confidentiality properties under a passive-adversarial model and collusions between up to $N-1$ parties, for unbounded $N$. On the contrary, MPC-based solutions limit the number of parties among which the trust is split (typically, 2, 3, or 4 servers) and assume an honest majority among them.\\
  \textbf{Scalability.} \sys's communication is linear in the number of parties, whereas MPC-based solutions scale quadratically.
    \item Unlike differential privacy-based approaches for federated learning~\cite{Nvidia_Fed,shokri2015privacy,McMahan2018}, \sys does not degrade the utility of the data, and the impact on the model's accuracy is negligible.
 %   \item \sys provides stronger security guarantees than existing solutions relying on homomorphic encryption for training NNs~\cite{Karthik2019,cryptoDL}, as these do not employ realistic cryptographic parameters. Furthermore, it achieves better performance: \sys achieves an improvement of $10-900\times$ in similar evaluation settings. \sinem{here: is it ok to state this although other two HE solutions use local bootstrapping?}.\ap{this is risky: can we compare with them if they do not support the N-party setting?}
\end{itemize}

\noindent To the best of our knowledge, \sys is the first system that enables quantum-resistant distributed learning on neural networks with $N$ parties in a federated learning setting, and that preserves the privacy of the parties' confidential data, the intermediate model updates, and the final model weights. 
\vspace{-0.5em}
\section{Related Work}\label{sec:related}
\descr{Privacy-Preserving Machine Learning (PPML).} 
%\sinem{check if I mistakenly cite 2 or 3-server solutions that also address NN training }
% Early work on PPML has focused on secure clustering \cite{cluster1,cluster2}, regularized linear models, and generalized linear models
Previous PPML works focus exclusively on the training of (generalized) linear models~\cite{aono2016scalable,jiang2019securelr,bonte2018privacy,crawford2018doing,kim2018logistic,kim2018secure}. They rely on \emph{centralized} solutions where the learning task is securely outsourced to a server, notably using homomorphic encryption (HE) techniques. As such, these works do not solve the problem of privacy-preserving \emph{distributed} ML, where multiple parties collaboratively train an ML model on their data.
% as the data records of individual data providers have to be transferred to another server which might be difficult due to the data protection regulations. 
To address the latter,
% problem of privacy-preserving distributed machine-learning,
several works propose multi-party computation (MPC) solutions where several tasks, such as clustering and regression, are distributed among 2 or 3 servers~\cite{cluster1,cluster2,nikolaenko2013privacy,gascon2017privacy,giacomelli2018privacy,Akavia_WAHC,schoppmann2019make,bogdanov2016rmind,Cho_GWAS}.
% These works employ MPC, or a combination of both HE and MPC techniques (hybrid). 
Although such solutions enable multiple parties to collaboratively \changed{learn} on their data, the trust distribution is limited to the number of computing servers that train the model, and they rely on assumptions such as non-collusion, or an honest majority among the servers.
% either two non-colluding servers or honest majority in 3 server settings. 
There exist only a few works that extend the distribution of ML computations to $N$ parties ($N\geq4$) and that remove the need for outsourcing~\cite{corrigan2017prio, zheng2019helen,Drynx,spindle}. For instance, Zheng et al. propose Helen, a system for privacy-preserving learning of linear models that combines HE  with MPC techniques~\cite{zheng2019helen}. However, the use of the Paillier additive HE scheme~\cite{Paillier} makes their system vulnerable to post-quantum attacks. To address this issue, Froelicher et al. introduce SPINDLE~\cite{spindle}, a system that provides support for generalized linear models and security against post-quantum attacks. These works have paved the way for PPML computations in the N-party setting, but none of them addresses the challenges associated with the privacy-preserving training of and inference on neural networks (NNs).

\descr{Privacy-Preserving Inference on Neural Networks.} In this research direction, the majority of works operate on the following setting: a central server holds a trained NN model and clients communicate their evaluation data to obtain predictions-as-a-service~\cite{CryptoNets,MiniONN,Gazelle}. Their aim is to protect both the confidentiality of the server's model and the clients' data. Dowlin et al. propose the use of a ring-based leveled HE scheme to enable the inference phase on encrypted data~\cite{CryptoNets}. Other works rely on hybrid approaches by employing two-party computation (2PC) and HE~\cite{Gazelle,MiniONN}, or
%Liu et al. propose MiniONN \cite{MiniONN} that relies on leveled HE, in which the non-linear activation functions are enabled by secure two-party computation (2PC) - garbled circuits while Juvekar et al. propose Gazelle that employs HE supporting basic single instruction, multiple data (SIMD) operations, and garbled circuits \cite{Gazelle}.
% Patra and Suresh rely on 
secret sharing and garbled circuits to enable privacy-preserving inference on NNs~\cite{blaze,riazi2019xonn,mishra2020}. For instance, Riazi et al. use garbled circuits to achieve constant round communication complexity during the evaluation of binary neural networks~\cite{riazi2019xonn}, whereas Mishra et al. propose a similar hybrid solution that outperforms previous works in terms of efficiency, by tolerating a small decrease in the model's accuracy~\cite{mishra2020}.
% or improving its accuracy at a cost of performance.

Boemer et al. develop a deep-learning graph compiler for multiple HE cryptographic libraries~\cite{boemer2019ngraph,boemer2018ngraph}, such as SEAL~\cite{seal}, HElib~\cite{helib}, and
Palisade~\cite{palisade}. Their work enables the deployment of a model, which is trained with well-known frameworks (e.g., Tensorflow~\cite{tensorflow}, PyTorch~\cite{pytorch}), and enables predictions on encrypted data. Dalskov et al. use quantization techniques to enable efficient privacy-preserving inference on models trained with Tensorflow~\cite{tensorflow} by using MP-SPDZ~\cite{mpspdz} and demonstrate benchmarks for a wide range of adversarial models~\cite{Dalskov}.

All aforementioned solutions enable only privacy-preserving inference on NNs, whereas our work focuses on both the privacy-preserving training of and the inference on NNs, protecting the training data, the resulting model, and the evaluation data.

\descr{Privacy-Preserving Training of Neural Networks.} A number of works focus on \emph{centralized} solutions to enable privacy-preserving learning of NNs~\cite{Song2013,abadi2016deep,Yu2019,Vizitu2020,Karthik2019,cryptoDL}. Some of them, e.g.,~\cite{Song2013,abadi2016deep,Yu2019}, employ differentially private techniques to execute the stochastic gradient descent while training a NN in order to derive models that are protected from inference attacks~\cite{shokri2017membership}. However, they assume that the training data is available to a \emph{trusted} party that applies the noise required during the training steps. Other works, e.g.,~\cite{Vizitu2020,Karthik2019,cryptoDL}, rely on HE to outsource the training of multi-layer perceptrons to a central server. These solutions either employ cryptographic parameters that are far from realistic~\cite{Vizitu2020,Karthik2019}, or yield impractical performance~\cite{cryptoDL}. Furthermore, they do not support the training of NNs in the $N$-party setting, which is the main focus of our work.

A number of works that enable privacy-preserving \emph{distributed} learning of NNs employ MPC approaches where the parties' confidential data is distributed among two~\cite{SecureML,quotient}, three~\cite{mohassel2018aby,wagh2019securenn,falcon,DTI,Chen}, or four servers~\cite{flash,trident} (2PC, 3PC, and 4PC, resp.). For instance, in the 2PC setting, Mohassel and Zhang describe a system where data owners process and secret-share their data among two non-colluding servers to train various ML models~\cite{SecureML}, and Agrawal et al. propose a framework that supports discretized training of NNs by ternarizing the weights~\cite{quotient}. 
%These two servers can train various ML models without learning any information beyond the trained model.
Then, Mohassel and Rindal extend~\cite{SecureML} to the 3PC setting and introduce new fixed-point multiplication protocols for shared decimal numbers~\cite{mohassel2018aby}. Wagh et al. further improve the efficiency of privacy-preserving NN training on secret-shared data~\cite{wagh2019securenn} and provide security against malicious adversaries, assuming an honest majority among 3 servers~\cite{falcon}.
% Hie et al. demonstrate the application of privacy-preserving NN training with 3PC on a predictive model for drug-target interactions~\cite{DTI}, while Chen and Zhong enable the privacy-preserving backpropagation algorithm using ElGamal encryption~\cite{elgamal1985public} for NN training over vertically partitioned data in 2PC setting ~\cite{Chen}. Their work enabled the privacy-preserving NN training over vertically partitioned data among 2 servers, in a semi-honest setting. 
More recently, 4PC honest-majority malicious frameworks for PPML have been proposed~\cite{flash,trident}. These works split the trust between more servers and achieve better round complexities than previous ones, yet they do not address NN training among $N$-parties. Note that 2PC, 3PC, and 4PC solutions fall under the \textit{cloud outsourcing} model, as the data of the parties has to be transferred to several servers among which the majority has to be trusted. Our work, however, focuses on a distributed setting where the data owners maintain their data locally and iteratively update the collective model, yet data and model confidentiality is ensured in the existence of a dishonest majority in a semi-honest setting, thus withstanding passive adversaries and up to $N-1$ collusions between them. We provide a comparison with these works in Section~\ref{sec:evalComparison}.
% approach that was recently proposed by Google

Another widely employed approach for training NNs in a distributed manner is that of federated learning~\cite{federatedLearning1,Konecny2016,Konency2016fed}. The main idea is to train a global model on data that is distributed across multiple clients, with the assistance of a server that coordinates model updates on each client and averages them. This approach does not require clients to send their local data to the central server, but several works show that the clients' model updates \textit{leak} information about their local data~\cite{Hitaj2017,Wang2019,NIPS2019_9617}. To counter this, some works focus on secure aggregation techniques for distributed NNs, based on
% where the aggregation of the local updates is done in a secure way by using 
HE~\cite{Phong2017,Phong2018} or MPC~\cite{Bonawitz2016}. Although encrypting the gradient values prevents the leakage of parties' confidential data to the central server, these solutions do not account for potential leakage from the \emph{aggregate} values themselves. In particular, parties that decrypt the received model before the next iteration are able to infer information about other parties' data from its parameters~\cite{Hitaj2017,Melis2019,Nasr2019,NIPS2019_9617}. Another line of research relies on differential privacy (DP) to enable privacy-preserving federated learning for NNs. Shokri and Shmatikov~\cite{shokri2015privacy} apply DP to the parameter update stages, and Li et al. design a privacy-preserving federated learning system for medical image analysis where the parties exchange differentially private gradients~\cite{Nvidia_Fed}. McMahan et al. propose differentially private federated learning~\cite{McMahan2018}, by employing the moments accountant method~\cite{abadi2016deep}, to protect the privacy of all the records belonging to a user. Finally, other works combine MPC with DP techniques to achieve better privacy guarantees~\cite{jayaraman2018distributed,truex2019hybrid}. While DP-based learning aims to mitigate inference attacks, it significantly degrades model utility, as training accurate NN models requires high privacy budgets~\cite{Rahman2018dp}. As such, it is hard to quantify the level of privacy protection that can be achieved with these approaches~\cite{jayaraman2019evaluating}. To account for these issues, our work employs multiparty homomorphic encryption techniques to achieve \textit{zero-leakage} training of NNs in a distributed setting where the parties' intermediate updates and the final model remain under encryption.

%\sinem{make it more strong, mention zero-leakage if the model is not revealed at the end}
% \sinem{JP: Rephrase the first two sentences}

% \descr{Motivation for $N$-party framework.} While the solutions proposed for 2PC, 3PC, and 4PC settings have the advantage of providing efficient computations for training NNs as they employ secret-sharing techniques combined with several optimizations, the data providers, for the machine-learning task in these frameworks, have to trust all or the majority of the servers for the confidentiality of their data and the model parameters. These solutions fall into the cloud outsourcing model and increasing the number of parties is impractical because of the quadratic increase in the communication~\cite{Ishai2018}. However, in our $N$-party setting, the trust is distributed among more servers and the communication scales linearly with the number of parties. We strongly motivate the federated learning-based approach to provide a stronger trust assumption. In our solution, the confidentiality of the parties' data and the model's parameters is ensured in the existence of a dishonest majority in a semi-honest setting, allowing $N-1$ passive adversaries and collusions between parties.
\section{Preliminaries} \label{sec:background}
 We introduce here the background information about NNs as well as the multiparty homomorphic encryption (MHE) scheme on which \sys relies to achieve privacy-preserving training of and inference on NN models in a federated $N$-party setting.
 \vspace{-0.4em}
\subsection{Neural Networks}
Neural networks (NNs) are machine learning algorithms that extract complex non-linear relationships between the input and output data. They are used in a wide range of fields such as pattern recognition, data/image analysis, face recognition, forecasting, and data validation in the medicine, banking, finance, marketing, and health industries~\cite{ABIODUN2018}. Typical NNs are composed of a pipeline of layers where feed-forward and backpropagation steps for linear and non-linear transformations (activations) are applied to the input data iteratively~\cite{NeuralBook}. Each training iteration is composed of one forward pass and one backward pass, and the term epoch refers to processing once all the samples in a dataset.

% For the design and evaluation of our system, we focus on fully-connected deep NN structures or

Multilayer perceptrons (MLPs) are fully-connected deep neural network structures which are widely used in the industry, e.g., they constitute $61\%$ of Tensor Processing Units' workload in Google's datacenters~\cite{Jouppi2017}. MLPs are composed of an input layer, one or more hidden layer(s), as well as an output layer, and each neuron is connected to all the neurons in the following layer. At iteration $k$, the weights between layers $j$ and $j+1$, are denoted by a matrix $W_j^k$, whereas the matrix $L_j$ represents the activation of the neurons in the $j^{th}$ layer.
%where the entry $w_{ij}$ is the weight between the neuron $i$ and the neuron $j$ in the following layer.
The forward pass requires first the linear combination of each layer's weights with the activation values of the previous layer, i.e., $U_j= W_j^k \times L_{j-1}$. Then, an activation function is applied to calculate the values of each layer as $L_j = \varphi(U_j)$.

% This process is repeated for all layers for $j=1,...,\ell$, iteratively.
%Note that the first hidden layer activation is computed as $L_1 = \varphi(X \times W_1)$ where $X_j$ is the $i^{th}$ sample data from the training dataset, as the first layer is the input layer.

Backpropagation, a method based on gradient descent, is then used to update the weights during the backward pass. Here, we describe the update rules for mini-batch gradient descent where a random batch of sample inputs of size $B$ is used in each iteration. The aim is to minimize each iteration's error based on a cost function $E$ (e.g., mean squared error) and update the weights accordingly. The update rule is $W_j^{k+1} = W_{j}^k - \frac{\eta}{B} \nabla W_{j}^k$, where $\eta$ is the learning rate and $\nabla W_j^k$ denotes the gradient of the cost function with respect to the weights and calculated as $\nabla W_j^k=\frac{\partial E}{\partial W_{j}^k}$. We note that backpropagation requires several transpose operations applied to matrices/vectors and we denote transpose of a matrix/vector as $W^T$.
%Given a cost function, e.g., the mean squared error where the last layer's error is $E_l = \frac{1}{2}(L_l-Y)^2$ for its outputs (predictions) $L_l$ and true class labels $Y$, the error term, $\delta_j$ for each layer $j$, is calculated iteratively. Starting from the last layer $\delta_l= \frac{\partial E}{\partial L_{j}}\odot\frac{\partial L_j}{\partial U_{j}}$, where $\frac{\partial L_j}{\partial U_{j} }$ is the derivative of the activation function ($\varphi'$) and $\odot$ denotes element-wise multiplication. Following the chain rule  $\delta_j^k= (\delta_{j+1}^k\times (W_j^k)^T)\odot\varphi'(U_j)$ for each hidden layer $j$, and thus, $\nabla W_j^k= \delta_j^k L_j$.
%These steps are the local gradient descent computations per party and are described in Protocol~\ref{gradprotocol}.

Convolutional neural networks (CNNs) follow a very similar sequence of operations, i.e., forward and backpropagation passes, and typically consist of convolutional  (CV), pooling, and fully connected (FC) layers. It is worth mentioning that CV layer operations can be expressed as FC layer operations by representing them as matrix multiplications; in our protocols, we simplify CV layer operations by employing this representation~\cite{wagh2019securenn,conversionFC}. Finally, pooling layers are downsampling layers where a kernel, i.e., a matrix that moves over the input matrix with a stride of $a$, is convoluted with the current sub-matrix. For a kernel of size $k\times k$, the minimum, maximum, or average (depending on the pooling type) of each $k\times k$ sub-matrix of the layer's input is computed. 

%$\frac{\partial E}{\partial W_{j}}=\frac{\partial E}{\partial L_{j}}\odot\frac{\partial L_i}{\partial U_{j}}\times\frac{\partial U_i}{\partial W_{j}} = \frac{\partial E}{\partial L_{j}}\odot\frac{\partial L_i}{\partial U_{j}} U_i=\delta_{j+1} U_i$, where $\delta_l =...$ and for the output layer and $\delta_i=...$ for hidden layers.
%\subsubsection{Widely Used Activation Functions}
%\label{sec:activations}
%Here we give the definitions of widely used activation functions for NN training: ReLU ($f(x)=\relu$), sigmoid ($f(x)=\sigmoid$), softmax ($f(x)=\softmax$), and softplus ($f(x)=\softplus$). 
 \vspace{-0.4em}
\subsection{Distributed Deep Learning}
\label{sec:distributedDL}
We employ the well-known MapReduce abstraction to describe the training of data-parallel NNs in a distributed setting where multiple data providers hold their respective datasets~\cite{parallel_SGD,MapReduce_ML}. We rely on a variant of the parallel stochastic gradient descent (SGD)~\cite{parallel_SGD} algorithm, where each party performs $b$ local iterations and calculates each layer's partial gradients. These gradients are aggregated over all parties and the reducer performs the model update with the average of gradients~\cite{MapReduce_ML}. This process is repeated for $m$ global iterations. Note that averaging the gradients from $N$ parties is equivalent to performing batch gradient descent with a batch size of $b\times N$. Thus, we differentiate between the local batch size as $b$ and the global batch size $B = b\times N$. 
%Lines 6-17 of Protocol~\ref{protocol:collectivetrain} show the MapReduce steps for distributed mini-batch gradient descent.
 \vspace{-0.4em} 
\subsection{Multiparty Homomorphic Encryption (MHE)}
\label{sec:distributedHomomorphic}

In our system, we rely on the Cheon-Kim-Kim-Song (CKKS)~\cite{cheon2017homomorphic} variant of the MHE scheme proposed by Mouchet et al.~\cite{mouchet2019distributedbfv}. In this scheme, a public collective key is known by all parties while the corresponding secret key is distributed among them. As such, decryption is only possible with the participation of \emph{all} parties. Our motivations for choosing this scheme are: (i) It is well suited for floating point arithmetic, (ii) it is based on the ring learning with errors (RLWE) problem~\cite{Lyubashevsky2010}, making our system secure against post-quantum attacks~\cite{Acar_2018}, (iii) it enables secure and flexible collaborative computations between parties without sharing their respective secret key, and (iv) it enables a secure collective key-switch functionality, that is, changing the encryption key of a ciphertext without decryption. Here, we provide a brief description of the cryptographic scheme's functionalities that we use throughout our protocols.
The cyclotomic polynomial ring of dimension $\mathcal{N}$, where $\mathcal{N}$ is a power-of-two integer, defines the plaintext and ciphertext space as $R_{Q_{L}}=\mathbb{Z}_{Q_{L}}[X]/(X^\mathcal{N}+1)$, with $ Q_{L} = \prod_{0}^{L} q_{i}$ in our case. Each $q_{i}$ is a unique prime, and $Q_{L}$ is the ciphertext modulus at an initial level $L$. Note that a plaintext encodes a vector of up to ${\mathcal{N}/2}$ values. Below, we introduce the main functions that we use in our system. We denote by $\bm{c}=(c_{0}, c_{1}) \in R^{2}_{Q_{L}}$ and $p\in R_{Q_{L}}$, a ciphertext (indicated as boldface) and a plaintext, respectively. $\bar{p}$ denotes an encoded(packed) plaintext. We denote by $L_{\bm{c}}$, $S_{\bm{c}}$, $L$, and $S$, the current level of a ciphertext $\bm{c}$, the current scale of $\bm{c}$, the initial level, and the initial scale (precision) of a fresh ciphertext respectively, and we use the equivalent notations for plaintexts. The functions below that start with 'D' are distributed, and executed among all the secret-key-holders, whereas the others can be executed locally by anyone with the public key.
\vspace{-0.1em}
\begin{itemize}
    \item $\textsf{SecKeyGen}(1^\lambda)$: Returns the set of secret keys $\{sk_i\}$, i.e., $sk_i$ for each party $P_i$, for a security parameter $\lambda$.
    
    \item $\textsf{DKeyGen}(\{sk_i\})$: Returns the collective public key $pk$.
    
    \item $\textsf{Encode}(msg)$ : Returns a plaintext $\bar{p} \in R_{Q_{L}}$ with scale $S$, encoding $msg$.
    
    \item $\textsf{Decode}(\bar{p})$ : For $\bar{p} \in R_{Q_{L_{p}}}$ and scale $S_{p}$, returns the decoding of $p$.

    \item $\textsf{DDecrypt}(\bm{c}, \{sk_i\})$: For $\bm{c} \in R^2_{Q_{L_{\bm{c}}}}$ and scale $S_{\bm{c}}$, returns the plaintext $p \in R_{Q_{L_{\bm{c}}}}$ with scale $S_{\bm{c}}$.
    
    \item $\textsf{Enc}(pk, \bar{p})$: Returns $\bm{c}_{pk}\in R^2_{Q_{L}}$ with scale $S$ such that $\textsf{DDecrypt}(\bm{c}_{pk}, \{sk_i\}) \approx \bar{p}$.
     
    \item $\textsf{Add}(\bm{c}_{pk}, \bm{c'}_{pk})$: Returns $(\bm{c} + \bm{c'})_{pk}$ at level $ \min(L_{\bm{c}},L_{\bm{c'}})$ and scale $\max(S_{\bm{c}},S_{\bm{c'}})$.
     
    \item $\textsf{Sub}(\bm{c}_{pk}, \bm{c'}_{pk})$: Returns $(\bm{c} - \bm{c'})_{pk}$ at level $ \min(L_{\bm{c}},L_{\bm{c'}})$ and scale $\max(S_{\bm{c}},S_{\bm{c'}})$.
     
    \item $\textsf{Mul}_{\textsf{pt}}$($\bm{c}_{pk}, \bar{p})$: Returns $(\bm{c}p)_{pk}$ at level $\min(L_{\bm{c}},L_{p})$ with scale $(S_{\bm{c}}\times S_{p})$.
     
    \item $\textsf{Mul}_{\textsf{ct}}$($\bm{c}_{pk}, \bm{c'}_{pk})$: Returns $(\bm{c}\bm{c'})_{pk}$ at level $\min(L_{\bm{c}},L_{\bm{c'}})$ with scale $(S_{\bm{c}}\times S_{\bm{c'}})$.
     
    \item $\textsf{RotL/R}(\bm{c}_{pk}, k)$: Homomorphically rotates $\bm{c}_{pk}$ to the left/right by $k$ positions.
     
    \item $\textsf{Res}(\bm{c}_{pk})$: Returns $\bm{c}_{pk}$ with scale $S_{\bm{c}}/q_{L_{\bm{c}}}$ at level $L_{\bm{c}}-1$.
     
    \item $\textsf{SetScale}(\bm{c}_{pk}, S)$: Returns $\bm{c}_{pk}$ with scale $S$ at level $L_{\bm{c}}-1$.
     
    \item  $\textsf{KS}(\bm{c}_{pk} \in R^{3})$: Returns $\bm{c}_{pk} \in R^{2}$.
     
    \item $\textsf{DKeySwitch}(\bm{c}_{pk}, pk', \{sk_i\})$ : Returns $\bm{c}_{pk'}$.
    
    \item $\textsf{DBootstrap}(\bm{c}_{pk},L_{\bm{c}},S_{\bm{c}}, \{sk_i\})$: Returns $\bm{c}_{pk}$ with initial level $L$ and scale $S$.

\end{itemize}
\vspace{-0.2em}
We note that $\textsf{Res}(\cdot)$ is applied to a resulting ciphertext after each multiplication. Further, for a ciphertext at an initial level $L$, at most an $L$-depth circuit can be evaluated. To enable more homomorphic operations to be carried on, the ciphertext must be re-encrypted to its original level $L$. This is done by the bootstrapping functionality ($\textsf{DBootstrap}(\cdot)$). $\textsf{Encode}(\cdot)$ enables us to pack several values into one ciphertext and operate on them in parallel. 

For the sake of clarity, we differentiate between the functionality of the collective key-switch ($\textsf{DKeySwitch}(\cdot)$), that requires interaction between all the parties, and a local key-switch ($\textsf{KS}(\cdot)$) that uses a special public-key. The former is used to decrypt the results or change the encryption key of a ciphertext. The latter, which does not require interactivity, is used during the local computation for slot rotations or relinearization after each multiplication. 
We provide the frequently used symbols and notations in Table \ref{table:notations}, Appendix \ref{sec:notations}.
\section{System Overview}\label{sec:overview}

We introduce \sys's system and threat model, as well as its objectives (Sections~\ref{sec:system-threat} and~\ref{sec:objectives}). Moreover, we provide a high level description of its functionality (Sections~\ref{sec:overview} and~\ref{sec:Protocols}).
 \vspace{-0.3em}
\subsection{System and Threat Model}\label{sec:system-threat}
\begin{figure}[t]
    \vspace{-1.6em}
    \centering
    \small
    \includegraphics[width=0.5\columnwidth]{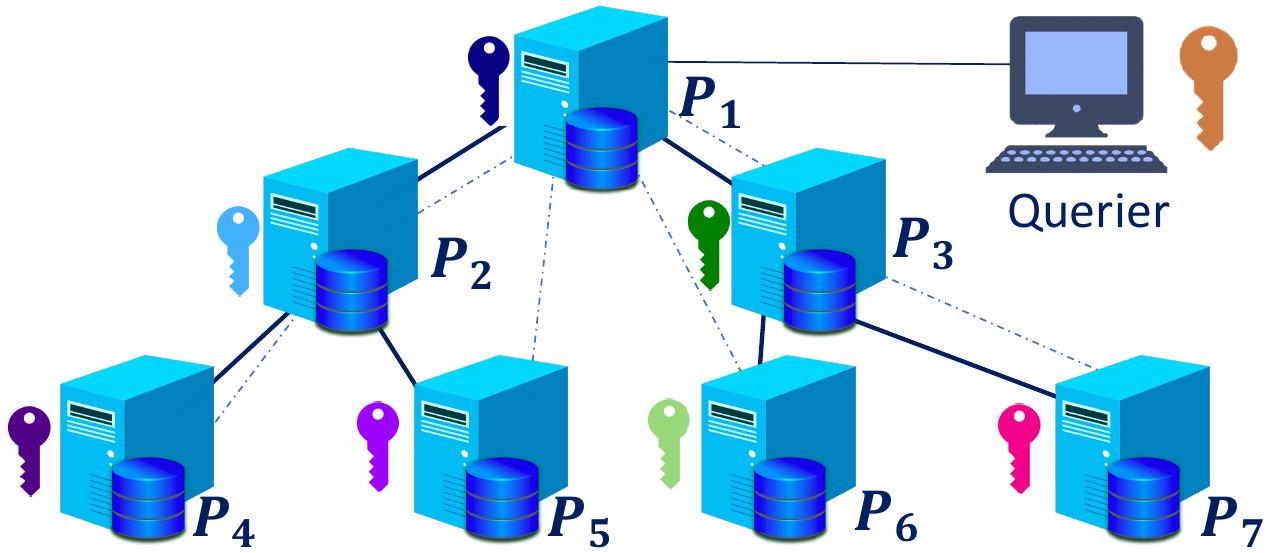}
    \caption{\sys's System Model.}
    \label{fig:system}
\end{figure}

 \vspace{-0.3em}
We introduce \sys's system and threat model below. \\
 \vspace{-0.3em}
\descr{System Model.} We consider a setting where $N$ parties, each locally holding its own data $X_i$, and one-hot vector of labels $y_i$, collectively train a neural network (NN) model. At the end of the training process, a querier -- which can be one of the $N$ parties or an external entity -- queries the model and obtains prediction results $y_q$ on its evaluation data $X_q$. The parties involved in the training process are interested in preserving the privacy of their local data, the intermediate model updates, and the resulting model. The querier obtains prediction results on the trained model and keeps its evaluation data confidential. We assume that the parties are interconnected and organized in a tree-network structure for communication efficiency, as shown in Figure~\ref{fig:system} (thick lines). However, our system is fully distributed and does not assume any hierarchy, therefore remaining agnostic of the network topology, e.g., we can consider a fully-connected network, or a star topology in which each party communicates with a central server (dotted lines in Figure~\ref{fig:system}).

% For the inference phase, the data of the querier is protected against any malicious $P_i$ by default, as it is encrypted under querier's public key and only the querier is able to decrypt the result.

\descr{Threat Model.} We consider a \textit{passive-adversary model} with collusions of up to $N-1$ parties: i.e., the parties follow the protocol but up to $N-1$ parties might share among them their inputs and observations during the training phase of the protocol, %e information they observe during its execution,
to extract information about the other parties' inputs through membership inference or federated learning attacks~\cite{Melis2019,Nasr2019,Hitaj2017,NIPS2019_9617,Wang2019}, prevented by our work. Inference attacks on the model's \textit{prediction} phase, such as membership~\cite{shokri2017membership} or model inversion~\cite{fredrikson2015model}, exploit the final prediction result, and are out-of-the-scope of this work. We discuss complementary security mechanisms that can bound the information a querier infers from the prediction results and an extension to the active-adversary model in Appendix~\ref{securityExtensions}.
%\sinem{make it more clear, mention membership inference FOR pred. is out of scope but we thwart the one in training due to FL attacks}
%We consider two settings for the inference on the trained protocol where the querier is sending the encrypted data to get the inference result from the trained model: (i) The querier is an honest-party, (ii) The querier is malicious and can try to make model attacks such as membership inference or model inversion attacks. For the latter threat model, we propose using complementary security mechanisms for the inference e.g., differential privacy or bounding the number of queries, to bound the information that the querier can infer from the prediction results (see Appendix~\ref{securityExtensions}). Note that in both settings, the data of the querier is protected against any malicious $P_i$ by default, as it is encrypted under querier
%'s public key and only the querier is able to decrypt the result.  
\vspace{-0.4em}
\subsection{Objectives}\label{sec:objectives}

\sys's main objective is to enable the privacy-preserving training of and the evaluation on NNs in the above system and threat model. During the training process, \sys protects both the intermediate updates and the final model weights --- that can potentially leak information about the parties' input data~\cite{Hitaj2017,Melis2019,Nasr2019,NIPS2019_9617} --- from any party. In the inference step, the parties holding the protected model should not learn the querier's data, or the prediction results, and the querier should not obtain the model's weights. Therefore, \sys's objective is to protect the parties' and querier's \textbf{data confidentiality}, as well as the trained \textbf{model confidentiality}, as defined below:
\begin{itemize}
    
	\item \descr{Data Confidentiality.} During training and prediction, no party $P_i$ (including the querier $P_q$) should learn more information about the input data $X_j$ of any other honest party $P_j$ ($j \neq i$, including the querier $P_q$), other than what can be deduced from its own input data $X_i, y_i$ (or the input $X_q$ and output $y_q$, for the querier).
	\item \descr{Model Confidentiality.} During training and prediction, no party $P_i$ (including the querier $P_q$) should gain more information about the trained model weights, other than what can be deduced from its own input data $X_i, y_i$ (or $X_q, y_q$ for the querier).
	
	%any different honest party $P_j$ ($j\neq i$, including the querier $P_q$), other than what can be deduced from its own inputs and outputs $X_i, y_i$.
	%During training, no party $P_i$ should gain more information about the trained model weights, other than what can be deduced from its own input data $X_i, y_i$. During prediction, the querier should not obtain more information about the trained model weights, other than what can be extracted from its evaluation data $X_q$ and the prediction outputs $y_q$.
\end{itemize}

\subsection{Overview of \sys}\label{sec:overview}

% In our system, no entity reveals its local data to any other entity and the model, as well as the intermediate values always kept encrypted. Each party $P_i$ holds their respective secret key $sk_i$, that is detailed under Section~\ref{sec:distributedHomomorphic}, for the protocol computations.

% In \sys, each party $P_i$ holds their respective plaintext data and we rely 
\sys achieves its objectives by exploiting the MHE scheme described in Section~\ref{sec:distributedHomomorphic}. In particular, the model weights are kept encrypted, with the parties' collective public key, throughout the training process. The operations required for the communication-efficient training of neural networks are enabled by the scheme's computation homomorphic properties, which enables the parties to perform operations between their local data and the encrypted model weights. To enable oblivious inference on the trained model, \sys utilizes the scheme's key-switching functionality that allows the parties to collectively re-encrypt the prediction results with the querier's public key.

\sys employs several packing schemes to enable Single Instruction, Multiple Data (SIMD) operations on the weights of various network layers (e.g., fully connected or convolutional ones) and uses approximations that enable the evaluation of multiple activation functions (e.g., Sigmoid, Softmax, ReLU) under encryption. Furthermore, to account for the complex operations required for the forward and backward passes performed during the training of a neural network, \sys uses the scheme's distributed (collective) bootstrapping capability that enables us to refresh ciphertexts. In the following subsection, we provide a high-level description of \sys' phases, the cryptographic operations and optimizations are described in Section~\ref{sec:systemDesign}.

We present \sys as a synchronous distributed learning protocol throughout the paper. An extension to asynchronous distributed NNs is presented in Appendix~\ref{sec:learnextensions}.
\subsection{High-Level Protocols}
\label{sec:Protocols}

To describe the distributed training of and evaluation on NNs, we employ the extended MapReduce abstraction for privacy-preserving machine learning computations introduced in SPINDLE~\cite{spindle}. The overall learning procedure is composed of four phases: \textbf{PREPARE}, \textbf{MAP}, \textbf{COMBINE}, and \textbf{REDUCE}. Protocol~\ref{protocol:collectivetrain} describes the steps required for the federated training of a neural network with $N$ parties. The bold terms denote encrypted values and $\bm{W}_{j,i}^k$ represents the weight matrix of the $j^{th}$ layer, at iteration $k$, of the party $P_i$. When there is no ambiguity or when we refer to the global model, we replace the sub-index $i$ with $\cdot$ and denote weights by $\bm{W}_{j,\cdot}^k$. Similarly, we denote the local gradients at party $P_i$ by $\bm{\nabla W}_{j,i}^k$, for each network layer $j$ and iteration $k$. Throughout the paper, the $n^{th}$ row of a matrix that belongs to the $i^{th}$ party is represented by $X_i[n]$ and its encoded (packed) version as $\bar{X_i}[n]$.

% We combine the distributed training of neural networks with the protection of the weights and the gradients communicated between parties by using distributed FHE scheme.

\descr{1) PREPARE:} In this offline phase, the parties collectively agree on the learning parameters: the number of hidden layers ($l$), the number of neurons ($h_j$) in each layer $j, \forall j \in \{1,2,...,l\}$, the learning rate ($\eta$), the number of global iterations ($m$), the activation functions to be used in each layer ($\varphi(\cdot)$) and their approximations (see  Section~\ref{sec:activationsOpt}), and the local batch size ($b$). Then, the parties generate their secret keys $sk_i$ and collectively generate the public key $pk$. Subsequently, they collectively normalize or standardize their input data with the secure aggregation protocol described in~\cite{Drynx}. Each $P_i$ encodes (packs) its input data samples $X_i$ and output labels $y_i$ (see Section~\ref{sec:AlternatingPacking}) as $\bar{X}_i,\bar{y}_i$. Finally, the root of the tree ($P_1$) initializes and encrypts the global weights.
    
\descrit{Weight Initialization.} To avoid exploding or vanishing gradients, we rely on commonly used techniques: (i) Xavier initialization for the sigmoid or tanh activated layers: $W_{j}=r\times h_{j-1}$ where $r$ is a random number sampled from a uniform distribution in the range $[-1,1]$~\cite{Glorot2010}, and (ii) He initialization~\cite{He2015} for ReLU activated layers, where the Xavier-initialized weights are multiplied twice by their variance. 
%effects are moved to approx. activation section
\vspace{-0.4em}
\begin{protocolalg}[t]
\caption{\textsf{Collective Training}}
\label{protocol:collectivetrain}
\begin{algorithmic}[1]
\Require $X_i, y_i$ for $i \in \{1,2,...,N\}$
\Ensure $\bm{W}_{1,\cdot}^m,\bm{W}_{2,\cdot}^m, \dots, \bm{W}_{\ell,\cdot}^m$ 
\Statex \textbf{PREPARE:}
\State \quad Parties collectively agree on $\ell,h_1,...,h_{\ell},\eta,\varphi(\cdot),m,b$
\State \quad Each $P_i$ generates $sk_i \leftarrow \textsf{SecKeyGen}(1^\lambda)$ 
\State \quad Parties collectively generate $pk \leftarrow \textsf{DKeyGen}(\{sk_i\})$ 
\State \quad Each $P_i$ encodes its local data as $\bar{X}_i$, $\bar{y}_i$
% , and generates masks
\State \quad $P_1$ initializes $\bm{W}_{1,\cdot}^0,\bm{W}_{2,\cdot}^0,...,\bm{W}_{\ell,\cdot}^0$
\For{$k=0 \rightarrow m-1$}
\Statex \quad \textbf{MAP:}
\State $P_1$ sends $\bm{W}_{1,\cdot}^k,\bm{W}_{2\cdot}^k,...,\bm{W}_{\ell,\cdot}^k$ down the tree
\State Each $P_i$ does:
%\ForAll{$P_{i} \,\, \text{ with } \,\, i \in \{1,2,...,N\}$}
\State \quad {Local Gradient Descent Computation: }
  \State \quad $\bm{\nabla W}_{1,i}^k, \bm{\nabla W}_{2,i}^k, \dots, \bm{\nabla W}_{\ell,i}^k$
% \EndFor
 \Statex \quad \textbf{COMBINE:} 
 \State \text{Parties collectively aggregate:} $\bm{\nabla W}_{1,\cdot}^k, \dots, \bm{\nabla W}_{\ell,\cdot}^k \leftarrow$
 \quad $\sum_{i=1}^{N} \bm{\nabla W}_{1,i}^k, \dots, \bm{\nabla W}_{\ell,i}^k$
 \State $P_1 \text{ obtains } \bm{\nabla W}_{1,\cdot}^k, \bm{\nabla W}_{2,\cdot}^k, \dots, \bm{\nabla W}_{\ell,\cdot}^k$
 \Statex \quad \textbf{REDUCE} (performed by $P_1$) :
 \For{$j=1 \rightarrow \ell$}
 \State  $\bm{W}_{j,\cdot}^{k+1} += \eta\frac{\bm{\nabla W}_{j,\cdot}^k}{b\times N}$
 \EndFor
  \EndFor
\end{algorithmic}
\end{protocolalg}
\setlength{\textfloatsep}{3pt}
% \vspace{-0.4em}

\descr{2) MAP:} The root of the tree $P_1$ communicates the current encrypted weights, to every other party for their local gradient descent (LGD) computation.

\descrit{LGD Computation:} Each $P_i$ performs $b$ forward and backward passes, i.e., to compute and aggregate the local gradients, by processing each sample of its respective batch. Protocol~\ref{gradprotocol} describes the LGD steps performed by each party $P_i$, at iteration $k$; $\odot$ represents an element-wise product and $\varphi'(\cdot)$ the derivative of an activation function. As the protocol refers to one local iteration for a specific party, we omit $k$ and $i$ from the weight and gradient indices. This protocol describes the typical operations for the forward pass and backpropagation using stochastic gradient descent (SGD) with the $L2$ loss  (see Section~\ref{sec:background}). We note that the operations in this protocol are performed over encrypted data.
%, but the cryptographic operations, such as $\textsf{Mul}_{\textsf{pt}}$($\cdot)$, are omitted for the sake of clarity; we introduce the detailed steps and operations in the next section. 
%We here note that the operations required for convolutional or downsampling layers are omitted in Protocol~\ref{gradprotocol}, for the sake of clarity.

% All values indicated in bold are encrypted and we describe the necessary operations that enable the efficient computations on the encrypted values in Protocol~\ref{gradprotocol} in Section \ref{sec:cryptoOptimization}. 

\descr{3) COMBINE:} In this phase, each party communicates its encrypted local gradients to their parent, and each parent homomorphically sums the received gradients with their own ones. At the end of this phase, the root of the tree ($P_1$) receives the globally aggregated gradients.

\descr{4) REDUCE:} $P_1$ updates the global model weights by using the averaged aggregated gradients. The averaging is done with respect to the global batch size $|B| = b \times N$, as described in Section~\ref{sec:distributedDL}.

\begin{protocolalg}[t]
\caption{\textsf{Local Gradient Descent (LGD) Computation}}
\label{gradprotocol}
\begin{algorithmic}[1]
\Require $\bm{ {W}}_{1,\cdot}^k,\bm{ {W}}_{2,\cdot}^k, \dots, \bm{ {W}}_{\ell,\cdot}^k$ 
\Ensure $\bm{ {\nabla W}}_{1,i}^k, \bm{ {\nabla W}}_{2,i}^k, \dots, \bm{ {\nabla W}}_{\ell,i}^k$. Note that $i$ and $k$ indices are omitted in this protocol.
\tikzmark{right}
\For{$t = 1 \rightarrow b$} \Comment{Forward Pass}
\State $L_0 = \bar{X}[t]$  
\For{$j = 1 \rightarrow \ell$}
\State $\bm{U}_j = \bm{L}_{j-1} \times \bm{W}_j$  %\textcolor{red}{$\;\;\;\;\;\;\rightarrow l_n - (d+1)$}
\State $\bm{L}_j = \varphi(\bm{U}_j)$
\EndFor
 \tikzmark{right}
  \State $\bm{E}_\ell = \bar{y}[t]-\bm{L_}\ell$ \Comment{Backpropagation}
    \State $\bm{E}_\ell = \varphi'(\bm{U}_\ell) \odot \bm{E}_\ell$
  \State $\bm{\nabla W}_{\ell} += \bm{L}_{{\ell-1}}\bm{^T} \times \bm{ E}_\ell$ 
  \For{$j = \ell-1 \rightarrow 1$}
  \State $\bm{E}_j = \bm{E}_{j+1}\times \bm{W}_{j+1}\bm{^T}$
  \State $\bm{E}_j = \varphi'(\bm{U}_j) \odot \bm{E}_j$
  \State $\bm{\nabla W}_{j} += \bm{L}_{j-1}\bm{^T} \times \bm{E}_j$
\EndFor 
 \EndFor
\end{algorithmic}
\end{protocolalg}
    
% \subsection{Model Convergence and Termination}
% \label{sec:ModelConverge}
\descrit{Training Termination:} In our system, we stop the learning process after a predefined number of epochs. We discuss other well-known techniques for the termination of NN training and how to integrate them in \sys in Appendix~\ref{sec:learnextensions}.

% \subsection{Model Release and Oblivious Inference}
% \label{sec:ModelRelease}

% perform oblivious inference defined, we adapt the in~\cite{spindle} to be able to evaluate the forward pass with encrypted weights ($\bm{W_{i}}$) and encrypted queriers' data ($\bm{ X_q}$) under the collective public key ($pk$).

\descrit{Prediction:} At the end of the training phase, the model is kept in an encrypted form such that no individual party or the querier can access the model weights. To enable oblivious inference, the querier encrypts its evaluation data $X_q$ with the parties' collective key. We note that an oblivious inference is equivalent to one forward pass (see Protocol~\ref{gradprotocol}), except that the first plaintext multiplication ($\textsf{Mul}_{\textsf{pt}}(\cdot)$) of $L_0$ with the first layer weights is substituted with a ciphertext multiplication ($\textsf{Mul}_{\textsf{ct}}(\cdot)$). At the end of the forward pass, the parties collectively re-encrypt the result with the querier's public key by using the key-switch functionality of the underlying MHE scheme. Thus, only the querier is able to decrypt the prediction results. Note that any party $P_i$ can perform the oblivious inference step, but the collaboration between all the parties is required to perform the distributed bootstrap and key-switch functionalities. 
 \vspace{-0.4em}
\section{Cryptographic Operations and Optimizations}\label{sec:systemDesign}

We first present the alternating packing (AP) approach that we use for packing the weight matrices of NNs (Section~\ref{sec:AlternatingPacking}). We then explain how we enable activation functions on encrypted values (Section~\ref{sec:activationsOpt}) and introduce the cryptographic building blocks and functions employed in \sys (Section~\ref{sec:cryptoBuildingBlock}), together with their execution pipeline and their complexity (Sections~\ref{sec:executionPipeline} and~\ref{sec:theoryEval}). Finally, we formulate a constrained optimization problem that depends on a cost function for choosing the parameters of the cryptoscheme (Section~\ref{sec:paramSelect}).

% Finally, we explain several parallelizations for \sys.
\vspace{-0.4em}
\subsection{Alternating Packing (AP) Approach}\label{sec:AlternatingPacking}

For the efficient computation of the forward pass and backpropagation described in Protocol~\ref{gradprotocol}, we rely on the packing capabilities of the cryptoscheme that enables Single Instruction, Multiple Data (SIMD) operations on ciphertexts. Packing enables coding a vector of values in a ciphertext and to parallelize the computations across its different slots, thus significantly improving the overall performance.

% We observe that the existing packing strategies for enabling secure distributed machine learning in \textsc{SPINDLE}~\cite{spindle}, which are row-based~\cite{kim2018logistic} and diagonal approach~\cite{helib},
Existing packing strategies that are commonly used for machine learning operations on encrypted data~\cite{spindle}, e.g., the row-based~\cite{kim2018logistic} or diagonal~\cite{helib}, require a high-number of rotations for the execution of the matrix-matrix multiplications and matrix transpose operations, performed during the forward and backward pass of the local gradient descent computation (see Protocol~\ref{gradprotocol}). We here remark that the number of rotations has a significant effect on the overall training time of a neural network on encrypted data, as they require costly key-switch operations (see Section~\ref{sec:theoryEval}). As an example, the diagonal approach scales linearly with the size of the weight matrices, when it is used for batch-learning of neural networks, due to the matrix transpose operations in the backpropagation.
% and they do not show how to apply transpose operations for the backpropagation (see Protocol~\ref{gradprotocol}). 
%  In addition, the transpose operations required for the backpropagation limit the efficient use of the diagonal-approach for batch-learning of neural networks when the whole mini-batch is diagonalized and multiplied with the weight matrix.
% When the weight matrices are encrypted, as it is the case for \sys, the diagonal-approach scales linearly with their size. 
We follow a different packing approach and process each batch sample one by one, making the execution embarrassingly parallelizable. This enables us to optimize the number of rotations, to eliminate the transpose operation applied to matrices in the backpropagation, and to scale logarithmically with the dimension and number of neurons in each layer.

% We pack all rows or columns of a layer's weight matrix into one ciphertext.
We propose an "alternating packing (AP) approach" that combines row-based and column-based packing, i.e., rows or columns of the matrix are vectorized and packed into one ciphertext. In particular, the weight matrix of every FC layer in the network is packed following the opposite approach from that used to pack the weights of the previous layer. With the AP approach, the number of rotations scales logarithmically with the dimension of the matrices, i.e., the number of features ($d$), and the number of hidden neurons in each layer ($h_i$). To enable this, we pad the matrices with zeros to get power-of-two dimensions. In addition, the AP approach reduces the cost of transforming the packing between two consecutive layers.

Protocol~\ref{protocolAlternating} describes a generic way for the initialization of encrypted weights for an $\ell$-layer MLP by $P_1$ and for the encoding of the input matrix ($X_i$) and labels ($y_i$) of each party $P_i$. It takes as inputs the NN parameters: the dimension of the data ($d$) that describes the shape of the input layer, the number of hidden neurons in the $j^{th}$ layer ($h_j$), and the number of outputs ($h_\ell$). We denote by $gap$ a vector of zeros, and by $|\cdot|$ the size of a vector or the number of rows of a matrix. $\textsf{Replicate}(v,k,gap)$ returns a vector that replicates $v$, $k$ times with a $gap$ in between each replica. $\textsf{Flatten}(W,gap,dim)$, flattens the rows or columns of a matrix $W$ into a vector and introduces $gap$ in between each row/column. If a vector is given as input to this function, it places $gap$ in between all of its indices. The argument $dim$ indicates flattening of rows ($\textrm{'}r\textrm{'}$) or columns ($\textrm{'}c\textrm{'}$) and $dim=\textrm{'}\cdot\textrm{'}$ for the case of vector inputs.

We observe that the rows (or columns) packed into one ciphertext, must be aligned with the rows (or columns) of the following layer for the next layer multiplications in the forward pass and for the alignment of multiplication operations in the backpropagation, as depicted in Table~\ref{table:representation} (e.g., see steps F1, F6, B3, B5, B6).
We enable this alignment by adding $gap$ between rows or columns and using rotations, described in the next section. Note that these steps correspond to the weight initialization and to the input preparation steps of the \textbf{PREPARE} (offline) phase.

\begin{protocolalg}[t]
\caption{\textsf{Alternating Packing (AP) Protocol}}
\label{protocolAlternating}
\begin{algorithmic}[1]
\Require $X_i,y_i,d,\{h_1,h_2,...,h_{\ell}\},\ell$
\Ensure $\bm{W}_{1,\cdot}^0, \bm{W}_{2,\cdot}^0, ...,\bm{W}_{\ell,\cdot}^0,\bar{X}_i, \bar{y}_i$ 
\For{$i=1 \rightarrow N$ \text{each} $P_i$}
\tikzmark{right}
\State Initialize $|gap|=max(h_1-d,0)$ \Comment{Input Preparation}
%\item \quad \quad \quad Initialize $|gap|=h_1-d$
%\EndIf
\For{$n=1 \rightarrow |X_i|$}
\State ${X}_i[n]=\textsf{Replicate}(X_i[n],h_1, gap)$
%\For {$k=0 \rightarrow h_1$}
%\State $v_k = append(v_k,X_i)$ 
%\State  $v_k = append(v_k, gap)$
%\EndFor
\State ${ \bar{X}_{i}[n]} = \textsf{Encode}(X_i[n])$ 
\EndFor
% \tikzmark{right}
 %\For{ $i=1 \rightarrow d$} \Comment{Labels Preparation}
 \tikzmark{right}
 \If{$\ell \%2 != 0$}\Comment{Labels Preparation}

\State Initialize  $|gap|=h_{\ell}$
\State $y_i=\textsf{Flatten}(y_i, gap, \textrm{'}\cdot\textrm{'})$
%\For{$k=0 \rightarrow h_\ell$}
%\State $v_k = append(v_k,y_i[k])$ 
%\State$v_k = append(v_k, gap)$
%\EndFor
%\Else
%\State $v_k = y_i$
\EndIf
\State ${ \bar{y}_{i}} = \textsf{Encode}(y_{i})$
\If{i==1} \Comment{$P_1$ performs Weight Initialization:}
\State Initialize $W_{1,\cdot}^0,W_{2,\cdot}^0,...,W_{\ell,\cdot}^0$
\For{$j=1 \rightarrow {\ell}$}
\tikzmark{right}
\If {$j\%2==0$} \Comment{Row Packing}
%\item \quad\quad \textbf{Row-Packing:}
\If{ $h_{j-2}>h_{j}$}
\State Initialize $|gap|=h_{j-2}-h_{j}$
\EndIf 
\State $W_{j,\cdot}^0=\textsf{Flatten}(W_{j,\cdot}^0, gap, \textrm{'}r\textrm{'})$
%\For{$k=0 \rightarrow h_i$}
%\State $v_k = append(v_k, r_k^i)$
%\State $v_k = append(v_k, gap)$
%\EndFor
\State $\bm{ { W}}_{j,\cdot}^0 = \textsf{Enc}(pk,W_{j,\cdot}^0)$
\tikzmark{right}
\Else \Comment{Column Packing}
%\item \quad\quad \textbf{Column-Packing:}
\If {$h_{j+1}>h_{j-1}$}
\State Initialize $|gap|=h_{j+1}-h_{j-1}$
\EndIf
\State $W_{j,\cdot}^0=\textsf{Flatten}(W_{j,\cdot}^0, gap, \textrm{'}c\textrm{'})$
 %\For{ $k=0 \rightarrow h_{i+1}$}
%\State $v_k = append(v_k, c_k^i)$
%\State $v_k = append(v_k, gap)$
%\EndFor
\State $\bm{ { W}}_{j,\cdot}^0 = \textsf{Enc}(pk,W_{j,\cdot}^0)$
\EndIf 
 \EndFor
\EndIf
\EndFor
\end{algorithmic}
\end{protocolalg}

\descr{Convolutional Layer Packing.} To optimize the SIMD operations for CV layers, we decompose the $n^{th}$ input sample $X_i[n]$ into $t$ smaller matrices according to the kernel size $h=f\times f$. We pack these decomposed flattened matrices into one ciphertext, with a $gap$ in between each matrix that is defined based on the number of neurons in the next layer ($h_2-h_1$), similarly to the AP approach. The weight matrix is then replicated $t$ times with the same $gap$ between each replica. 
%Protocol~\ref{protocolConv} in Appendix~\ref{sec:convAppendix} shows how to pack a CV layer weight matrix and the input data in case of a convolutional layer. 
If the next layer is another convolutional or downsampling layer, the $gap$ is not needed and the values in the slots are re-arranged during the training execution (see Section~\ref{sec:cryptoBuildingBlock}). Lastly, we introduce the average-pooling operation to our bootstrapping function (\textsf{DBootstrapALT}($\cdot$), see Section~\ref{sec:cryptoBuildingBlock}), and we re-arrange almost for free the slots for any CV layer that comes after average-pooling.

We note that high-depth kernels, i.e., layers with a large number of kernels, require a different packing optimization. In this case, we alternate row and column-based packing (similar to the AP approach), replicate the decomposed matrices, and pack all kernels in one ciphertext. This approach introduces $k$ multiplications in the \textbf{MAP} phase, where $k$ is the number of kernels in that layer, and comes with reduced communication overhead; the latter would be $k$ times larger for \textbf{COMBINE}, \textbf{MAP}, and \textsf{DBootstrap}($\cdot$), if the packing described in the previous paragraph was employed.

\descr{Downsampling (Pooling) Layers.} As there is no weight matrix for downsampling layers, they are not included in the offline packing phase. The cryptographic operations for pooling are described in Section~\ref{sec:executionPipeline}.
 \vspace{-0.4em}
\subsection{Approximated Activation Functions}
\label{sec:activationsOpt}
For the encrypted evaluation of non-linear activation functions, such as Sigmoid or Softmax, we use least-squares approximations and rely on the optimized polynomial evaluation that, as described in~\cite{spindle}, consumes $\ceil{\log(d_a+1)}$ levels for an approximation degree $d_a$. For the piece-wise function ReLU, we approximate the smooth approximation of ReLU, softplus (SmoothReLU), $\varphi(x)=\ln(1+e^x)$ with least-squares. Lastly, we use derivatives of the approximated functions. We discuss possible alternatives to these approximations in Appendix~\ref{sec:approxAlternative}.

To achieve better approximation with the lowest possible degree, we apply two approaches to keep the input range of the activation function as small as possible, by using (i) different weight initialization techniques for different layers (i.e., Xavier or He initialization), and (ii) collective normalization of the data by sharing and collectively aggregating statistics on each party's local data in a privacy-preserving way~\cite{Drynx}. Finally, the interval and the degree of the approximations are chosen based on the heuristics on the data distribution in a privacy-preserving way, as described in~\cite{cryptoDL}.

%We observe that the combination of data normalization with the aforementioned weight initialization techniques allows us to control the range of the activation function outputs. As such, we are able to approximate them with a low approximation degree and high accuracy (see Section~\ref{sec:activationsOpt}.
\begin{table*}[h!]
%\hspace{-1em}
\small
\begin{tabular}{lll}
\toprule
%\textbf{\textsc{PREPARE:} }&  &   \\ \cmidrule(l{0.5em}r{2em}){1-2}
& \textbf{AP Approach} & \textbf{Representation}\\
\midrule
\textbf{\textsc{PREPARE:} }\\ 1. \begin{tabular}[c]{@{}l@{}}  Each $P_i$ prepares\\  $X_i[n],y_i[n]$\end{tabular}  & \begin{tabular}[c]{@{}l@{}}Encode $X_i[n]$, $y_i[n] \rightarrow \bar{X}_i[n]$, $\bar{y}_i[n] $\\$\bar{L}_0=\bar{X}_i[n]$\end{tabular} & \hspace{-2.5em} \raisebox{30pt}{\multirow{44}{*}{\includegraphics[width=0.455\textwidth, height =0.68\textheight ]{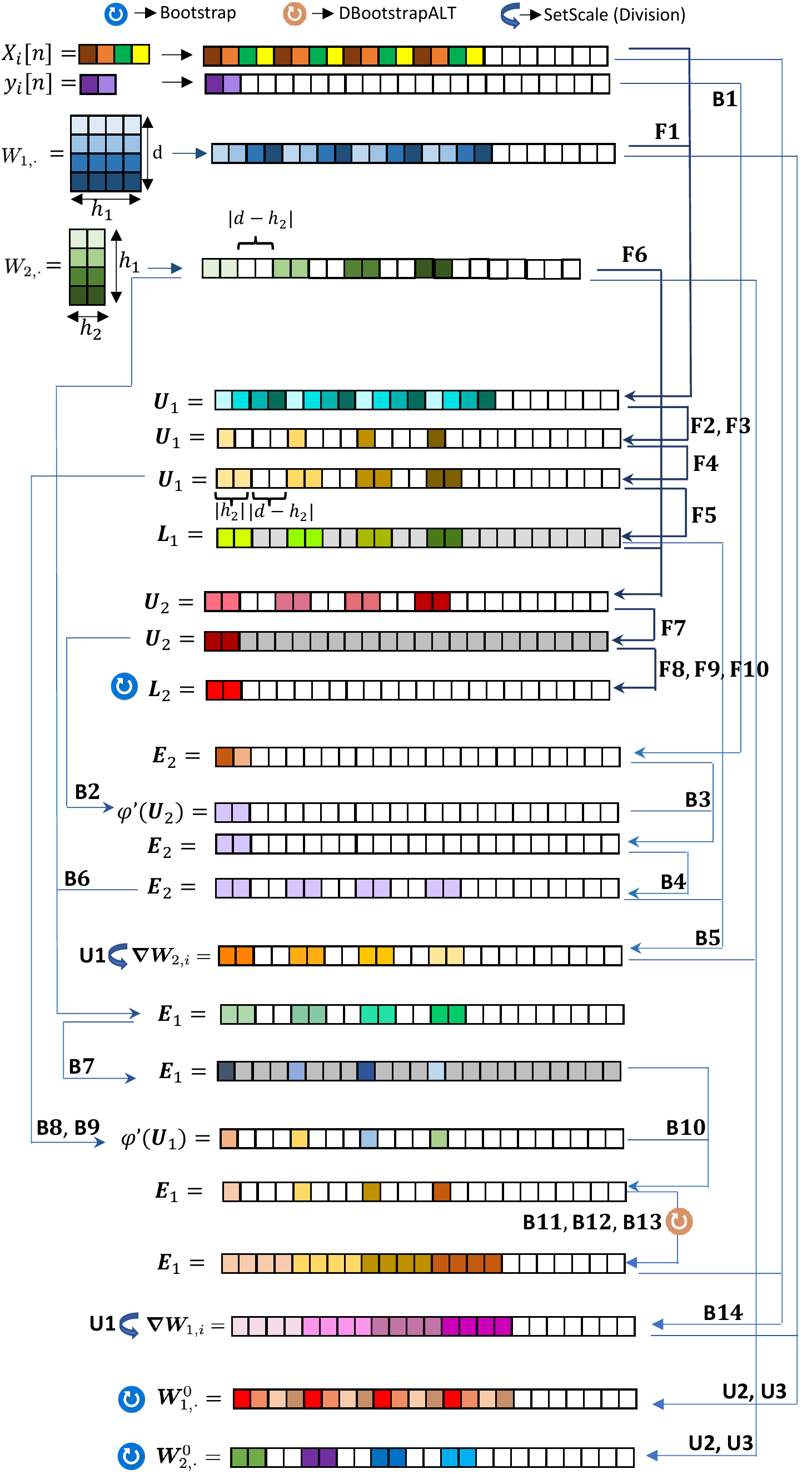}}} \\ \cmidrule(l{0.5em}r{2em}){1-2}
2. $P_1$ initializes $W_{1,\cdot}$ & \begin{tabular}[c]{@{}l@{}} Vectorize columns, pack with $|gap|=0$\\ $\bm{W}_{1,\cdot}^0=\textsf{Flatten}(W_{1,\cdot}^0, gap, \textrm{'}c\textrm{'})$\end{tabular}   &  \\ \cmidrule(l{0.5em}r{2em}){1-2}
3. $P_1$ initializes $W_{2,\cdot}$ & \begin{tabular}[c]{@{}l@{}l@{}}Vectorize rows, pack with $|gap|=d-h_\ell$\\$\bm{W}_{2,\cdot}^0=\textsf{Flatten}(W_{2,\cdot}^0, gap, \textrm{'}r\textrm{'})$\end{tabular} & \\ \cmidrule(l{0.5em}r{2em}){1-2}
4. \begin{tabular}[c]{@{}l@{}} Each $P_i$ generates\\  masks $\bar{m_1},\bar{m_2}$\end{tabular}  & \begin{tabular}[c]{@{}l@{}l@{}}$\bar{m_1} = [1,0,0,0,1,0,0,0,1,0,0,0,1,...]$\\ $\bar{m_2} = [1,1,0,0,0,0,0,0,0,0,0,0,0,...]$ \end{tabular} & \\ \cmidrule(l{0.5em}r{2em}){1-2}
%\textbf{Forward Pass:} &  &   \\

\begin{tabular}[c]{@{}l@{}}
\textbf{Forward Pass (Each $P_i$):}\\
\\
1. $\bm{U}_1 = \bar{L}_0 \times \bm{W}_{1,\cdot}$ \\
2. $\bm{L}_1 = \varphi(\bm{U}_1)$ 
\end{tabular}
  & \begin{tabular}[c]{@{}l@{}}F1. $\bm{U}_1 = \textsf{Mul}_{\textsf{pt}} (\bar{L}_0,\bm{W}_{1,\cdot}$), $\textsf{Res}(\bm{U}_1)$\\ F2. $\bm{U}_1= \textsf{RIS}(\bm{U}_1,1,d) $\\ F3.
  $\bm{U}_1 = \textsf{Mul}_{\textsf{pt}}(\bm{U}_1,\bar{m_1})$,  $\textsf{Res}(\bm{U}_1)$ \\F4. $\bm{U}_1= \textsf{RR}(\bm{U}_1,1,h_\ell)$ \\F5. $\bm{L}_1 = \varphi(\bm{U}_1)$ \\  \end{tabular} &  \\ 
\cmidrule(l{.5em}r{2em}){1-2}
\begin{tabular}[c]{@{}l@{}}
3. $\bm{U}_2 = \bm{L}_1 \times \bm{W}_{2,\cdot}$ \\ 
4. $\bm{L}_2 = \varphi(\bm{U}_2)$ 
\end{tabular} &
\begin{tabular}[c]{@{}l@{}}F6. $\bm{U}_2 = \textsf{Mul}_{\textsf{ct}} (\bm{L}_1,\bm{W}_{2,\cdot}$), $\textsf{Res}(\bm{L}_2)$\\ F7. $\bm{U}_2= \textsf{RIS}(\bm{U}_1,d,h_1)$ \\F8. $\bm{ L}_2 = \textsf{Mul}_{\textsf{pt}} (\bm{ L}_2,\bar{m_2}$), $\textsf{Res}(\bm{ L}_2)$ \\ F9. $\textsf{DBootstrap}({ \bm{U}_{2}})$\\F10. $\bm{ L}_2=\varphi(\bm{U}_2)$ \\ \end{tabular} &   \\
\cmidrule(l{0.5em}r{2em}){1-2}
\textbf{Backpropagation (Each $P_i$):}\\
1. $\bm{E}_2 = \bar{y}_i[n]-\bm{ L}_2$ & B1. $\bm{E}_2 = \textsf{Sub}(\bar{y}_i[n],\bm{ L}_2)$ &  \\
\cmidrule(l{0.5em}r{2em}){1-2}
2. $\bm{E}_2 = (\varphi'(\bm{U}_2)) \odot \bm{E}_2$ & \begin{tabular}[c]{@{}l@{}}B2. $\bm{{d}} = \varphi'(\bm{U}_2)$\\ B3. $\bm{E}_2=\textsf{Mul}_{\textsf{ct}}(\bm{E}_2,\bm{ {d}}$), $\textsf{Res}(\bm{E}_2)$\\ B4.  $\bm{E_2}= \textsf{RR}(\bm{E_2},d,h_1)$\end{tabular} &\\
\cmidrule(l{0.5em}r{2em}){1-2}
3. $\bm{\nabla W}_{2,i} = \bm{L}\bm{^T}_1 \times \bm{E}_l$ & \begin{tabular}[c]{@{}l@{}}B5. $\bm{\nabla W}_{2,i}=\textsf{Mul}_{\textsf{ct}}(\bm{L}_{1},\bm{E}_2$), $\textsf{Res}(\bm {\nabla W}_{2,i})$\\ \end{tabular} &   \\
\cmidrule(l{0.5em}r{2em}){1-2}
4. $\bm{E}_1 = \bm{E}_2\times \bm{W}_{2,\cdot}\bm{^T}$ & \begin{tabular}[c]{@{}l@{}}B6. $\bm{E}_1 =\textsf{Mul}_{\textsf{ct}}(\bm{E}_2, \bm{W}_{2,i})$, $\textsf{Res}(\bm{E}_1$)\\ B7. $\bm{E}_1= \textsf{RIS}(\bm{E}_1,1,h_\ell)$ \end{tabular} &  \\
\cmidrule(l{0.5em}r{2em}){1-2}
5. $\bm{E}_1 = (\varphi'(\bm{U}_1) \odot \bm{E}_1)$ & \begin{tabular}[c]{@{}l@{}}B8. $\bm{{d}} = \varphi'(\bm{U}_1)$\\B9. $\bm{ {d}} =\textsf{Mul}_{\textsf{pt}}(\bm{ {d}},\bar{m_1})$ \\ B10. $\bm{E}_1 =\textsf{Mul}_{\textsf{ct}}(\bm{E}_1,\bm{ {d}})$, $\textsf{Res}(\bm{E}_1)$\\\textcolor{orange}{B11. $\textsf{DBootstrapALT}(\bm{E}_1$)}\end{tabular} &\\
\cmidrule(l{0.5em}r{2em}){1-2}
6. $\bm {\nabla W}_{1,i} = \bar{L}_{0}^T \times \bm{E}_1$ & \begin{tabular}[c]{@{}l@{}} \textcolor{orange}{B12. $\bm{E}_{1} = \textsf{Mul}_{\textsf{pt}}(\bm{E}_{1},\bar{m_1}) $, $\textsf{Res}(\bm{ { E_{1}}})$}\\ \textcolor{orange}{B13. $\bm{E_1}= \textsf{RR}(\bm{E}_1,1,d)$ }\\ B14.  $\bm {\nabla W}_{1,i} = \textsf{Mul}_{\textsf{pt}}(\bar{L}_0,\bm{E}_1 $), $\textsf{Res}(\bm {\nabla W}_{1,i})$\\\end{tabular} &  \\
\cmidrule(l{0.5em}r{2em}){1-2}
%\textbf{Update (at root):} &  &  &   \\
\begin{tabular}[c]{@{}l@{}}
\textbf{Update (at $P_1$):}\\
1. $\bm{W}_{j,\cdot} += \eta \frac{{}{\bm{ {\nabla W_{j,\cdot}}}}}{b\times N} $\\ \quad $\forall j \in \{1,2,..,l\}$ 
\end{tabular}
&\begin{tabular}[c]{@{}l@{}} U1. $\textsf{SetScale}(\bm{\nabla W}_{j,\cdot}, S_{\bm{\nabla W}_{j,\cdot}} \times (b\times N))/\eta)$ \\ U2. $ \bm{W}_{j,\cdot}= \textsf{Add}(\bm{W}_{j,\cdot},\bm{\nabla W}_{j,\cdot}) $\\ U3. $\textsf{DBootstrap}(\bm{ W}_{j,\cdot})$ \end{tabular}
& \\
 \midrule
\bottomrule 
\end{tabular}
\captionsetup{width=\linewidth}
\caption{Execution pipeline for a 2-layer MLP network with Alternating Packing (AP). \textcolor{orange}{Orange steps} indicate the operations introduced to $\textsf{DBootstrapALT}(\cdot)$.}
\label{table:representation}
\vspace{-1.8em}
\end{table*}
\vspace{-0.4em}
\subsection{Cryptographic Building Blocks}
\label{sec:cryptoBuildingBlock}
We present each cryptographic function that we employ to enable the privacy-preserving training of NNs with $N$ parties. We also discuss the optimizations employed to avoid costly transpose operations in the encrypted domain. \\
\descr{Rotations.} As we rely on packing capabilities, computation of the inner-sum of vector-matrix multiplications and transpose operation implies a restructuring of the vectors, that can only be achieved by applying slot rotations. Throughout the paper, we use two types of rotation functions: (i) Rotate For Inner Sum ($\textsf{RIS}(\bm{c},p,s)$) is used to compute the inner-sum of a packed vector $\bm{c}$ by homomorphically rotating it to the left with $\textsf{RotL}(\bm{c},p)$ and by adding it to itself iteratively $\log_{2}(s)$ times, and (ii) Rotate For Replication ($\textsf{RR}(\bm{c},p,s)$) replicates the values in the slots of a ciphertext by rotating the ciphertext to the right with $\textsf{RotR}(\bm{c},p)$ and by adding to itself, iteratively $\log_{2}(s)$ times. For both functions, $p$ is multiplied by two
at each iteration, thus both yield $\log_2(s)$ rotations.
As rotations are costly cryptographic functions (see Table~\ref{table:theoryAnalysis}), and the matrix operations required for NN training require a considerable amount of rotations, we minimize the number of executed rotations by leveraging a modified bootstrapping operation, that automatically performs some of the required rotations.

\descr{Distributed Bootstrapping with Arbitrary Linear Transformations.} To execute the high-depth homomorphic operations required for training NNs, bootstrapping is required several times to refresh a ciphertext, depending on the initial level $L$. In \sys, we use a distributed version of bootstrapping~\cite{mouchet2019distributedbfv}, as it is several orders of magnitude more efficient than the traditional centralized bootstrapping. Then we modify it, to leverage on the interaction to automatically perform some of the rotations, or pooling operations, embedded as transforms in the bootstrapping.

%We here propose a protocol that extends the $\textsf{DBootstrap}(\cdot)$ operation to perform arbitrary transformations during bootstrapping, to reduce the cost of rotations and the cost of pooling layers. 

%Mouchet et al.~\cite{mouchet2019distributedbfv} introduce a protocol for the collective refresh of a Brakerski/Fan-Vercauteren (BFV)~\cite{bfv} ciphertext. This protocol is several orders of magnitude more efficient than the traditional centralized bootstrapping that would require a circuit of depth 15 to 20 and large parameters to enable the computation of a circuit having this depth.
%details of bootstrap
%high-level explanation of the DbootstrapALT
Mouchet et al. replace the expensive bootstrap circuit by a one-round protocol where the parties collectively switch a Brakerski/Fan-Vercauteren (BFV)~\cite{bfv} ciphertext to secret-shares in $\mathbb{Z}^{\mathcal{N}}_{t}$.
Since the BFV encoding and decoding algorithms are linear transformations, they can be performed without interaction on a secret-shared plaintext. %Although the encoding and decoding algorithms of the CKKS scheme are also linear transformations, 
Despite its properties, the protocol that Mouchet et al. propose for the BFV scheme cannot be directly applied to CKKS, as CKKS is a \textit{leveled} scheme):
The re-encryption process extends the residue number system (RNS) basis from $Q_{\ell}$ to $Q_{L}$. %No modular reduction of the masks should happen in $Q_{\ell}$, or the re-encryption of the message in $Q_{L}$ will not be a correct encryption.
Modular reduction of the masks in $Q_{\ell}$ will result in an incorrect encryption. 
Our solution to this limitation is to collectively switch the ciphertext to a secret-shared plaintext with statistical indistinguishability.

We define this protocol as $\textsf{DBootstrapALT}(\cdot)$ (Protocol~\ref{algorithm:cbootwithpermute}) that takes as inputs a ciphertext $\bm{c}_{pk}$ at level $\ell$ encrypting a message $msg$ and returns a ciphertext $\bm{c}'_{pk}$ at level $L$ encrypting $\phi(msg)$, where $\phi(\cdot)$ is a linear transformation over the field of complex numbers. We denote by $||a||$ the infinity norm of the vector or polynomial $a$. As the security of the RLWE is based on computational indistinguishability, switching to the secret-shared domain does not hinder security. We refer to Appendix~\ref{sec:DBootstrapAltDetails} for technical details and the security proof of our protocol.

\begin{protocolalg}[t]
\caption{$\textsf{DBootstrapALT}(\cdot)$}
\label{algorithm:cbootwithpermute}
\begin{algorithmic}[1]
\Require $\bm{c}_{pk} = (c_{0}, c_{1}) \in R_{Q_{\ell}}^{2}$ encrypting $msg$, $\lambda$ a security parameter, $\phi(\cdot)$ a linear transformation over the field of complex numbers, $a$ a common reference polynomial, $s_{i}$ the secret-key of each party $P_i$, $\chi_{err}$ a distribution over $R$, where each coefficient is independently sampled from Gaussian distribution with the standard deviation $\sigma = 3.2$, and bound $\lfloor{6\sigma}\rfloor$.
\Statex \textbf{Constraints:} $Q_{\ell} > (N+1) \cdot ||msg|| \cdot 2^{\lambda}$.
\Ensure $\bm{c}'_{pk} = (c'_{0}, c'_{1})\in R^{2}_{Q_{L}}$
\ForAll{$P_{i}$}
    \State $M_{i} \leftarrow R_{||msg|| \cdot 2^{\lambda}}$, $e_{0, i}, e_{1, i} \leftarrow \chi_{err}$
    \State $M'_{i} \leftarrow \textsf{Encode}(\phi(\textsf{Decode}(M_{i})))$
    \State $h_{0, i} \leftarrow s_{i}c_{1}+M_{i}+e_{0, i} \mod \ Q_{\ell}$
    \State $h_{1, i} \leftarrow -s_{i}a-M_{i}+e_{1, i} \mod \ Q_{L}$
\EndFor
\State $h_{0} \leftarrow \sum h_{0, i}, h_{1} \leftarrow \sum h_{1, i}$
\State $c'_{0} \leftarrow \textsf{Encode}(\phi(\textsf{Decode}(c_{0} + h_{0} \mod Q_{\ell})))$
\State \Return $\bm{c}'_{pk} = (c'_{0} + h_{1} \mod \ Q_{L}, a) \in R^{2}_{Q_{L}}$

\end{algorithmic}
\end{protocolalg}

\descr{Optimization of the Vector-Transpose Matrix Product. }
The backpropagation step of the local gradient computation at each party requires several multiplications of a vector (or matrix) with the transposed vector (or matrix) (see Lines 11-13 of Protocol~\ref{gradprotocol}). The naïve multiplication of a vector $\bm{{v}}$ with a transposed weight matrix $\bm{{W\bm{^T}}}$ that is fully packed in one ciphertext, requires converting $\bm{W}$ of size $g \times k$, from column-packed to row-packed. This is equivalent to applying a permutation of the plaintext slots, that can be expressed with a plaintext matrix $W_{gk\times gk}$ and homomorphically computed by doing a matrix-vector multiplication. As a result, a naïve multiplication requires $\sqrt{g\times k}$ rotations followed by $\log_{2}(k)$ rotations to obtain the inner sum from the matrix-vector multiplication. We propose several approaches to reduce the number of rotations when computing the multiplication of a packed matrix (to be transposed) and a vector:
\textbf{(i)} For the mini-batch gradient descent, we do not perform operations on the batch matrix. Instead, we process each batch sample in parallel, because having separate vectors (instead of a matrix that is packed into one ciphertext) enables us to reorder them at a lower cost. This approach translates ${\ell}$ matrix transpose operations to be transposes in vectors (the transpose of the vectors representing each layer activations in the backpropagation, see Line 13, Protocol~\ref{gradprotocol}), \textbf{(ii)} Instead of taking the transpose of $\bm{W}$, we replicate the values in the vector that will be multiplied with the transposed matrix (for the operation in Line 11, Protocol~\ref{gradprotocol}), leveraging the gaps between slots with the AP approach. That is, for a vector $\bm{v}$ of size $k$ and the column-packed matrix $\bm{{W}}$ of size $g \times k$, $\bm{{v}}$ has the form $[a,0,0,0…,b,0,0,0,…,c,0,0,0,…]$ with at least $k$ zeros in between values (due to Protocol
~\ref{protocolAlternating}). Hence, any resulting ciphertext requiring the transpose of the matrix that will be subsequently multiplied, will also include gaps in between values. We apply $\textsf{RR}(\bm{v},1,k)$ that consumes $\log_2(k)$ rotations to generate $[a,a,a,...0...,b,b,b,..,0...,c,c,c,...,0,...]$. Finally, we compute the product $\bm{\mathcal{P}}=\textsf{Mul}_\textsf{ct}(\bm{v},\bm{{W}})$ and apply $\textsf{RIS}(\bm{\mathcal{P}},1,g)$ to get the inner sum with $\log_{2}(g)$ rotations, and \textbf{(iii)} We further optimize the performance by using $\textsf{DBootstrapALT}(\cdot)$ (Protocol~\ref{algorithm:cbootwithpermute}): If the ciphertext before the multiplication must be bootstrapped, we embed the $\log_{2}(k)$ rotations as a linear transformation performed during the bootstrapping.
\subsection{Execution Pipeline}\label{sec:executionPipeline}

Table~\ref{table:representation} depicts the pipeline of the operations for processing one sample in LGD computation for a 2-layer MLP. These steps can be extended to an $\ell$-layer MLP by following the same operations for multiple layers. The weights are encoded and encrypted using the AP approach, and the shape of the packed ciphertext for each step is shown in the representation column. Each forward and backward pass on a layer in the pipeline consumes one Rotate For Inner Sum ($\textsf{RIS}(\cdot)$) and one Rotate For Replication ($\textsf{RR}(\cdot)$) operation, except for the last layer, as the labels are prepared according to the shape of the $\ell^{th}$ layer output. In Table~\ref{table:representation}, we assume that the initial level $L=7$. When a bootstrapping function is followed by a masking (that is used to eliminate unnecessary values during multiplications) and/or several rotations, we perform these operations embedded as part of the distributed bootstrapping ($\textsf{DBootstrapALT}(\cdot)$) to minimize their computational cost. The steps highlighted in orange are the operations embedded in the $\textsf{DBootstrapALT}(\cdot)$. The complexity of each cryptographic function is analyzed in Section~\ref{sec:theoryEval}. \\
\descr{Convolutional Layers.} As we flatten, replicate, and pack the kernel in one ciphertext, a CV layer follows the exact same execution pipeline as a FC layer. However, the number of $\textsf{RIS}(\cdot)$ operations for a CV layer is smaller than for a FC layer. That is because the kernel size is usually smaller than the number of neurons in a FC layer. For a kernel of size $h=f\times f$, the inner sum is calculated by $log_2(f)$ rotations. Note that when a CV layer is followed by a FC layer, the output of the $i^{th}$ CV layer ($L_i$) already gives the flattened version of the matrix in one ciphertext. We apply $\textsf{RR}(\bm{L}_i,1,h_{i+1})$ for the preparation of the next layer multiplication. When a CV layer is followed by a pooling layer, however, the $\textsf{RR}(\cdot)$ operation is not needed, as the pooling layer requires a new arrangement of the slots of $\bm{L}_i$. We avoid this costly operation by passing $\bm{L}_i$ to $\textsf{DBootstrapALT}(\cdot)$, and by embedding both the pooling and its derivative in $\textsf{DBootstrapALT}(\cdot)$. \\
\descr{Pooling Layers.} In \sys, we evaluate our system based on average pooling as it is the most efficient type of pooling that can be evaluated under encryption~\cite{CryptoNets}. To do so, we exploit our modified collective bootstrapping to perform arbitrary linear transformations. Indeed, the average pooling is a linear function, and so is its derivative (note that this is not the case for the max pooling). Therefore, in the case of a CV layer followed by a pooling layer, we apply $\textsf{DBootstrapALT}(\cdot)$ and use it both to rearrange the slots and to compute the convolution of the average pooling in the forward pass and its derivative, that is used later in the backward pass. For a $h=f\times f$ kernel size, this saves $\log_2(h)$ rotations and additions ($\textsf{RIS}(\cdot)$) and one level if masking is needed. For max/min pooling, which are non-linear functions, we refer the reader to Appendix~\ref{sec:maxpool} and highlight that evaluating these functions by using encrypted arithmetic remains impractical due to the need of high-precision approximations.
\begin{table*}[t]
\footnotesize
\centering
\begin{tabular}{lllll}
\toprule
 & Computational Complexity & \#Levels Used & Communication & Rounds \\
 \midrule
FORWARD P. (\textsf{FP}) &$(\log_2(h_{i-1})+\log_2(h_{i+1})) \cdot \textsf{KS}+\textsf{Mul}_{\textsf{ct} }+\textsf{Mul}_{\textsf{pt} }+\varphi$ & $2+\lceil\log_{2}(d_a + 1)\rceil$&\multicolumn{1}{c}{$-$} &\multicolumn{1}{c}{$-$}  \\
\midrule
BACKWARD P. (\textsf{BP}) & $(\log_2(h_{i-1})+\log_{2}(h_{i+1}))\cdot \textsf{KS}+2\textsf{Mul}_{\textsf{ct}}+\textsf{Mul}_{\textsf{pt} }+\varphi'$ &$3+\lceil\log_{2}(d_a)\rceil$ &\multicolumn{1}{c}{$-$}&\multicolumn{1}{c}{$-$}    \\
\midrule
\textbf{MAP} & $\ell(\textsf{FP} + \textsf{BP})-2\log_{2}(h_\ell)$ &$\ell(5+\lceil\log_{2}(d_a+1)+\lceil\log_{2}(d_a)\rceil)$&$z(N-1)|c|$ &\multicolumn{1}{c}{$1/2$}   \\
\midrule
\textbf{COMBINE} & $-$ & $-$ & $z(N-1)|c|$&\multicolumn{1}{c}{$1/2$} \\
\midrule
\textbf{REDUCE} &$\ell(\textsf{Mul}_{\textsf{pt}}+\textsf{DB})$  &$-$& \multicolumn{1}{c}{$-$} &\multicolumn{1}{c}{$-$} \\
\midrule
DBootstrap (\textsf{DB}) & $N\log_{2}(N)(L + 1)+N\log_{2}(N)(L_{\bm{c}} + 1)$ & $-$& $  (N-1)|c|$ &\multicolumn{1}{c}{$1$}\\
\midrule
\textsf{Mul Plaintext} $(\textsf{Mul}_{\textsf{pt}})$ &$2N(L_{\bm{c}}+1)$ &$1$& \multicolumn{1}{c}{$-$}  &\multicolumn{1}{c}{$-$} \\
\midrule
\textsf{Mul Ciphertext} $(\textsf{Mul}_{\textsf{ct}})$ & $4N(L_{\bm{c}}+1) + \textsf{KS}$ & $1$ & \multicolumn{1}{c}{$-$}  &\multicolumn{1}{c}{$-$} \\
\midrule
Approx. Activation Function ($\varphi$) & $(2^{\kappa} + \textsf{m} - \kappa - 3 + \lceil(d_{a}+1)/2^{\kappa}\rceil) \cdot \textsf{Mul}_{\textsf{ct}}$ & $\lceil\log_{2}(d_{a}+1)\rceil$ & \multicolumn{1}{c}{$-$}  &\multicolumn{1}{c}{$-$} \\
\midrule
$\textsf{RIS}(\bm{c},p,s)$,  $\textsf{RR}(\bm{c},p,s)$ &$\log_{2}(s)\cdot\textsf{KS}$&$-$&\multicolumn{1}{c}{$-$}& \multicolumn{1}{c}{$-$}   \\
\midrule
 Key-switch (\textsf{KS}) & $ \mathcal{O}(\mathcal{N}\log_{2}(\mathcal{N})L_{\bm{c}}\beta)$ & $-$ & \multicolumn{1}{c}{$-$}  &\multicolumn{1}{c}{$-$} \\
 \bottomrule
\end{tabular}
\captionsetup{width=\linewidth}
\caption{Complexity analysis of \sys's building blocks. $\mathcal{N},\alpha,L, L_{\bm{c}},d_a$ stand for the cyclotomic ring size, the number of secondary moduli used during the key-switching, maximum level, current level, and the approximation degree, respectively. $\beta=\ceil{L_{\bm{c}} + 1/\alpha}$, $\textsf{m} = \lceil\log(d_{a}+1)\rceil$, $\kappa = \lfloor \textsf{m}/2 \rfloor$.} %$\textsf{RIS}(\bm{c},p,s)$, $\textsf{RR}(\bm{c},p,s)$ are the RotateInnerSum, RotateReplication functions with $log_2(s)$ number of rotations.}
\label{table:theoryAnalysis}
%\vspace{-1.9em}
\end{table*}

%\vspace{-0.36em}
\subsection{Complexity Analysis}
\label{sec:theoryEval}
Table~\ref{table:theoryAnalysis} displays the communication and \textit{worst-case} computational complexity of \sys's building blocks. This includes the MHE primitives, thus facilitating the discussion on the parameter selection in the following section. We define the complexity in terms of key-switch $\textsf{KS}(\cdot)$ operations and recall that this is a different operation than $\textsf{DKeySwitch}(\cdot)$, as explained in Section
~\ref{sec:distributedHomomorphic}. We note that $\textsf{KS}(\cdot)$ and $\textsf{DBootstrap}(\cdot)$ are 2 orders of magnitude slower than an addition operation, rendering the complexity of an addition negligible.

We observe that \sys's communication complexity depends solely on the number of parties ($N$), the number of total ciphertexts sent in each global iteration ($z$), and the size of one ciphertext ($|\bm{c}|$). The building blocks that do not require communication are indicated as $-$.

In Table~\ref{table:theoryAnalysis}, forward and backward passes represent the per-layer complexity for FC layers, so they are an \textit{overestimate} for CV layers. Note that the number of multiplications differs in a forward pass and a backward pass, depending on the packing scheme, e.g., if the current layer is row-packed, it requires 1 less $\textsf{Mul}_{\textsf{ct}}(\cdot)$ in the backward pass, and we have 1 less $\textsf{Mul}_{\textsf{pt}}(\cdot)$ in several layers, depending on the masking requirements. Furthermore, the last layer of forward pass and the first layer of backpropagation take 1 less $\textsf{RR}(\cdot)$ operation that we gain from packing the labels in the offline phase, depending on the NN structure (see Protocol~\ref{protocolAlternating}). Hence, we save $2\log_2(h_\ell)$ rotations per one LGD computation. 

In the \textbf{MAP} phase, we provide the complexity of the local computations per $P_i$, depending on the total number of layers $\ell$. In the \textbf{COMBINE} phase, each $P_i$ performs an addition for the collective aggregation of the gradients in which the complexity is negligible. To update the weights, \textbf{REDUCE}  is done by one party ($P_1$) and divisions do not consume levels when performed with $\textsf{SetScale}(\cdot)$. The complexity of an activation function ($\varphi(\cdot)$) depends on the approximation degree $d_a$. We note that the derivative of the activation function ($\varphi'(\cdot)$) has the same complexity as $\varphi(\cdot)$ with degree $d_a-1$.

For the cryptographic primitives represented in Table \ref{table:theoryAnalysis}, we rely on the CKKS variant of the MHE cryptosystem in \cite{mouchet2019distributedbfv}, and we report the dominating terms. The distributed bootstrapping takes 1 round of communication and the size of the communication scales with the number of parties ($N$) and the size of the ciphertext (see~\cite{mouchet2019distributedbfv} for details). 
%For a ciphertext $c = (c_0, c_1)$, the DP sends $c_1$, i.e., half of the ciphertext, to all parties, each party performs the local operations and sends the shares back that are approximately half of the ciphertext size. 
\vspace{-0.4em}
\subsection{Parameter Selection}
\label{sec:paramSelect}
We first discuss several details to optimize the number of $\textsf{Res}(\cdot)$ operations and give a cost function which is computed by the complexities of each functionality presented in Table~\ref{table:theoryAnalysis}. Finally, relying on this cost function we formulate an optimization problem for choosing \sys' parameters.

As discussed in Section~\ref{sec:distributedHomomorphic}, we assume that each multiplication is followed by a $\textsf{Res}(\cdot)$ operation. The number of total rescaling operations, however, can be further reduced by checking the scale of the ciphertext. When the initial scale $S$ is chosen such that $Q/S=r$ for a ciphertext modulus $Q$, the ciphertext is rescaled after $r$ consecutive multiplications. This reduces the level consumption and is integrated into our cost function hereinafter.

\descr{Cryptographic Parameters Optimization.} We define the overall complexity of an $\ell$-layer MLP aiming to formulate a constrained optimization problem for choosing the cryptographic parameters. We first introduce the total number of bootstrapping operations ($\mathcal{B}$) required in one forward and backward pass, depending on the multiplicative depth as
\vspace{0.6em}
\begin{equation*}
	\mathcal{B}=\frac{\ell(5+\lceil\log_{2}(d_a+1)+\lceil\log_{2}(d_a)\rceil)}{(L-\tau)r},
%	\label{eq:numBootstrap}
\end{equation*}
%\vspace{0.6em}
where $r=Q/S$, for a ciphertext modulus $Q$ and an initial scale $S$. The number of total bootstrapping operations is calculated by the total number of consumed levels (numerator), the level requiring a bootstrap ($L-\tau$) and $r$ which denotes how many consecutive multiplications are allowed before rescaling (denominator). The initial level of a fresh ciphertext $L$ has an effect on the design of the protocols, as the ciphertext should be bootstrapped before the level $L_{\bm{c}}$ reaches a number ($L-\tau$) that is close to zero, where $\tau$ depends on the security parameters. For a cyclotomic ring size $\mathcal{N}$, the initial level of a ciphertext $L$, and for the fixed neural network parameters such as the number of layers $\ell$, the number of neurons in each layer ${h_1,h_2,...,h_\ell}$, and for the number of global iterations $m$, the overall complexity is defined as
\begin{align}
%\begin{split}
  C(\mathcal{N},L)= m({\sum_{i=1}^{\ell} \{(2\log_2(h_{i-1})+\log_2(h_{i+1})) \cdot \textsf{KS}
  +3\textsf{Mul}_{\textsf{ct} }+2\textsf{Mul}_{\textsf{pt}}
  +\varphi+\varphi'}\}-2\log_2(h_{\ell})
  +\mathcal{B}\cdot \textsf{DB}).
%\end{split}
\end{align}
%\vspace{0.6em}

Note that the complexity of each $\textsf{KS}(\cdot)$ operation depends on the level of the ciphertext that it is performed on (see Table~\ref{table:theoryAnalysis}), but we use the initial level $L$ in the cost function for the sake of clarity. The complexity of $\textsf{Mul}_{\textsf{ct}}$,$\textsf{Mul}_{\textsf{pt}}$,$\textsf{DB}$, and $\textsf{KS}$ is defined in Table~\ref{table:theoryAnalysis}. Then, the optimization problem for a fixed scale (precision) $S$ and a security level $\lambda$, which defines the security parameters, can be formulated as
\begin{align}
\label{constrainedOpt}
%\begin{split}
   &\min_{\mathcal{N},L} C(\mathcal{N},L)\\\nonumber
  \text{subject to } &mc = \{q_1,...,q_L\};\,%\\
   L = |mc|;\,%\\
    Q=\prod_{i=1}^{L} q_i;\,%\\
    Q = \textit{k}S, \text{ } \textit{k} \in \mathbb{R}^+;\\\nonumber
    &Q_{L-\tau} > 2^\lambda |plaintext|N;\,%\\
   \mathcal{N} \leftarrow \text{postQsec}(Q,\lambda),
%\end{split}
\end{align}
 where \text{postQsec}($Q,L,\lambda$) gives the necessary cyclotomic ring size $\mathcal{N}$, depending on the ciphertext modulus ($Q$) and on the desired security level ($\lambda$), according to the homomorphic encryption standard whitepaper~\cite{HEStandardPaper}. 
%To ease the process of deciding where to bootstrap and to introduce several rotations to bootstrap, we first set our level $L$ to $6$ or $7$. Next, we chose the scale $<\bm{ct}>.S$ depending on the minimum precision that is needed for input data and weights. The ring dimension $N$ is then chosen based on the learning parameters, i.e., the sized of each weight matrix and the dimension of the data. As $N/2$ is the available slots in a ciphertext, it represents the maximum number of values that can be packed in one ciphertext. At this stage, we introduce two approaches to choose $N$ for our proposed AP approach:
Eq.~\eqref{constrainedOpt} gives the optimal $\mathcal{N}$ and $L$ for a given NN structure. We then pack each weight matrix into one ciphertext. It is worth mentioning that the solution might give an $\mathcal{N}$ that has fewer slots than the required number to pack the big weight matrices in the neural network. In this case, we use a multi-cipher approach where we pack the weight matrix using more than one ciphertext and do the operations in parallel. \\
%\begin{itemize}
%\item \textbf{One-cipher approach:} We find the largest weight matrix to be packed and calculate the necessary number of slots to fit this matrix into one ciphertext, including the gaps introduced in between rows or columns. When optimization problem~\ref{constrainedOpt} finds optimal $\mathcal{N}$ such that the number of slots fit into one ciphertext, we follow one-cipher approach for the evaluation.
\textbf{Multi-cipher Approach.} In the case of a big weight matrix, we divide the flattened weight vector into multiple ciphertexts and carry out the neural network operations on several ciphertexts in parallel. E.g., for a weight matrix of size $1,024\times64$ and $\mathcal{N}/2=4,096$ slots, we divide the weight matrix into $1,024\times64/4,096 = 16$ ciphers.
%\end{itemize}

\section{Security Analysis}
\label{sec:securityAnalysis}

We demonstrate that \sys achieves the Data and Model Confidentiality properties defined in Section~\ref{sec:objectives}, under a passive-adversary model with up to $N-1$ colluding parties. We follow the real/ideal world simulation paradigm~\cite{lindell2017simulate} for the confidentiality proofs.

The semantic security of the CKKS scheme is based on the hardness of the decisional RLWE problem~\cite{cheon2017homomorphic,Lyubashevsky2010,Lindner2011}. The achieved practical bit-security against state-of-the-art attacks can be computed using Albrecht's LWE-Estimator~\cite{HEStandardPaper,Albrecht2015OnTC}.
%According to Lemma 1 in~\cite{cheon2017homomorphic}, a CKKS ciphertext generated with parameters ($\mathcal{N},Q_L,S$) that ensures a post-quantum security level of $\lambda$, is a valid encryption (i.e., indistinguishable from random data) due to the semantic security of RLWE-based encryptions. 
%Moreover, following Lemmas 2, 3, and 4 in~\cite{cheon2017homomorphic}, any local homomorphic operation on a CKKS ciphertext, e.g., addition, multiplication, or rescaling, yields a valid encryption of the result of the operation.
The security of the used distributed cryptographic protocols, i.e., $\textsf{DKeyGen}(\cdot)$ and $\textsf{DKeySwitch}(\cdot)$, relies on the proofs by Mouchet et al.~\cite{mouchet2019distributedbfv}. They show that these protocols are secure in a passive-adversary model with up to $N-1$ colluding parties, under the assumption that the underlying RLWE problem is hard~\cite{mouchet2019distributedbfv}. The security of $\textsf{DBootstrap}(\cdot)$, and its variant $\textsf{DBootstrapALT}(\cdot)$ is based on Lemma~\ref{lemmaDBoot} which we state and prove in Appendix~\ref{sec:DBootstrapAltDetails}. 

%For the following propositions, we assume that the parties participating in \sys have established their common public key $pk$ following the $\textsf{DKeyGen}(
%\cdot)$ protocol.

% We rely on the proofs given by Mouchet et al.~\cite{mouchet2019distributedbfv} for the security of the distributed cryptographic protocols. 
%During the training, the  interaction between the parties comes from the $\textsf{DKeyGen}(\cdot)$, $\textsf{DKeySwitch}(\cdot)$, $\textsf{Dbootstrap}(\cdot)$ and its variant, $\textsf{DbootstrapALT}(\cdot)$.
% Mouchet et al. show that $\textsf{DKeyGen}(\cdot)$, $\textsf{DKeySwitch}(\cdot)$, and $\textsf{DBootstrap}(\cdot)$ protocols   For the security proofs below, assume that a simulator has its own public key and secret key that ensure post-quantum security under CKKS scheme.

% \label{lemma1}
% Assume that any intermediate value communicated between the parties is encrypted under CKKS with the scheme parameters ($\mathcal{N},Q_L,S$) that ensure post-quantum security with $\lambda$ security-level. Then, \sys achieves (a) privacy-preserving training and prediction on NNs by satisfying \emph{Data Confidentiality} of each $P_i$ and the querier's data; under a passive-adversary setting with up to $N-1$ collusions.
% \label{lemma1}

% The \textbf{PREPARE} phase of \sys applies $\textsf{DKeyGen}(\cdot)$ to generate the collective public-key $pk$ and respective secret-key $sk_i$ of each $P_i$ that ensures $\lambda$ security-level with CKKS scheme. 
\descr{Remark 1.} Any encryption broadcast to the network in Protocol~\ref{protocol:collectivetrain} is re-randomized to avoid leakage about parties' confidential data by two consecutive broadcasts. We omit this operation in Protocol~\ref{protocol:collectivetrain} for clarity.

\begin{proposition}
\label{lemma1}
Assume that \sys's encryptions are generated using the CKKS cryptosystem with parameters ($\mathcal{N}, Q_L, S$) ensuring a post-quantum security level of $\lambda$. Given a passive adversary corrupting at most $N-1$ parties, \sys achieves \emph{Data and Model Confidentiality} during training.
\end{proposition}
%\vspace{-0.3em}
\descr{Proof (Sketch).} Let us assume a real-world simulator $\mathcal{S}_t$ that simulates the view of a computationally-bounded adversary corrupting $N-1$ parties, as such having access to the inputs and outputs of $N-1$ parties. 
%Without loss of generality, let us assume one training iteration in \sys. 
%During the \textbf{MAP, COMBINE, REDUCE} phases, each party receives the global model weights encrypted with $pk$, performs the local gradient descent computation (via homomorphic operations), and outputs its encrypted gradients. During the \textbf{COMBINE} phase, the gradients of each party are aggregated with homomorphic addition operations, and during \textbf{REDUCE} $P_1$ updates the global model by averaging the aggregated gradients.
As stated above, any encryption under CKKS with parameters that ensure a post-quantum security level of $\lambda$ is semantically secure. During \sys's training phase, the model parameters that are exchanged in between parties are encrypted, and all phases rely on the aforementioned  CPA-secure-proven protocols. Moreover, as shown in Appendix~\ref{sec:DBootstrapAltDetails}, the $\textsf{DBootstrap}(\cdot)$ and $\textsf{DBootstrapALT}(\cdot)$ protocols are simulatable. Hence, $\mathcal{S}_t$ can simulate all of the values communicated during \sys's training phase by using the parameters ($\mathcal{N}, Q_L, S$) to generate random ciphertexts such that the real outputs cannot be distinguished from the ideal ones. The sequential composition of all cryptographic functions remains simulatable by $\mathcal{S}_t$ due to using different random values in each phase and due to Remark 1. As such, there is no dependency between the random values that an adversary can leverage on. Moreover, the adversary is not able to decrypt the communicated values of an honest party because decryption is only possible with the collaboration of \textit{all} the parties. Following this, \sys protects the data confidentiality of the honest party/ies.

Analogously, the same argument follows to prove that \sys protects the confidentiality of the trained model, as it is a function of the parties' inputs, and its intermediate and final weights are always under encryption. Hence, \sys eliminates federated learning attacks~\cite{Hitaj2017,Melis2019,Nasr2019,NIPS2019_9617}, that aim at extracting private information about the parties from the intermediate parameters or the final model.

\begin{proposition}
\label{lemma2}
Assume that \sys's encryptions are generated using the CKKS cryptosystem with parameters ($\mathcal{N}, Q_L, S$) ensuring a post-quantum security level of $\lambda$. Given a passive adversary corrupting at most $N-1$ parties, \sys achieves \emph{Data} and \emph{Model Confidentiality} during prediction.
\end{proposition}
%\vspace{-0.3em}
\descr{Proof (Sketch).} (a) Let us assume a real-world simulator $\mathcal{S}_p$ that simulates the view of a computationally-bounded adversary corrupting $N-1$ computing nodes (parties). The \emph{Data Confidentiality} of the honest parties and \emph{Model Confidentiality} is ensured following the arguments of Proposition~\ref{lemma1}, as the prediction protocol is equivalent to a forward-pass performed during a training iteration by a computing party. Following similar arguments to Proposition~\ref{lemma1}, the encryption of the querier's input data (with the parties common public key $pk$) can be simulated by $\mathcal{S}_p$. The only additional function used in the prediction step is $\textsf{DKeySwitch}(\cdot)$ that is proven to be simulatable by $\mathcal{S}_p$~\cite{mouchet2019distributedbfv}. Thus, \sys ensures \emph{Data Confidentiality} of the querier.
\noindent
(b) Let us assume a real-world simulator $\mathcal{S'}_p$ that simulates a computationally-bounded adversary corrupting $N-2$ parties and the querier. \emph{Data Confidentiality} of the querier is trivial, as it is controlled by the adversary. The simulator has access to the prediction result as the output of the process for $P_q$, so it can produce all the intermediate (indistinguishable) encryptions that the adversary sees (based on the simulatability of the key-switch/collective decrypt protocol \cite{mouchet2019distributedbfv}). Following this and the arguments of Proposition~\ref{lemma1}, \emph{Data and Model Confidentiality} are ensured during prediction. We remind here that the membership inference~\cite{shokri2017membership} and model inversion~\cite{fredrikson2015model} are out-of-the-scope attacks (see~\ref{securityExtensions} for complementary security mechanisms against these attacks).
\vspace{-0.3em}
\section{Experimental Evaluation}\label{sec:evaluation}
\vspace{-0.3em}
In this section, we experimentally evaluate \sys's performance and present our empirical results. We also compare \sys to other state-of-the-art privacy-preserving solutions.

\vspace{-0.5em}
\subsection{Implementation Details}\label{sec:hardwareSpec}
\noindent
\vspace{-0.2em}
We implement \sys in Go~\cite{Go}, building on top of the Lattigo lattice-based library~\cite{lattigo} for the multiparty cryptographic operations. We make use of Onet~\cite{onet} and build a decentralized system where the parties communicate over TCP with secure channels (TLS).

\vspace{-0.5em}
\subsection{Experimental Setup}\label{sec:experimentSetup}
\vspace{-0.2em}
\noindent
We use Mininet~\cite{mininet} to evaluate \sys in a virtual network with an average network delay of 0.17ms and 1Gbps bandwidth. All the experiments are performed on 10 Linux servers with Intel Xeon E5-2680 v3 CPUs running at 2.5GHz with 24 threads on 12 cores and 256 GB RAM. Unless otherwise stated, in our \emph{default} experimental setting, we instantiate \sys with $N=10$ and $N=50$ parties. 
%When we run experiments with more than $N=10$ parties, we employ multiple cores on the same 10 Linux machines. 
As for the parameters of the cryptographic scheme, we use a precision of 32 bits, number of levels $L=6$, and $\mathcal{N}=2^{13}$ for the datasets with $d<32$ or $32 \times 32$ images, and $\mathcal{N}=2^{14}$ for those with $d>32$, following the multi-cipher approach (see Section~\ref{sec:paramSelect}).

\vspace{-0.5em}
\subsection{Datasets}\label{sec:datasets}
\vspace{-0.3em}
\noindent
For the evaluation of \sys's performance, we use the following real-world and publicly available datasets: (a) the Breast Cancer Wisconsin dataset (BCW)~\cite{BCw} with $n=699, \,\,d=9, \,\,h_\ell=2$, (b) the hand-written digits (MNIST) dataset~\cite{MNIST} with $n=70,000, \,\,d=28 \times 28, \,\,h_\ell=10$, %for modelling hand-written digits, 
(c) the Epileptic seizure recognition (ESR) dataset~\cite{ESR} with $n=11,500, \,\,d=179, \,\,h_\ell=2$, 
%that is used to model seizure, 
(d) the default of credit card clients (CREDIT) dataset~\cite{creditCard} with $n=30,000, \,\,d=23, \,\,h_\ell=2$, %where the goal is to model the status of the clients' default payment, 
(d) the street view house numbers (SVHN) dataset ~\cite{svhn} with colored images (3 channels), $n=600,000, \,\,d=3\times 32 \times 32, \,\,h_\ell=10$, and (e) the CIFAR-10 and CIFAR-100~\cite{cifarPaper} datasets with colored images (3 channels), $n=60,000, \,\,d=3 \times 32 \times 32,  \,\,h_\ell=10$, and $h_\ell=100$, respectively. Recall that $h_\ell$ represents the number of neurons in the last layer of a neural network (NN), i.e., the number of output labels. We convert SVHN to gray-scale to reduce the number of channels. Moreover, since we pad with zeros each dimension of a weight matrix to the nearest power-of-two (see Section~\ref{sec:AlternatingPacking}), for the experiments using the CREDIT, ESR, and MNIST datasets, we actually perform the NN training with $d=32$, $256$, and $1,024$ features, respectively. For SVHN, the number of features for a flattened gray-scale image is already a power-of-two ($32 \times 32 = 1,024)$. To evaluate the scalability of our system, we generate synthetic datasets and vary the number of features or samples. Finally, for our experiments we evenly and randomly distribute all the above datasets among the participating parties. We note that the data and label distribution between the parties, and its effects on the model accuracy is orthogonal to this paper (see Appendix~\ref{sec:learnextensions} for extensions related to this issue).

\vspace{-0.4em}
\subsection{Neural Network Configuration}
\label{sec:NNConfig}
\vspace{-0.2em}
\noindent
For the BCW, ESR, and CREDIT datasets, we deploy a 2-layer fully connected NN with 64 neurons per layer, and we use the same NN structure for the synthetic datasets used to test \sys's scalability. For the MNIST and SVHN datasets, we train a 3-layer fully connected NN with 64 neurons per-layer. For the CIFAR-10, we train two models: (i) a CNN with 2 CV and 2 average-pooling with kernel size of $2 \times 2$, and 2 FC layers with 128 neurons and 10 neurons, labeled as N1, and (ii) a CNN with 4 CV with kernel size of $3 \times 4$, 2 average-pooling with kernel size of $2 \times 2$ and 2 FC layers with 128 and 10 neurons labeled as N2. For CIFAR-100, we train a CNN with 6 CV with kernel size of $3 \times 4$, 2 average-pooling with kernel size of $2 \times 2$ and 2 FC layers with 128 neurons each. For all CV layers, we vary the number of filters between 3 to 16. We use the approximated sigmoid, SmoothReLU, or tanh activation functions (see Section~\ref{sec:activationsOpt}), depending on the dataset. We train the above models for 100, 600, 500, 1,000, 18,000, 25,000, 16,800, and 54,000 global iterations for the BCW, ESR, CREDIT, MNIST, SVHN, CIFAR-10-N1, CIFAR-10-N2, and CIFAR100 datasets, respectively. For the SVHN and CIFAR datasets, we use momentum-based gradient descent or Nesterov's accelerated gradient descent, which introduces an additional multiplication to the update rule (in the \textbf{MAP} phase). Finally, we set the local batch size $b$ to $10$ and, as such, the global batch size is $B=100$ in our default setting with 10 parties and $B=500$ with 50 parties. For a fixed number of layers, we choose the learning parameters by grid search with 3-fold cross validation on clear data with the \textit{approximated} activation functions. In a practical FL setting, however, the parties can collectively agree on these parameters by using secure statistics computations~\cite{Drynx,UnLynx}.

\begin{table}[t]
\centering
\footnotesize
\setlength{\tabcolsep}{2.33pt}
\begin{tabular}{l ccccc cc}
\toprule
Dataset & \multicolumn{5}{c}{Accuracy} & \multicolumn{2}{c}{Execution time (s)} \\
& \multicolumn{1}{c}{C1} & \multicolumn{1}{c}{C2} & \multicolumn{1}{c}{L} & \multicolumn{1}{c}{D} & \sys & \multicolumn{1}{c}{Training} & Inference \\
\midrule
BCW &$97.8\%$ &$97.4\%$&$93.9\%$ &$97.4\%$  & $96.9\%$ & 91.06 & 0.21 \\
\midrule
ESR & $93.6 \%$ &$91.2\%$& $89.9\%$& $91.1\%$ & $90.4\%$ & 851.84 & 0.30 \\
\midrule
CREDIT & $81.4\%$& $80.9\%$&$79.6\%$ & $80.6\%$ & $80.2\%$ & 516.61 & 0.26 \\
\midrule
MNIST &$92.1 \%$ & $91.3\%$&$87.8\%$ & $90.6\%$ &$89.9\%$ & 5,283.1 & 0.38 \\
\bottomrule
\end{tabular}
\captionsetup{width=\linewidth}
\caption{\sys's accuracy and execution times for $N=10$ parties. The model accuracy is compared to several non-private approaches.}
%: centralized with exact activation functions (C1), centralized with approximated activation functions (C2), and decentralized with approximated activation functions (D).}
\label{table:baseline}
%\vspace{-1.9em}
\end{table}

\vspace{-0.4em}
\subsection{Empirical Results}\label{sec:empiricalRes}
\vspace{-0.2em}
We experimentally evaluate \sys in terms of accuracy of the trained model, execution time for both training and prediction phases, and communication overhead. We also evaluate \sys's scalability with respect to the number of parties $N$, as well as the number of data samples $n$ and features $d$ in a dataset. We further provide microbenchmark timings and communication overhead for the various functionalities and operations for FC, CV, and pooling layers in Appendix~\ref{sec:microbenchmarks} that can be used to extrapolate \sys' execution time for different NN structures. We further give per-global-iteration execution times of various NN architectures in Appendix~\ref{sec:supplementaryResults} and various CNN architectures in Appendix~\ref{sec:supplementaryResultsCNN}.

% \descr{Baseline Comparison.}

% The table shows the execution time of overall training, oblivious inference on querier's encrypted data, and the accuracy of the model.

% \begin{table}[t]
% \small
% \begin{tabular}{llllll}
% \toprule
% Dataset & \multicolumn{2}{l}{Execution time (s)} & \multicolumn{3}{l}{Accuracy} \\
%  & \multicolumn{1}{l}{Training} & Inference & \multicolumn{1}{l}{C} & \multicolumn{1}{l}{D} & \sys \\
%  \midrule
% BCW & 488.0 &0.21 & $97.85\%$ &$97.42\%$  & $96.99\%$  \\
% \midrule
% ESR & \sinem{ic c} & \sinem{ic c} & $88.40\%$ & $88.00\%$ & $88.00\%$ \\
% \midrule
% Credit & 1,168.0 & 0.28  & $82.54\%$ & $82.32\%$ & $81.88\%$ \\
% \midrule
% MNIST & 5,283.1 & 0.38 & $91.3\%$ & $90.6\%$ &$90.2\%$ \\
% \bottomrule
% \end{tabular}
% \captionsetup{width=\linewidth}
% \caption{\sys's execution times for different datasets in the default setting where each dataset is distributed among 10 parties. The trained model's accuracy is compared to centralized (C) and a decentralized (D) non-private approaches.}
% \label{table:baseline}
% \end{table}

%\sinem{get average of different 10 splits}

\descr{Model Accuracy.} Tables~\ref{table:baseline} and~\ref{table:baseline2} display \sys's accuracy results on the used real-world datasets with 10 and 50 parties, respectively. The accuracy column shows four baselines with the following approaches: two approaches where the data is collected to a central party in its clear form: centralized with original activation functions (C1), and centralized with approximated activation functions (C2); one approach where each party trains the model only with its local data (L), and a decentralized approach with approximated activation functions (D), where the data is distributed among the parties, but the learning is performed on cleartext data, i.e., without any protection of the gradients communicated between the parties. For all baselines, we use the same NN structure and learning parameters as \sys, but adjust the learning rate ($\eta$) or use adaptive learning rate to ensure the range of the approximated activation functions is minimized, i.e., a smaller interval for an activation-function approximation requires smaller $\eta$ to prevent divergence while bigger intervals make the choice of $\eta$ more flexible. These baselines enable us to evaluate \sys's accuracy loss due to the approximation of the activation functions, distribution, encryption and the impact of privacy-preserving federated learning. We exclude the (D) column from Table~\ref{table:baseline2} for the sake of space; the pattern is similar to Table~\ref{table:baseline} and \sys's accuracy loss is negligible. To obtain accuracy results for the CIFAR-10 and CIFAR-100 datasets, we simulate \sys in Tensorflow~\cite{tensorflow} by using its approximated activation functions and a fixed-precision.
We observe that the accuracy loss between C1, C2, D, and \sys is $0.9-3\%$ when 32-bits precision is used. For instance, \sys achieves $90.4\%$ training accuracy on the ESR dataset, a performance that is equivalent to a decentralized (D) non-private approach and only slightly lower compared to centralized approaches. Note that the accuracy difference between non-secure solutions and \sys can be further reduced by increasing the number of training iterations, however, we use the same number of iterations for the sake of comparison. Moreover, we remind that CIFAR-100 has 100 class labels (i.e., a random guess baseline of $1\%$ accuracy) and is usually trained with special NN structures (ResNet) or special layers (batch normalization) to achieve higher accuracy than the reported ones: we leave these NN types as future work (see Appendix~\ref{sec:learnextensions}).

We compare \sys's accuracy with that achieved by one party using its local dataset (L), that is $1/10$ (or $1/50$) of the overall data, with \emph{exact} activation functions. We compute the accuracy for the (L) setting by averaging the test accuracy of the 10 and 50 locally trained models (Tables~\ref{table:baseline} and~\ref{table:baseline2}, respectively). We observe that even with the accuracy loss due to approximation and encryption, \sys still achieves $1-3\%$ increase in the model accuracy due to privacy-preserving collaboration (Table~\ref{table:baseline}). This increase is more significant when the data is partitioned across 50 parties (Table~\ref{table:baseline2}) as the number of training samples \textit{per-party} is further reduced and is not sufficient to learn an accurate model.

\begin{table}[t]
\centering
\footnotesize
\setlength{\tabcolsep}{2.0pt}
\begin{tabular}{l cccc ccc}
\toprule
\added{Dataset} & \multicolumn{4}{c}{\added{Accuracy}} & \multicolumn{3}{c}{\added{Execution time (hrs)}} \\
& \multicolumn{1}{c}{\added{C1}} & \multicolumn{1}{c}{\added{C2}} & \multicolumn{1}{c}{\added{L}} & \added{\sys} & \multicolumn{1}{c}{\added{One-GI}}& {\added{Training}} & \added{Inference} \\
\midrule
\added{SVHN} &\added{$68.4\%$} & \added{$68.1\%$} & \added{$35.1\%$}& \added{$67.8\%$}& \added{0.0034} & \added{61.2} & \added{$8.89 \times 10^{-5}$} \\
\midrule
\added{CIFAR-10-N1} & \added{$54.6\%$}&\added{$52.1\%$}&\added{$26.8\%$}&\added{$51.8\%$}& \added{0.007} &\added{175}&\added{0.001}\\ 
\midrule
\added{CIFAR-10-N2} & \added{$63.6\%$}&\added{$62.0\%$}&\added{$28.0\%$}&\added{$61.1\%$}& \added{0.011} &\added{184.8}&\added{0.004}\\
\midrule
\added{CIFAR-100} &  \added{$43.6\%$}&\added{$41.8\%$}&\added{$8.2\%$}&\added{$41.1\%$}& \added{0.026} &\added{1404}&\added{0.006}\\
\bottomrule
\end{tabular}
\captionsetup{width=\linewidth}
\caption{\sys's accuracy and execution times for $N=50$ parties (extrapolated). One-GI indicates the execution time of one global iteration.}
%: centralized with exact activation functions (C1), centralized with approximated activation functions (C2), and decentralized with approximated activation functions (D).}
\label{table:baseline2}
%\vspace{-1.9em}
\end{table}

\descr{Execution Time.} As shown on the right-hand side of Table~\ref{table:baseline}, \sys trains the BCW, ESR, and CREDIT datasets in less than 15 minutes and the MNIST in 1.4 hours, when each dataset is evenly distributed among 10 parties. Note that \sys's overall training time for MNIST is less than an hour when the dataset is split among 20 parties that use the same local batch size. We extrapolate the training times of \sys on more complex datasets and architectures for one global iteration (one-GI) in Table~\ref{table:baseline2}; these can be used to estimate the training times of these structures with a larger number of global iterations. For instance, CIFAR-10 is trained in 175 hours with 2CV, 2 pooling, 2 FC layers and with dropouts (adding one more multiplication in the dropout layer). Note that it is possible to increase the accuracy with higher run-time or fine-tuned architectures, but we aim at finding a trade-off between accuracy and run-time. For example, \sys's accuracy on SVHN reaches $75\%$ by doubling the training epochs and thus its execution time.
The per-sample inference times presented in Tables~\ref{table:baseline} and~\ref{table:baseline2} include the forward pass, the $\textsf{DKeySwitch}(\cdot)$ operations that reencrypt the result with the querier's public key, and the communication among the parties. We note that as all the parties keep the model in encrypted form, any of them can process the prediction query. Hence, taking the advantage of parallel query executions and multi-threading, \sys achieves a throughput of 864,000 predictions per hour on the MNIST dataset with the chosen NN structure.

\begin{figure*}[t]
	\centering
	\footnotesize
	\begin{subfigure}[t]{0.245\textwidth}
		\centering
			\footnotesize
		\includegraphics[width=1\columnwidth]
		{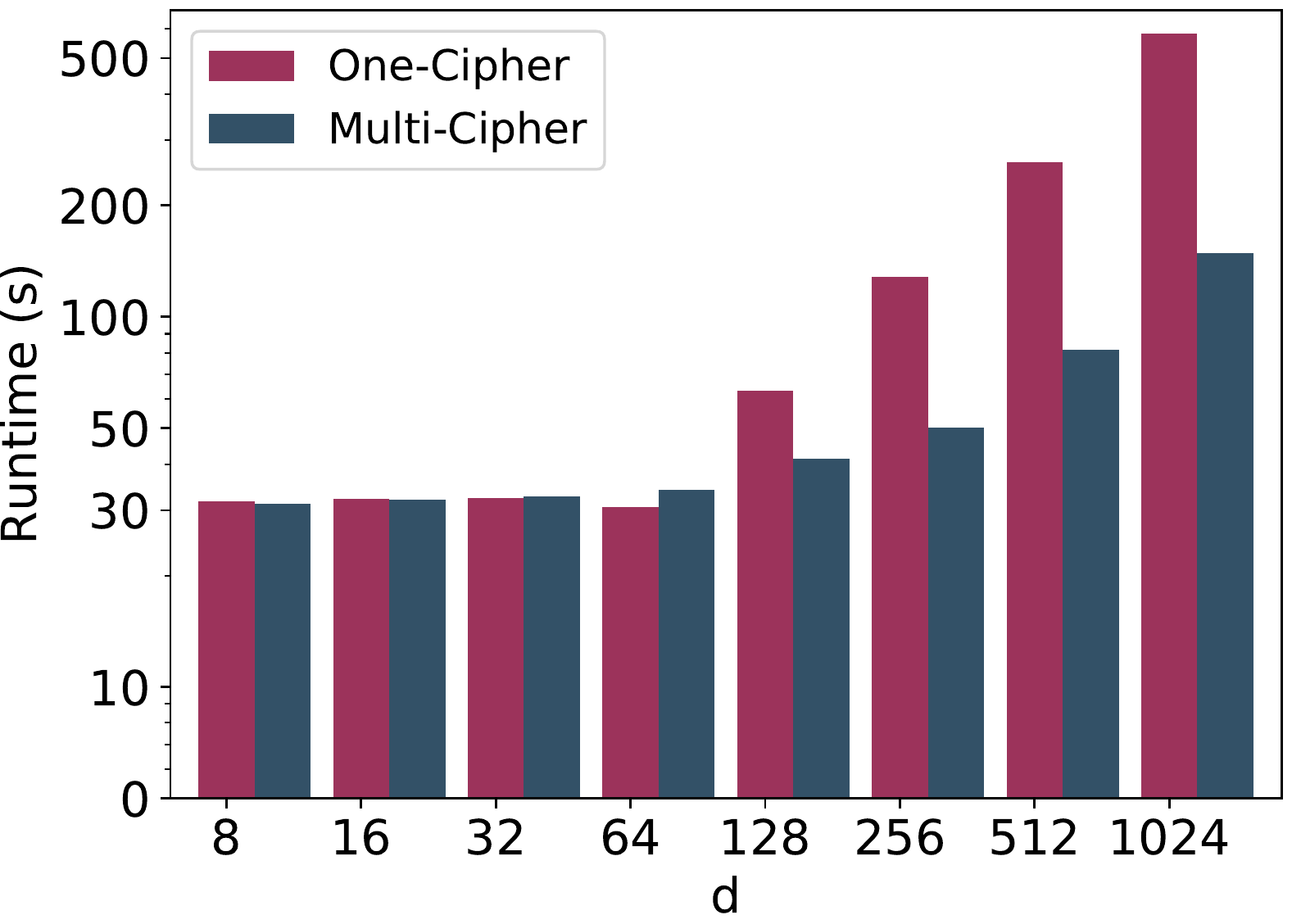}
		\vspace{-1em}
					\captionsetup{width=0.975\linewidth}
		\caption{Increasing number of features ($d$), $N=10$, and $n=2,000*N$.}
		\label{fig:performanceFeature}
	%	\vspace{-1.9em}
\end{subfigure}
	\begin{subfigure}[t]{0.245\textwidth}
		\centering
			\footnotesize
		\includegraphics[width=0.98\columnwidth]
		{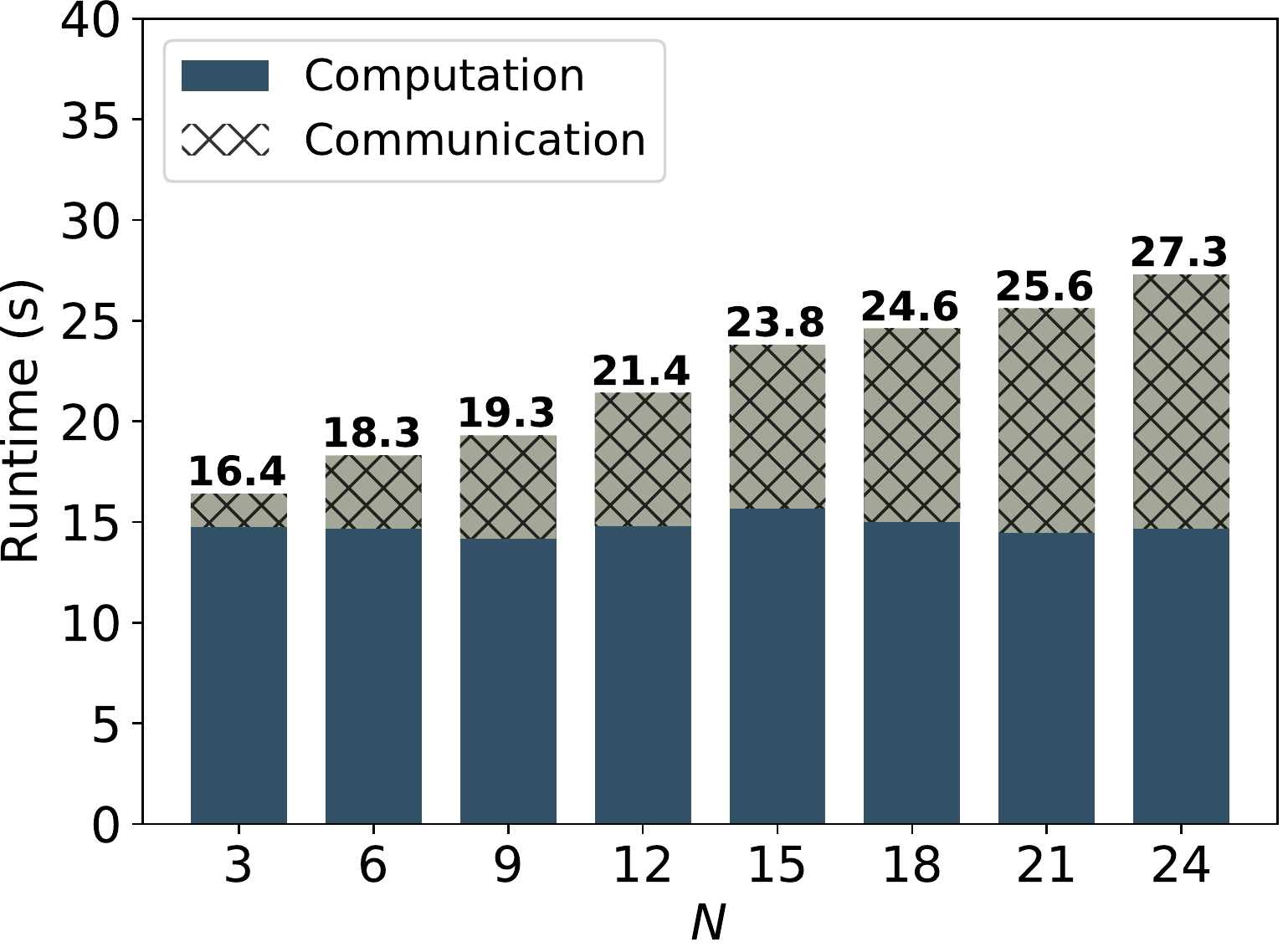}
					\captionsetup{width=0.975\linewidth}

		\caption{Increasing number of parties ($N$), each having 200 samples ($n$).}
		\label{fig:performanceDP1}
	\end{subfigure}
	\begin{subfigure}[t]{0.245\textwidth}
		\centering
			\footnotesize
		\includegraphics[width=0.99\columnwidth]
		{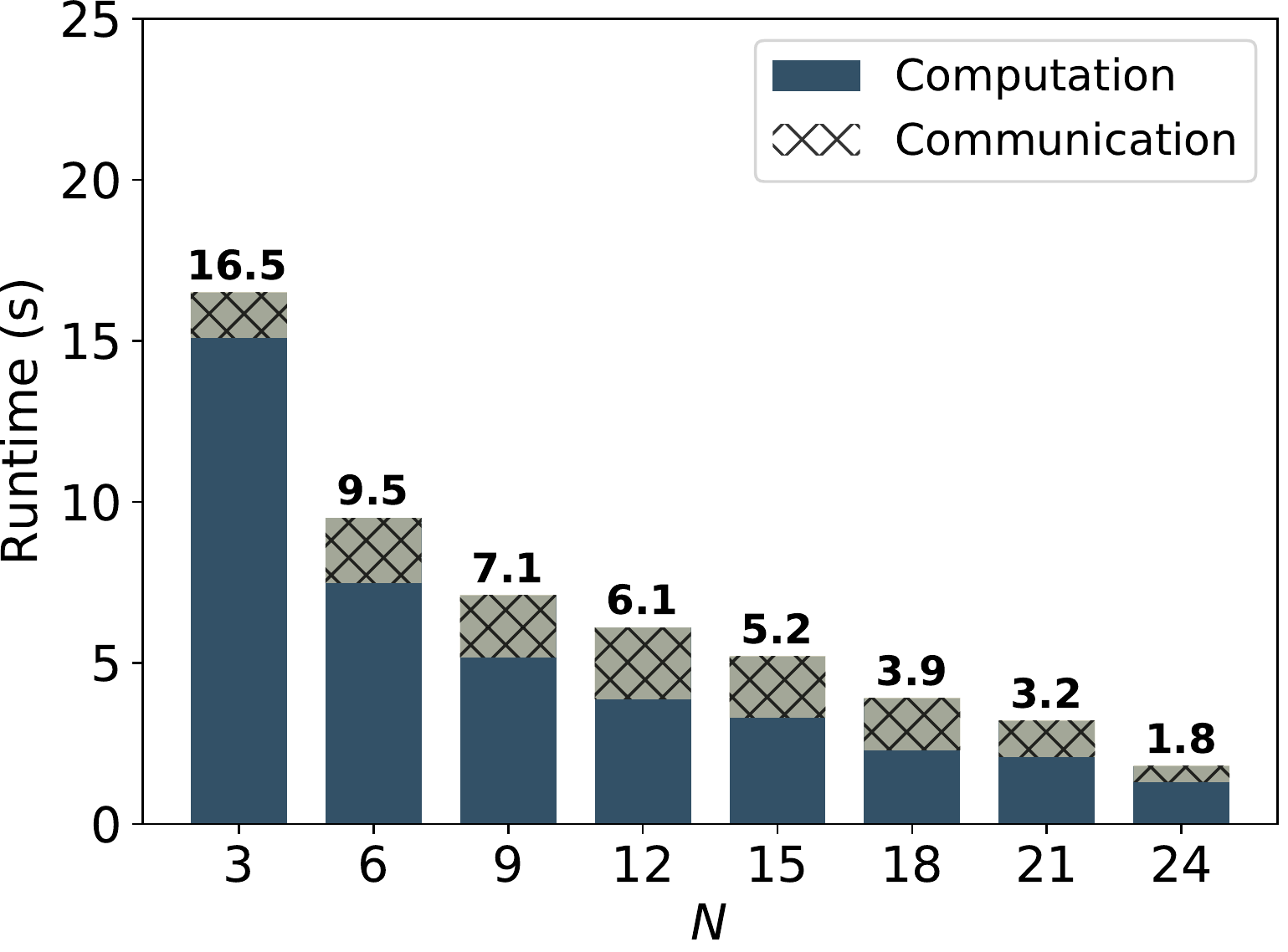}
		\vspace{-1em}
					\captionsetup{width=0.975\linewidth}

		\caption{Increasing number of parties ($N$) when $n$ (number of data samples) is fixed to 600.}
		\label{fig:performanceDP2}
	\end{subfigure}
	\begin{subfigure}[t]{0.245\textwidth}
		\centering
			\footnotesize
		\includegraphics[width=1.05\columnwidth]
		{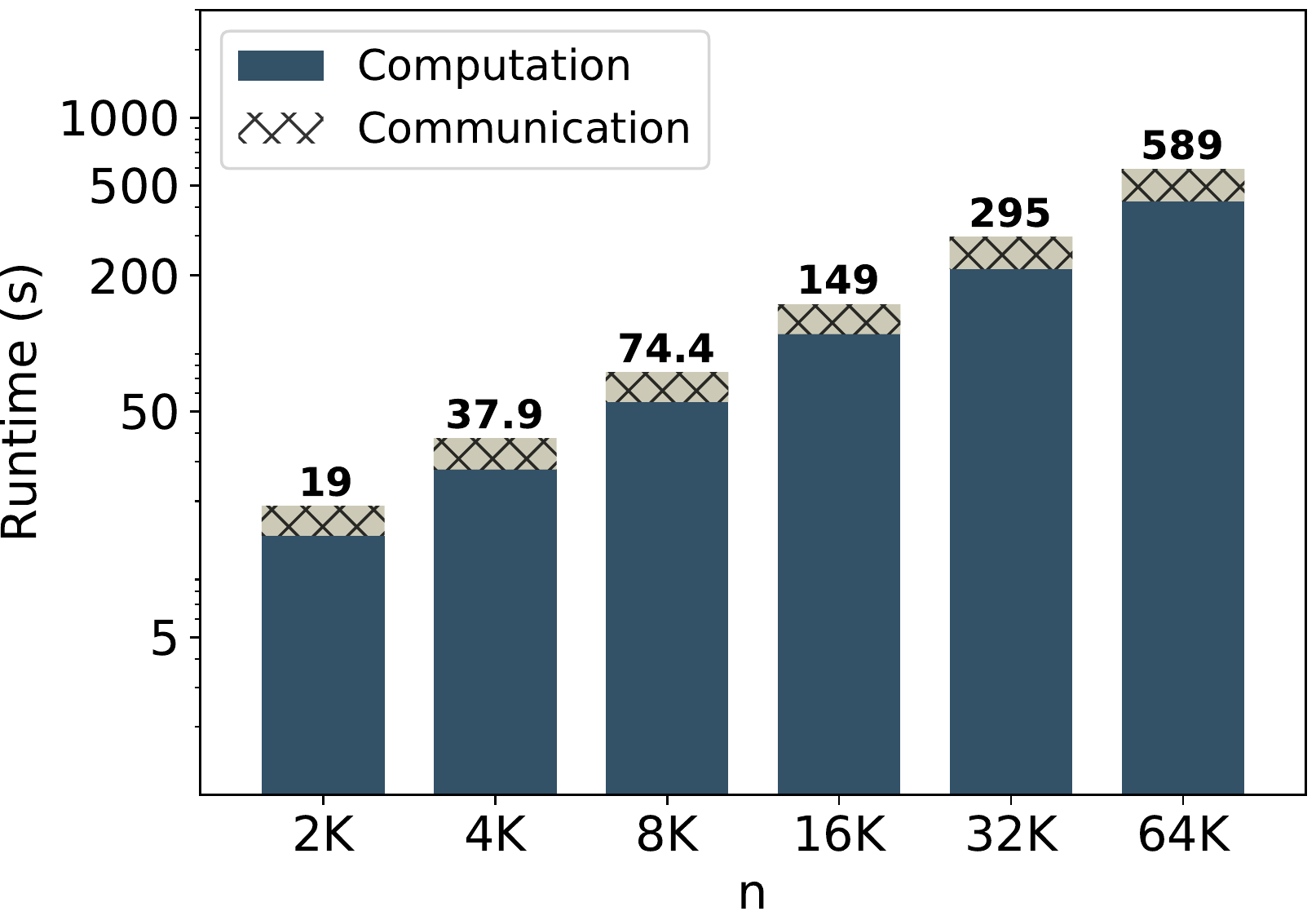}
		\vspace{-1em}
					\captionsetup{width=0.975\linewidth}

		\caption{Increasing number of data samples ($n$) when $N$ (number of parties) is fixed to 10.}
		\label{fig:performanceSamples}
	\end{subfigure}
	\captionsetup{width=\linewidth}
	\vspace{-0.2em}
	\caption{\sys's training execution time and communication overhead with increasing number of parties, features, and samples, for 1 training epoch.}
	\label{figPerformance}
%	\vspace{-1.5em}
\end{figure*}
\vspace{-0.2em}

\descr{Scalability.} Figure~\ref{fig:performanceFeature} shows the scaling of \sys with the number of features ($d$) when the one-cipher and multi-cipher with parallelization approaches are used for a 2-layer NN with $64$ hidden neurons. The runtime refers to one epoch, i.e., a processing of all the data from $N=10$ parties, each having 2,000 samples, and employing a batch size of $b=10$. For small datasets with a number of features between 1 and 64, we observe no difference in execution time between the one-cipher and multi-cipher approaches. This is because the weight matrices between layers fit in one ciphertext with $\mathcal{N}=2^{13}$. However, we observe a larger runtime of the one-cipher approach when the number of features increases further. This is because each power-of-two increase in the number of features requires an increase in the cryptographic parameters, thus introducing overhead in the arithmetic operations.

We further analyse \sys's scalability with respect to the number of parties ($N$) and the number of total samples in the distributed dataset ($n$), for a fixed number of features. Figures~\ref{fig:performanceDP1} and~\ref{fig:performanceDP2} display \sys's execution time, when the number of parties ranges from 3 to 24, and one training epoch is performed, i.e., all the data of the parties is processed once. For Figure~\ref{fig:performanceDP1}, we fix the number of data samples per party to 200 to study the effect of an increasing number of members in the federation. We observe that \sys's execution time is almost independent of $N$ and is affected only by increasing communication between the parties. When we fix the global number of samples ($n$), increasing $N$ results in a runtime decrease, as the samples are processed by the parties in parallel (see Figure~\ref{fig:performanceDP2}). Then, we evaluate \sys's runtime with an increasing number of data samples and a fixed number of parties $N=10$, in Figure~\ref{fig:performanceSamples}. We observe that \sys scales linearly with the number of data samples. Finally, we remark that \sys also scales proportionally with the number of layers in the NN structure, if these are all of the same type, i.e, FC, CV, or pooling, and if the number of neurons per layer or the kernel size is fixed. 
%(see Appendix~\ref{sec:supplementaryResults}).
\vspace{-0.6em}
\subsection{Comparison with Prior Work}\label{sec:evalComparison}
\vspace{-0.2em}
A quantitative comparison of our work with the state-of-the-art solutions for privacy-preserving NN executions is a non-trivial task. 
%To the best of our knowledge, \sys is the first system that enables privacy-preserving training of and evaluation on NNs in an $N$-party setting and protects the confidentiality of the input data and the resulting model. 
Indeed, the most recent cryptographic solutions for privacy-preserving machine learning in the $N$-party setting, i.e., Helen~\cite{zheng2019helen} and SPINDLE~\cite{spindle}, support the functionalities of only regularized~\cite{zheng2019helen} and generalized~\cite{spindle} linear models respectively. We provide a detailed qualitative comparison with the state-of-the-art privacy-preserving deep learning frameworks in Table~\ref{table:comparison} in Appendix and expand on it here.
%The MPC-setup row of the table denotes the number of parties responsible for the execution of the NN operations. The adversarial model for data confidentiality indicates the capabilities of the parties (active (A) or passive (P)), and collusion shows the maximum number of tolerated colluding parties. We note that several works allow an admissible adversary, i.e., collusions between one server and an arbitrary number of clients/data owners~\cite{SecureML}. For a fair comparison, we consider only collusions between the parties (servers) that are responsible for the training. 

\sys operates in a federated learning setting where the parties maintain their data locally. This is a substantially different setting compared to that envisioned by MPC-based solutions~\cite{SecureML,mohassel2018aby,wagh2019securenn,falcon,flash,trident}, for privacy-preserving NN training. In these solutions, the parties' data has to be communicated (i.e., secret-shared) outside their premises, and the data and model confidentiality is preserved as long as there exists an honest majority among a limited number of computing servers (typically, 2 to 4, depending on the setting). Hence, a similar experimental setting is hard to achieve. 
% For the similar settings with 3 or 4 parties, these solutions achieve better performance as they do not rely on HE, but requires parties to communicate their data to computing servers. 
Nonetheless, we compare \sys to SecureML~\cite{SecureML}, SecureNN~\cite{wagh2019securenn}, and FALCON~\cite{falcon}, when training a 3-layer NN with 128 neurons per layer for 15 epochs, as described in~\cite{SecureML}, on the MNIST dataset. We set $N=3$ to simulate a similar setting and use \sys's approximated activation functions. \sys trains MNIST in 73.1 hours whereas SecureML with 2-parties, SecureNN and FALCON with 3-parties, need 81.7, 1.03, and 0.56 hours, respectively. Depending on the activation functions, SecureML yields $93.1-93.4\%$ accuracy, SecureNN $93.4\%$, and FALCON $97.4\%$. \sys achieves $92.5\%$ accuracy with the approximated SmoothReLU and $96.2\%$ with approximated tanh activation functions. We remind that \sys operates under a different system (federated learning based) and threat model, it supports more parties, and scales linearly with $N$ whereas MPC solutions are based on outsourced learning with limited number of computing servers.
% is not designed for small number of parties, and it has  within a federated learning setting, with more parties where it 

% \changed{Table~\ref{table:comparison} displays a qualitative comparison of \sys with the state-of-the-art privacy-preserving neural network training and/or inference solutions.

Federated learning approaches based on differential privacy (DP), e.g.,~\cite{Nvidia_Fed,shokri2015privacy,McMahan2018}, train a NN while introducing some noise to the intermediate values to mitigate adversarial inferences. However, training an accurate NN model with DP requires a high privacy budget~\cite{Rahman2018dp}, hence it remains unclear what privacy protection is obtained in practice~\cite{jayaraman2019evaluating}. We note that DP-based approaches introduce a different tradeoff than \sys: they tradeoff privacy for accuracy, while \sys decouples accuracy from privacy and tradeoffs accuracy for complexity (i.e., execution time and communication overhead). Nonetheless and as an example, we compare \sys's accuracy results with those reported by Shokri and Shmatikov~\cite{shokri2015privacy} on the MNIST dataset. We focus on their results with the distributed selective SGD configured such that participants download/upload \emph{all} the parameters from/to the central server in each training iteration. 
% Their SGD updates are slightly different than ours, i.e., they rely on distributed selective SGD where participants download/upload a fraction of parameters from/to a central server in each iteration, thus, we focus on their results with fraction 1. 
We evaluate the same CNN structure used in~\cite{shokri2015privacy}, but with \sys's approximated activation functions and average-pooling instead of max-pooling. We compare the accuracy results presented in \cite[Figure 13]{shokri2015privacy} with $N=30, N=90,$ and $N=150$ participants. In all settings, \sys yields $>94\%$ accuracy whereas~\cite{shokri2015privacy} achieves similar accuracy only when the privacy budget per parameter is $\geq 10$. For more private solutions, where the privacy budget is $0.001$, $0.01$ or $0.1$, ~\cite{shokri2015privacy} achieves $\leq 90\%$ accuracy; smaller $\epsilon$ yields better privacy but degrades utility.

% \sys is the only solution that enables NN training in the aforementioned setting, with negligible accuracy loss and with linear scalability in the number of parties.

% We note here that there is no work with a similar setting to ours for a quantitative comparison. For the sake of comparison, however, we compare our solution to centralized privacy-preserving HE-based solutions hereinafter.

% Thus, we compare our work with the state-of-the-art privacy-preserving federated learning based solutions in Table~\ref{table:qualFLComparison}. \\
Finally, existing HE-based solutions~\cite{cryptoDL,Karthik2019,Vizitu2020}, focus on a centralized setting where the NN learning task is outsourced to a central server. These solutions, however, employ non-realistic cryptographic parameters~\cite{Vizitu2020,Karthik2019}, and their performance is not practical~\cite{cryptoDL} due to their costly homomorphic computations. Our system, focused on a federated learning-based setting and a multiparty homomorphic encryption scheme, improves the response time 3 to 4 orders of magnitude. The execution times produced by Nandakumar et al.~\cite{Karthik2019} for processing one batch of 60 samples in a single thread and 30 threads for a NN structure with $d=64$, $h_1=32$, $h_2=16$, $h_3=2$, are respectively 33,840s and 2,400s. When we evaluate the same setting, but with $N=10$ parties, we observe that \sys processes the same batch in 6.3s and 1s, respectively. We also achieve stronger security guarantees (128 bits) than~\cite{Karthik2019} (80 bits). Finally, for a NN structure with 2-hidden layers of 128 neurons each, and the MNIST dataset, CryptoDL~\cite{cryptoDL} processes a batch with $B=192$ in 10,476.3s, whereas our system in the distributed setting processes the same batch in 34.7s.
%We note that, to avoid costly bootstrapping function CryptoDL also requires communication between the server and the client.

Therefore, \sys is the only solution that performs both training and inference of NNs in an $N$-party setting, yet protects data and model confidentiality withstanding collusions up to $N-1$ parties.% Therefore, our work is inherently different from the aforementioned solutions.}

\section{Conclusion}
\label{sec:conclusion}
\vspace{-0.3em}
In this work, we presented \sys, a novel system for zero-leakage privacy-preserving federated neural network learning among $N$ parties. Based on lattice-based multiparty homomorphic encryption, our system protects the confidentiality of the training data, of the model, and of the evaluation data, under a passive adversary model with collusions of up to $N-1$ parties. By leveraging on packing strategies and an extended distributed bootstrapping functionality, \sys is the first system demonstrating that secure federated learning on neural networks is practical under multiparty homomorphic encryption. Our experimental evaluation shows that \sys significantly improves on the accuracy of individual local training, bringing it on par with centralized and decentralized non-private approaches. Its computation and communication overhead scales linearly with the number of parties that participate in the training, and is between 3 to 4 orders of magnitude faster than equivalent centralized outsourced approaches based on traditional homomorphic encryption. This work opens up the door of practical and secure federated training in passive-adversarial settings. 
Future work involves extensions to other scenarios with active adversaries and further optimizations to the learning process.
\section*{Acknowledgment}
%\vspace{-1.0em}
We would like to thank all of those who reviewed the
manuscript, in particular Sylvain Chatel and the anonymous reviewers. This work was partially supported by the grant \#2017-201 of the Strategic Focal Area “Personalized Health and Related Technologies (PHRT)” of the ETH Domain.%\vspace{-1.5em}

\bibliographystyle{abbrv}
\bibliography{bibfile}

\begin{thebibliography}{100}

\bibitem{seal}
{ Microsoft SEAL (release 3.3)}.
\newblock \url{https://github.com/Microsoft/SEAL}.
\newblock (Accessed: 2021-01-06).

\bibitem{HIPAA}
{Centers for Medicare \& Medicaid Services. The Health Insurance Portability
  and Accountability Act of 1996 (HIPAA)}.
\newblock
  \url{https://www.cms.gov/Regulations-and-Guidance/Administrative-Simplification/HIPAA-ACA/PrivacyandSecurityInformation
  }.
\newblock (Accessed: 2021-01-06).

\bibitem{conversionFC}
{Convolutional Neural Networks.}
\newblock \url{https://cs231n.github.io/convolutional-networks/}.
\newblock (Accessed: 2021-01-06).

\bibitem{onet}
{Cothority network library}.
\newblock \url{https://github.com/dedis/onet}.
\newblock (Accessed: 2021-01-06).

\bibitem{mpspdz}
{Data61. MP-SPDZ - Versatile framework for multi-party computation.}
\newblock \url{https://github.com/data61/MP-SPDZ}.
\newblock (Accessed: 2021-01-06).

\bibitem{Go}
{Go Programming Language}.
\newblock \url{https://golang.org}.
\newblock (Accessed: 2021-01-06).

\bibitem{GDPR}
{The EU General Data Protection Regulation}.
\newblock \url{https://gdpr-info.eu/}.
\newblock (Accessed: 2021-01-06).

\bibitem{abadi2016deep}
M.~Abadi, A.~Chu, I.~Goodfellow, H.~B. McMahan, I.~Mironov, K.~Talwar, and
  L.~Zhang.
\newblock Deep learning with differential privacy.
\newblock In {\em ACM CCS}, 2016.

\bibitem{tensorflow}
M.~Abadi et~al.
\newblock {TensorFlow}: Large-scale machine learning on heterogeneous systems,
  2015.
\newblock Software available from tensorflow.org.

\bibitem{ABIODUN2018}
O.~I. Abiodun, A.~Jantan, A.~E. Omolara, K.~V. Dada, N.~A. Mohamed, and
  H.~Arshad.
\newblock State-of-the-art in artificial neural network applications: A survey.
\newblock {\em Elsevier Heliyon}, 4(11):e00938, 2018.

\bibitem{Acar_2018}
A.~Acar, H.~Aksu, A.~S. Uluagac, and M.~Conti.
\newblock A survey on homomorphic encryption schemes: Theory and
  implementation.
\newblock {\em ACM Comput. Surv.}, 51(4), July 2018.

\bibitem{quotient}
N.~Agrawal, A.~S. Shamsabadi, M.~J. Kusner, and A.~Gasc{\'o}n.
\newblock {QUOTIENT: T}wo-party secure neural network training and prediction.
\newblock {\em ACM CCS}, 2019.

\bibitem{Akavia_WAHC}
A.~Akavia, H.~Shaul, M.~Weiss, and Z.~Yakhini.
\newblock Linear-regression on packed encrypted data in the two-server model.
\newblock In {\em ACM WAHC}, 2019.

\bibitem{HEStandardPaper}
M.~Albrecht et~al.
\newblock Homomorphic {E}ncryption {S}ecurity {S}tandard.
\newblock Technical report, HomomorphicEncryption.org, 2018.

\bibitem{Albrecht2015OnTC}
M.~R. Albrecht, R.~Player, and S.~Scott.
\newblock On the concrete hardness of learning with errors.
\newblock {\em Journal of Mathematical Cryptology}, 9:169 -- 203, 2015.

\bibitem{Algesheimer2002}
S.~V. Algesheimer~J., Camenisch~J.
\newblock Efficient computation modulo a shared secret with application to the
  generation of shared safe-prime products.
\newblock In {\em CRYPTO}, 2002.

\bibitem{aono2016scalable}
Y.~Aono, T.~Hayashi, L.~Trieu~Phong, and L.~Wang.
\newblock Scalable and secure logistic regression via homomorphic encryption.
\newblock In {\em ACM CODASPY}, 2016.

\bibitem{Badawi2019}
A.~A. Badawi, L.~Hoang, C.~F. Mun, K.~Laine, and K.~M.~M. Aung.
\newblock Privft: Private and fast text classification with homomorphic
  encryption.
\newblock {\em CoRR}, abs:1908.06972, 2019.

\bibitem{BCw}
Breast cancer wisconsin (original).
\newblock
  \url{https://archive.ics.uci.edu/ml/datasets/breast+cancer+wisconsin+(original)}.
\newblock (Accessed: 2021-01-06).

\bibitem{boemer2019ngraph}
F.~Boemer, A.~Costache, R.~Cammarota, and C.~Wierzynski.
\newblock ngraph-he2: A high-throughput framework for neural network inference
  on encrypted data.
\newblock In {\em ACM WAHC}, 2019.

\bibitem{boemer2018ngraph}
F.~Boemer, Y.~Lao, and C.~Wierzynski.
\newblock ngraph-he: {A} graph compiler for deep learning on homomorphically
  encrypted data.
\newblock {\em CoRR}, abs/1810.10121, 2018.

\bibitem{bogdanov2016rmind}
D.~Bogdanov, L.~Kamm, S.~Laur, and V.~Sokk.
\newblock Rmind: a tool for cryptographically secure statistical analysis.
\newblock {\em IEEE TDSC}, 15(3):481--495, 2018.

\bibitem{Bonawitz2016}
K.~Bonawitz, V.~Ivanov, B.~Kreuter, A.~Marcedone, H.~B. McMahan, S.~Patel,
  D.~Ramage, A.~Segal, and K.~Seth.
\newblock Practical secure aggregation for federated learning on user-held
  data.
\newblock In {\em NIPS PPML Workshop}, 2016.

\bibitem{bonte2018privacy}
C.~Bonte and F.~Vercauteren.
\newblock Privacy-preserving logistic regression training.
\newblock {\em BMC Medical Genomics}, 11, 2018.

\bibitem{cluster2}
P.~Bunn and R.~Ostrovsky.
\newblock Secure two-party k-means clustering.
\newblock In {\em ACM CCS}, 2007.

\bibitem{flash}
M.~Byali, H.~Chaudhari, A.~Patra, and A.~Suresh.
\newblock {FLASH: F}ast and robust framework for privacy-preserving machine
  learning.
\newblock {\em PETS}, 2020.

\bibitem{trident}
H.~Chaudhari, R.~Rachuri, and A.~Suresh.
\newblock {Trident: E}fficient 4pc framework for privacy preserving machine
  learning.
\newblock In {\em NDSS}, 2020.

\bibitem{Chen}
T.~{Chen} and S.~{Zhong}.
\newblock Privacy-preserving backpropagation neural network learning.
\newblock {\em IEEE Transactions on Neural Networks}, 20(10):1554--1564, Oct
  2009.

\bibitem{cheon2017homomorphic}
J.~H. Cheon, A.~Kim, M.~Kim, and Y.~Song.
\newblock Homomorphic encryption for arithmetic of approximate numbers.
\newblock In {\em ASIACRYPT}, 2017.

\bibitem{approxMax}
J.~H. Cheon, D.~Kim, D.~Kim, H.~H. Lee, and K.~Lee.
\newblock Numerical method for comparison on homomorphically encrypted numbers.
\newblock In {\em ASIACRYPT}, 2019.

\bibitem{Cho_GWAS}
H.~Cho, D.~Wu, and B.~Berger.
\newblock Secure genome-wide association analysis using multiparty computation.
\newblock {\em Nature Biotechnology}, 36:547--551, 2018.

\bibitem{MapReduce_ML}
C.-T. Chu, S.~K. Kim, Y.-A. Lin, Y.~yu, G.~Bradski, A.~Ng, and K.~Olukotun.
\newblock Map-reduce for machine learning on multicore.
\newblock In {\em NIPS}, 2006.

\bibitem{corrigan2017prio}
H.~Corrigan-Gibbs and D.~Boneh.
\newblock Prio: {P}rivate, {R}obust, and {C}omputation of {A}ggregate
  {S}tatistics.
\newblock In {\em USENIX NSDI}, 2017.

\bibitem{crawford2018doing}
J.~L. Crawford, C.~Gentry, S.~Halevi, D.~Platt, and V.~Shoup.
\newblock Doing real work with fhe: The case of logistic regression.
\newblock In {\em ACM WAHC}, 2018.

\bibitem{Dalskov}
A.~Dalskov, D.~Escudero, and M.~Keller.
\newblock Secure evaluation of quantized neural networks.
\newblock {\em PETS}, 2020.

\bibitem{Downpour_SGD}
J.~Dean, G.~Corrado, R.~Monga, K.~Chen, M.~Devin, M.~Mao, M.~A. Ranzato,
  A.~Senior, P.~Tucker, K.~Yang, Q.~V. Le, and A.~Y. Ng.
\newblock Large scale distributed deep networks.
\newblock In {\em NIPS}. 2012.

\bibitem{ESR}
Epileptic {S}eizure {R}ecognition {D}ataset.
\newblock
  \url{https://archive.ics.uci.edu/ml/datasets/Epileptic+Seizure+Recognition}.
\newblock (Accessed: 2021-01-06).

\bibitem{bfv}
J.~Fan and F.~Vercauteren.
\newblock Somewhat practical fully homomorphic encryption.
\newblock Cryptology ePrint Archive, Report 2012/144, 2012.

\bibitem{fredrikson2015model}
M.~Fredrikson, S.~Jha, and T.~Ristenpart.
\newblock Model inversion attacks that exploit confidence information and basic
  countermeasures.
\newblock In {\em ACM CCS}, 2015.

\bibitem{UnLynx}
D.~Froelicher, P.~Egger, J.~Sousa, J.~L. Raisaro, Z.~Huang, C.~Mouchet,
  B.~Ford, and J.-P. Hubaux.
\newblock Unlynx: A decentralized system for privacy-conscious data sharing.
\newblock {\em PETS}, 2017.

\bibitem{spindle}
D.~Froelicher, J.~R. Troncoso-Pastoriza, A.~Pyrgelis, S.~Sav, J.~S. Sousa,
  J.-P. Bossuat, and J.-P. Hubaux.
\newblock Scalable privacy-preserving distributed learning.
\newblock {\em PETS}, 2021.

\bibitem{Drynx}
D.~{Froelicher}, J.~R. {Troncoso-Pastoriza}, J.~S. {Sousa}, and J.~{Hubaux}.
\newblock Drynx: Decentralized, secure, verifiable system for statistical
  queries and machine learning on distributed datasets.
\newblock {\em IEEE TIFS}, 15:3035--3050, 2020.

\bibitem{gascon2017privacy}
A.~Gasc{\'o}n, P.~Schoppmann, B.~Balle, M.~Raykova, J.~Doerner, S.~Zahur, and
  D.~Evans.
\newblock Privacy-preserving distributed linear regression on high-dimensional
  data.
\newblock {\em PETS}, 2017.

\bibitem{giacomelli2018privacy}
I.~Giacomelli, S.~Jha, M.~Joye, C.~D. Page, and K.~Yoon.
\newblock Privacy-preserving ridge regression with only linearly-homomorphic
  encryption.
\newblock In {\em Springer ACNS}, 2018.

\bibitem{CryptoNets}
R.~Gilad-Bachrach, N.~Dowlin, K.~Laine, K.~Lauter, M.~Naehrig, and J.~Wernsing.
\newblock Cryptonets: Applying neural networks to encrypted data with high
  throughput and accuracy.
\newblock In {\em ICML}, 2016.

\bibitem{Glorot2010}
X.~Glorot and Y.~Bengio.
\newblock Understanding the difficulty of training deep feedforward neural
  networks.
\newblock In {\em AISTATS}, 2010.

\bibitem{Quantum1}
L.~{Gomes}.
\newblock Quantum computing: Both here and not here.
\newblock {\em IEEE Spectrum}, 55(4):42--47, 2018.

\bibitem{NeuralBook}
I.~Goodfellow, Y.~Bengio, and A.~Courville.
\newblock {\em Deep Learning}.
\newblock MIT Press, 2016.

\bibitem{helib}
S.~Halevi and V.~Shoup.
\newblock {HElib - An Implementation of homomorphic encryption}.
\newblock \url{https://github.com/shaih/HElib/}.
\newblock (Accessed: 2021-01-06).

\bibitem{He2015}
K.~{He}, X.~{Zhang}, S.~{Ren}, and J.~{Sun}.
\newblock Delving deep into rectifiers: Surpassing human-level performance on
  imagenet classification.
\newblock {\em IEEE ICCV}, 1502, 2015.

\bibitem{cryptoDL}
E.~Hesamifard, H.~Takabi, M.~Ghasemi, and R.~Wright.
\newblock Privacy-preserving machine learning as a service.
\newblock {\em PETS}, 2018.

\bibitem{DTI}
B.~Hie, H.~Cho, and B.~Berger.
\newblock Realizing private and practical pharmacological collaboration.
\newblock {\em Science}, 362(6412):347--350, 2018.

\bibitem{Hitaj2017}
B.~Hitaj, G.~Ateniese, and F.~Perez-Cruz.
\newblock Deep models under the {GAN}: Information leakage from collaborative
  deep learning.
\newblock In {\em ACM CCS}, 2017.

\bibitem{cluster1}
G.~Jagannathan and R.~N. Wright.
\newblock Privacy-preserving distributed k-means clustering over arbitrarily
  partitioned data.
\newblock In {\em ACM SIGKDD}, 2005.

\bibitem{jayaraman2019evaluating}
B.~Jayaraman and D.~Evans.
\newblock Evaluating differentially private machine learning in practice.
\newblock In {\em USENIX Security}, 2019.

\bibitem{jayaraman2018distributed}
B.~Jayaraman, L.~Wang, D.~Evans, and Q.~Gu.
\newblock Distributed learning without distress: Privacy-preserving empirical
  risk minimization.
\newblock In {\em NIPS}, 2018.

\bibitem{jiang2019securelr}
Y.~Jiang, J.~Hamer, C.~Wang, X.~Jiang, M.~Kim, Y.~Song, Y.~Xia, N.~Mohammed,
  M.~N. Sadat, and S.~Wang.
\newblock Securelr: Secure logistic regression model via a hybrid cryptographic
  protocol.
\newblock {\em IEEE/ACM TCBB}, 2019.

\bibitem{Jouppi2017}
N.~P. Jouppi et~al.
\newblock In-datacenter performance analysis of a tensor processing unit.
\newblock {\em ACM/IEEE ISCA}, 2017.

\bibitem{Gazelle}
C.~Juvekar, V.~Vaikuntanathan, and A.~Chandrakasan.
\newblock Gazelle: {A} low latency framework for secure neural network
  inference.
\newblock {\em USENIX Security}, 2018.

\bibitem{ndas2}
Why we shouldn’t disregard the {NDA}.
\newblock
  \url{https://www.keystonelaw.com/keynotes/why-we-shouldnt-disregard-the-nda}.
\newblock (Accessed: 2021-01-06).

\bibitem{kim2018logistic}
A.~Kim, Y.~Song, M.~Kim, K.~Lee, and J.~H. Cheon.
\newblock Logistic regression model training based on the approximate
  homomorphic encryption.
\newblock {\em BMC medical genomics}, 2018.

\bibitem{kim2018secure}
M.~Kim, Y.~Song, S.~Wang, Y.~Xia, and X.~Jiang.
\newblock Secure logistic regression based on homomorphic encryption: Design
  and evaluation.
\newblock {\em JMIR Medical Informatics}, 6(2):e19, 2018.

\bibitem{Konency2016fed}
J.~Kone{\v{c}}n{\`y}, H.~B. McMahan, D.~Ramage, and P.~Richt{\'a}rik.
\newblock Federated optimization: Distributed machine learning for on-device
  intelligence.
\newblock {\em CoRR}, abs:1610.02527, 2016.

\bibitem{Konecny2016}
J.~Konecn{\'{y}}, H.~B. McMahan, F.~X. Yu, P.~Richt{\'{a}}rik, A.~T. Suresh,
  and D.~Bacon.
\newblock Federated learning: Strategies for improving communication
  efficiency.
\newblock {\em CoRR}, abs/1610.05492, 2016.

\bibitem{cifarPaper}
A.~Krizhevsky.
\newblock Learning multiple layers of features from tiny images.
\newblock {\em Technical Report, University of Toronto}, 2012.

\bibitem{MNIST}
Y.~LeCun and C.~Cortes.
\newblock {MNIST} handwritten digit database.
\newblock 2010.

\bibitem{Nvidia_Fed}
W.~Li, F.~Milletar{\`i}, D.~Xu, N.~Rieke, J.~Hancox, W.~Zhu, M.~Baust,
  Y.~Cheng, S.~Ourselin, M.~J. Cardoso, and A.~Feng.
\newblock Privacy-preserving federated brain tumour segmentation.
\newblock In {\em Springer MLMI}, 2019.

\bibitem{li2020convergence}
X.~Li, K.~Huang, W.~Yang, S.~Wang, and Z.~Zhang.
\newblock On the convergence of {F}ed{A}vg on non-{IID} data.
\newblock In {\em ICLR}, 2020.

\bibitem{Lian2018}
X.~Lian, W.~Zhang, C.~Zhang, and J.~Liu.
\newblock Asynchronous decentralized parallel stochastic gradient descent.
\newblock In {\em ICML}, 2018.

\bibitem{lindell2017simulate}
Y.~Lindell.
\newblock How to simulate it--a tutorial on the simulation proof technique.
\newblock In {\em Springer Tutorials on the Foundations of Cryptography}. 2017.

\bibitem{Lindner2011}
R.~Lindner and C.~Peikert.
\newblock Better key sizes (and attacks) for {LWE}-based encryption.
\newblock In {\em Springer Topics in Cryptology}, 2011.

\bibitem{ndas1}
Why {NDA}s often don't work when expected to do so and what to do about it.
\newblock
  \url{https://www.linkedin.com/pulse/why-ndas-often-dont-work-when-expected-do-so-what-martin-schweiger}.
\newblock (Accessed: 2021-01-06).

\bibitem{MiniONN}
J.~Liu, M.~Juuti, Y.~Lu, and N.~Asokan.
\newblock Oblivious neural network predictions via {M}ini{ONN} transformations.
\newblock In {\em ACM CCS}, 2017.

\bibitem{Lyubashevsky2010}
V.~Lyubashevsky, C.~Peikert, and O.~Regev.
\newblock On ideal lattices and learning with errors over rings.
\newblock In {\em EUROCRYPT}, 2010.

\bibitem{federatedLearning1}
H.~B. McMahan, E.~Moore, D.~Ramage, and B.~A. y~Arcas.
\newblock Federated learning of deep networks using model averaging.
\newblock {\em CoRR}, abs/1602.05629, 2016.

\bibitem{McMahan2018}
H.~B. McMahan, D.~Ramage, K.~Talwar, and L.~Zhang.
\newblock Learning differentially private recurrent language models.
\newblock In {\em ICLR}, 2018.

\bibitem{deep-learning-apps}
Top 15 deep learning applications that will rule the world in 2018 and beyond.
\newblock
  \url{https://medium.com/breathe-publication/top-15-deep-learning-applications-that-will-rule-the-world-in-2018-and-beyond-7c6130c43b01}.
\newblock (Accessed: 2021-01-06).

\bibitem{Melis2019}
L.~{Melis}, C.~{Song}, E.~{De Cristofaro}, and V.~{Shmatikov}.
\newblock Exploiting unintended feature leakage in collaborative learning.
\newblock In {\em IEEE S\&P}, 2019.

\bibitem{lattigo}
Lattigo: A library for lattice-based homomorphic encryption in go.
\newblock \url{https://github.com/ldsec/lattigo}.
\newblock (Accessed: 2021-01-06).

\bibitem{mininet}
Mininet.
\newblock \url{http://mininet.org}.
\newblock (Accessed: 2021-01-06).

\bibitem{mishra2020}
P.~Mishra, R.~Lehmkuhl, A.~Srinivasan, W.~Zheng, and R.~A. Popa.
\newblock Delphi: A cryptographic inference service for neural networks.
\newblock In {\em USENIX Security}, 2020.

\bibitem{mohassel2018aby}
P.~Mohassel and P.~Rindal.
\newblock Aby 3: a mixed protocol framework for machine learning.
\newblock In {\em ACM CCS}, 2018.

\bibitem{SecureML}
P.~Mohassel and Y.~Zhang.
\newblock Secureml: A system for scalable privacy-preserving machine learning.
\newblock In {\em IEEE S\&P}, 2017.

\bibitem{mouchet2019distributedbfv}
C.~Mouchet, J.~R. Troncoso-pastoriza, J.-P. Bossuat, and J.~P. Hubaux.
\newblock Multiparty homomorphic encryption: From theory to practice.
\newblock In {\em Technical Report \url{https://eprint.iacr.org/2020/304}},
  2019.

\bibitem{Karthik2019}
K.~Nandakumar, N.~Ratha, S.~Pankanti, and S.~Halevi.
\newblock Towards deep neural network training on encrypted data.
\newblock In {\em IEEE CVPR Workshops}, 2019.

\bibitem{Nasr2019}
M.~{Nasr}, R.~{Shokri}, and A.~{Houmansadr}.
\newblock Comprehensive privacy analysis of deep learning: Passive and active
  white-box inference attacks against centralized and federated learning.
\newblock In {\em IEEE S\&P}, 2019.

\bibitem{Quantum4}
C.~Neill et~al.
\newblock A blueprint for demonstrating quantum supremacy with superconducting
  qubits.
\newblock {\em Science}, 2018.

\bibitem{svhn}
Y.~Netzer, T.~Wang, A.~Coates, A.~Bissacco, B.~Wu, and A.~Ng.
\newblock Reading digits in natural images with unsupervised feature learning.
\newblock {\em NIPS}, 2011.

\bibitem{nikolaenko2013privacy}
V.~Nikolaenko, U.~Weinsberg, S.~Ioannidis, M.~Joye, D.~Boneh, and N.~Taft.
\newblock Privacy-preserving ridge regression on hundreds of millions of
  records.
\newblock In {\em IEEE S\&P}, 2013.

\bibitem{Paillier}
P.~Paillier.
\newblock Public-key cryptosystems based on composite degree residuosity
  classes.
\newblock In {\em EUROCRYPT}, 1999.

\bibitem{pytorch}
A.~Paszke, S.~Gross, S.~Chintala, G.~Chanan, E.~Yang, Z.~DeVito, Z.~Lin,
  A.~Desmaison, L.~Antiga, and A.~Lerer.
\newblock Automatic differentiation in pytorch.
\newblock 2017.

\bibitem{blaze}
A.~Patra and A.~Suresh.
\newblock Blaze: Blazing fast privacy-preserving machine learning.
\newblock In {\em NDSS}, 2020.

\bibitem{Phong2017}
L.~T. Phong, Y.~Aono, T.~Hayashi, L.~Wang, and S.~Moriai.
\newblock Privacy-preserving deep learning: Revisited and enhanced.
\newblock In {\em Springer ATIS}, 2017.

\bibitem{Phong2018}
L.~T. {Phong}, Y.~{Aono}, T.~{Hayashi}, L.~{Wang}, and S.~{Moriai}.
\newblock Privacy-preserving deep learning via additively homomorphic
  encryption.
\newblock {\em IEEE TIFS}, 13(5):1333--1345, 2018.

\bibitem{Prechelt1998}
L.~Prechelt.
\newblock Early stopping - but when?
\newblock In {\em Springer Neural Networks: Tricks of the Trade}, 1998.

\bibitem{Rahman2018dp}
M.~A. Rahman, T.~Rahman, R.~Lagani{\`e}re, and N.~Mohammed.
\newblock Membership inference attack against differentially private deep
  learning model.
\newblock {\em Transactions on Data Privacy}, 11:61--79, 2018.

\bibitem{riazi2019xonn}
M.~S. Riazi, M.~Samragh, H.~Chen, K.~Laine, K.~E. Lauter, and F.~Koushanfar.
\newblock Xonn: Xnor-based oblivious deep neural network inference.
\newblock In {\em USENIX Security}, 2019.

\bibitem{palisade}
K.~Rohloff.
\newblock {The PALISADE Lattice Cryptography Library}.
\newblock \url{https://git.njit.edu/palisade/PALISADE}, 2018.

\bibitem{Schoenmakers2006}
T.~P. Schoenmakers~B.
\newblock Efficient computation modulo a shared secret with application to the
  generation of shared safe-prime products.
\newblock In {\em EUROCRYPT}, 2006.

\bibitem{schoppmann2019make}
P.~Schoppmann, A.~Gascon, M.~Raykova, and B.~Pinkas.
\newblock Make some room for the zeros: Data sparsity in secure distributed
  machine learning.
\newblock In {\em ACM CCS}, 2019.

\bibitem{shokri2015privacy}
R.~Shokri and V.~Shmatikov.
\newblock Privacy-preserving deep learning.
\newblock In {\em ACM CCS}, 2015.

\bibitem{shokri2017membership}
R.~Shokri, M.~Stronati, C.~Song, and V.~Shmatikov.
\newblock Membership inference attacks against machine learning models.
\newblock In {\em IEEE S\&P}, 2017.

\bibitem{Song2013}
S.~{Song}, K.~{Chaudhuri}, and A.~D. {Sarwate}.
\newblock Stochastic gradient descent with differentially private updates.
\newblock In {\em IEEE GlobalSIP}, 2013.

\bibitem{Stoica2017}
I.~Stoica, D.~Song, R.~A. Popa, D.~A. Patterson, M.~W. Mahoney, R.~H. Katz,
  A.~D. Joseph, M.~I. Jordan, J.~M. Hellerstein, J.~E. Gonzalez, K.~Goldberg,
  A.~Ghodsi, D.~E. Culler, and P.~Abbeel.
\newblock A {B}erkeley view of systems challenges for {AI}.
\newblock {\em CoRR}, abs/1712.05855, 2017.

\bibitem{Quantum3}
B.~Terhal.
\newblock Quantum supremacy, here we come.
\newblock {\em Nature Physics}, 14(06), 2018.

\bibitem{deep-learning-apps2}
Top applications of deep learning across industries.
\newblock
  \url{https://www.mygreatlearning.com/blog/deep-learning-applications/}.
\newblock (Accessed: 2021-01-06).

\bibitem{Florian2016}
F.~Tram\`{e}r, F.~Zhang, A.~Juels, M.~K. Reiter, and T.~Ristenpart.
\newblock Stealing machine learning models via prediction {API}s.
\newblock In {\em USENIX Security}, 2016.

\bibitem{truex2019hybrid}
S.~Truex, N.~Baracaldo, A.~Anwar, T.~Steinke, H.~Ludwig, R.~Zhang, and Y.~Zhou.
\newblock A hybrid approach to privacy-preserving federated learning.
\newblock In {\em ACM AISec}, 2019.

\bibitem{Vizitu2020}
A.~Vizitu, C.~Nită, A.~Puiu, C.~Suciu, and L.~Itu.
\newblock Applying deep neural networks over homomorphic encrypted medical
  data.
\newblock {\em Computational and Mathematical Methods in Medicine}, 2020:1--26,
  2020.

\bibitem{wagh2019securenn}
S.~Wagh, D.~Gupta, and N.~Chandran.
\newblock Securenn: 3-party secure computation for neural network training.
\newblock {\em PETS}, 2019.

\bibitem{falcon}
S.~Wagh, S.~Tople, F.~Benhamouda, E.~Kushilevitz, P.~Mittal, and T.~Rabin.
\newblock {FALCON: H}onest-majority maliciously secure framework for private
  deep learning.
\newblock {\em PETS}, 2020.

\bibitem{Wang2018}
J.~Wang and G.~Joshi.
\newblock Cooperative {SGD}: A unified framework for the design and analysis of
  communication-efficient {SGD} algorithms.
\newblock {\em CoRR}, abs:1808.07576, 2018.

\bibitem{Wang2019}
Z.~{Wang}, M.~{Song}, Z.~{Zhang}, Y.~{Song}, Q.~{Wang}, and H.~{Qi}.
\newblock Beyond inferring class representatives: User-level privacy leakage
  from federated learning.
\newblock In {\em IEEE INFOCOM}, 2019.

\bibitem{creditCard}
I.-C. Yeh and C.~hui Lien.
\newblock The comparisons of data mining techniques for the predictive accuracy
  of probability of default of credit card clients.
\newblock {\em Expert Systems with Applications}, 36(2):2473 -- 2480, 2009.

\bibitem{Yu2019}
L.~Yu, L.~Liu, C.~Pu, M.~Gursoy, and S.~Truex.
\newblock Differentially private model publishing for deep learning.
\newblock In {\em IEEE S\&P}, 2019.

\bibitem{Quantum2_GoogleAI}
A.~{Zalcman} et~al.
\newblock Quantum supremacy using a programmable superconducting processor.
\newblock {\em Nature}, 574:505--510, 10 2019.

\bibitem{Zhang_BigDataSecurity}
D.~Zhang.
\newblock Big data security and privacy protection.
\newblock In {\em ICMCS}, 2018.

\bibitem{zhao2018federated}
Y.~Zhao, M.~Li, L.~Lai, N.~Suda, D.~Civin, and V.~Chandra.
\newblock Federated learning with non-{IID} data.
\newblock {\em CoRR}, abs/1806.00582, 2018.

\bibitem{zheng2019helen}
W.~Zheng, R.~A. Popa, J.~E. Gonzalez, and I.~Stoica.
\newblock Helen: Maliciously secure coopetitive learning for linear models.
\newblock In {\em IEEE S\&P}, 2019.

\bibitem{NIPS2019_9617}
L.~Zhu, Z.~Liu, and S.~Han.
\newblock Deep leakage from gradients.
\newblock In {\em NIPS}. 2019.

\bibitem{Zhu2016}
X.~Zhu, C.~Vondrick, C.~C. Fowlkes, and D.~Ramanan.
\newblock Do we need more training data?
\newblock {\em Springer IJCV}, 119(1):76–92, Aug. 2016.

\bibitem{parallel_SGD}
M.~Zinkevich, M.~Weimer, L.~Li, and A.~J. Smola.
\newblock Parallelized stochastic gradient descent.
\newblock In {\em NIPS}, 2010.

\end{thebibliography}

\newcounter{subsubsubsection}
\appendices
\section{Symbols and Notations}\label{sec:notations}

Table~\ref{table:notations} summarizes the symbols and notation used in our paper.

\begin{table}[h!]
\centering
\small
\begin{tabular}{ll}
\toprule
\textbf{Notation} &\textbf{Description}\\ \toprule
\multicolumn{1}{c}{${P_i}$}  & $i^{th}$ Party\\ \midrule
\multicolumn{1}{c}{$Q$}  &  Querier\\ \midrule
\multicolumn{1}{c}{$X_i$}  & Input matrix of $P_i$   \\ \midrule
\multicolumn{1}{c}{$X_i[n]$}  & $n^{th}$ row of the input matrix    \\ \midrule
\multicolumn{1}{c}{$y_i$} & True labels of $P_i$\\ \midrule
\multicolumn{1}{c}{$N$} & Total number of parties  \\ \midrule
\multicolumn{1}{c}{$W_{j,i}^k$}  &  Weight matrix in $P_i$,\\
\multicolumn{1}{c}{} & for a layer $j$, at $k^{th}$ iteration \\ \midrule
\multicolumn{1}{c}{$n$} & Number of data samples \\ \midrule
\multicolumn{1}{c}{$d$} & Number of features  \\ \midrule
\multicolumn{1}{c}{$d_a$} & Degree of an approximated polynomial \\ \midrule
\multicolumn{1}{c}{$\ell$} & Total number of layers \\ \midrule
\multicolumn{1}{c}{$h_j$} & Number of neurons in $j^{th}$ layer  \\ \midrule
\multicolumn{1}{c}{$h_\ell$} & Number of output labels\\ \midrule
\multicolumn{1}{c}{$\eta$} & Learning rate    \\ \midrule
\multicolumn{1}{c}{$\varphi(\cdot)$} & Activation function  \\ \midrule
\multicolumn{1}{c}{$\varphi'(\cdot)$} & Derivative of the activation function  \\ \midrule
\multicolumn{1}{c}{$E_{j}^k$} & Error propagated in layer $j$, at $k^{th}$ iteration \\ \midrule
\multicolumn{1}{c}{$\nabla W_{j,i}^k$} & Gradient computed in $P_i$,\\
\multicolumn{1}{c}{} & for a layer $j$, at $k^{th}$ iteration \\ \midrule
\multicolumn{1}{c}{$b$} & Local batch size \\ \midrule
\multicolumn{1}{c}{$B$} & Global batch size \\ \midrule
\multicolumn{1}{c}{$m$} & Number of global iterations  \\ \midrule
\multicolumn{1}{c}{$\odot$} & Element-wise multiplication \\ \midrule
\multicolumn{1}{c}{$\times$} & Matrix or vector multiplication \\ \midrule
\midrule
\multicolumn{1}{c}{$\bm{{W}}$} & Encryption of $W$ (bold-face)\\ \midrule
\multicolumn{1}{c}{$\bar{{msg}}$} & Encoded (packed) plaintext vector $msg$ \\ \midrule
\multicolumn{1}{c}{$\mathcal{N}$} & Ring dimension \\ \midrule
\multicolumn{1}{c}{$\lambda$} & Security level \\ \midrule
\multicolumn{1}{c}{$A^T$} & Transpose of matrix $A$ \\ \midrule
\multicolumn{1}{c}{$L$} & Initial level of a ciphertext \\\midrule
\multicolumn{1}{c}{$S$} & Initial scale of a ciphertext \\\midrule
\multicolumn{1}{c}{$L_c$} & Current level of a ciphertext $c$ \\\midrule
\multicolumn{1}{c}{$\textsf{RIS}(\bm{c},p,s)$} &  RotateInnerSum  with $log_2(s)$ number of rotations. \\\midrule
\multicolumn{1}{c}{$\textsf{RR}(\bm{c},p,s)$} & RotateReplication  with $log_2(s)$ number of rotations. \\\midrule
\multicolumn{1}{c}{$S_c$} & Current scale of a ciphertext $c$ \\
\bottomrule
\end{tabular}
\caption{Frequently Used Symbols and Notations.}
\label{table:notations}
\end{table}

\section{Comparison to Other State-of-the-Art Solutions}\label{sec:qualComp}

Table~\ref{table:comparison} displays a qualitative comparison of \sys with the state-of-the-art privacy-preserving neural network training and/or inference solutions. The MPC-setup row of the table denotes the number of parties responsible for the execution of the NN operations. The adversarial model for data confidentiality indicates the capabilities of the parties (active (A) or passive (P)), and collusion shows the maximum number of possible colluding parties.

We note that several works allow as admissible adversary, i.e., collusions between one server and an arbitrary number of clients/data owners~\cite{SecureML}. For a fair comparison, we consider only the collusions permitted between the parties (servers) that are responsible for the training. To the best of our knowledge, \sys is the only solution that performs both training and inference of NNs, in an $N$-party setting, yet protects data and model confidentiality and withstands collusions up to $N-1$ parties. Therefore, our work differentiates itself from cloud outsourcing models and enables a privacy-preserving federated learning approach.

\begin{table*}[t!]
\scriptsize
\centering
\setlength{\tabcolsep}{2pt}
\begin{tabular}{llllllllllllllll}
\toprule
  &  & \begin{tabular}[c]{@{}l@{}}\textbf{XONN}\\ \cite{riazi2019xonn} \end{tabular} & \begin{tabular}[c]{@{}l@{}}\textbf{Gazelle}\\ \cite{Gazelle} \end{tabular} &
  \begin{tabular}[c]{@{}l@{}}\textbf{Blaze}\\ \cite{blaze} \end{tabular} &
  \begin{tabular}[c]{@{}l@{}}\textbf{MiniONN}\\ \cite{MiniONN}\end{tabular} & \begin{tabular}[c]{@{}l@{}}\textbf{ABY3}\\ \cite{mohassel2018aby} \end{tabular} & \begin{tabular}[c]{@{}l@{}}\textbf{SecureML}\\ \cite{SecureML}\end{tabular} & \begin{tabular}[c]{@{}l@{}}\textbf{SecureNN}\\ \cite{wagh2019securenn}\end{tabular} & \begin{tabular}[c]{@{}l@{}}\textbf{FALCON}\\ \cite{falcon}\end{tabular} & \begin{tabular}[c]{@{}l@{}}\textbf{FLASH}\\ \cite{flash}\end{tabular} & \begin{tabular}[c]{@{}l@{}}\textbf{TRIDENT}\\ \cite{trident}\end{tabular} &\begin{tabular}[c]{@{}l@{}}\textbf{CryptoNets}\\ \cite{CryptoNets}\end{tabular} &\begin{tabular}[c]{@{}l@{}}\textbf{CryptoDL}\\ \cite{cryptoDL}\end{tabular} & \cite{Karthik2019} &  \textbf{\sys} \\
\toprule
MPC Setup &  & 2PC & 2PC& 3PC & 2PC & 3PC & 2PC & 3PC & 3PC & 4PC &4PC& 1PC &1PC&1PC & \textbf{N-Party} \\
\midrule
Private Infer. & & \ding{52} &\ding{52}&\ding{52}  &\ding{52}  &\ding{52}  & \ding{52} &\ding{52}  &\ding{52}&\ding{52}  & \ding{52} &\ding{52}  &\ding{52} & \ding{52}&  \ding{52}\\
\midrule
 Private Train. & &\ding{55} &\ding{55}&\ding{55}   &\ding{55}  &\ding{52}  &\ding{52}  &\ding{52}  &\ding{52}  &\ding{52} &\ding{52} &\ding{55} &\ding{52}&\ding{52}  &\ding{52}  \\
 \midrule
 \multirow{3}{*}{\begin{tabular}[c]{@{}l@{}}Data Conf. \\ Adversarial Model*\\Collusion*\end{tabular}} \\& &$1$ P  &$1$ P &$1$ A  &$1$ P  & $1$ A/P & 1 P &$1$ A/P  &$1$ A/P &$1$ A&$1$ A/P  & $1$ P' &$1$ P'&$1$ P'& $N-1$ P\\
&  &No  &No & No &No  &  No& No & No  & No &No&No  &NA  &NA&NA& $N-1$\\
 \midrule
%Model Owner & &Server & Server& Data Owner  & Server  & ?  & Servers/None & Servers/None & None  &\ding{52} &\ding{52} &\ding{55} &\ding{52}&\ding{52}  &\ding{52}  \\
% \midrule
%\midrule
%\multirow{2}{*}{\begin{tabular}[c]{@{}l@{}}Adversarial\\ Model\end{tabular}} & Conf. & SH+MH  & SH& MH  & SH & SH+MH  & SH &SH+MH & SH+MH & MH  &SH+MH& SH' & SH' & SH' \\
%&Corr. & SH+MH & SH &SH+MH& SH & SH+MH & SH & SH & SH+MH & MH & SH+MH&SH'& SH' &  SH*\\
% \midrule
%Data Conf. & & NA & NA& NA & NA & \ding{52} &\ding{52}  &\ding{52}  &\ding{52}  & \ding{52}& \ding{52} & NA &\ding{52}&\ding{52}  \\
%\midrule
%\multirow{2}{*}{\begin{tabular}[c]{@{}l@{}}Model\\ Conf.\end{tabular}} & wrt. clients & \ding{71} & \ding{71} & \ding{52} &\ding{71}  & \ding{52}& \ding{52} &\ding{52}  & \ding{52}  & \ding{52} & \ding{52}& \ding{52} & \ding{52} & \ding{52} \\
%&wrt servers & NA & NA & \ding{71}& NA & \ding{71}& \ding{55} &\ding{71}  & \ding{71}  & \ding{71}& \ding{71} & NA & NA & \ding{71} \\
%\midrule
%Query Conf. &  & \ding{52} & \ding{52} & \ding{52} &  \ding{52}&  \ding{52}&  \ding{52}& \ding{52} &\ding{52}  &\ding{52} & \ding{52} & \ding{52} \\
\midrule
Techniques & &GC,SS  & HE,GC,SS & GC,SS & HE,GC,SS  & GC,SS & HE,GC,SS & SS & SS & SS &GC,SS& HE&HE &HE & HE     \\
\midrule
\multirow{3}{*}{\begin{tabular}[c]{@{}l@{}}Supported\\ Layers\end{tabular}} & Linear & \ding{52} &  \ding{52}  & \ding{52}& \ding{52} &  \ding{52} &  \ding{52} & \ding{52}&  \ding{52} &  \ding{52}  &   \ding{52} &\ding{52}&\ding{52}  & \ding{52} &  \ding{52} \\
 & Conv. & \ding{52} &  \ding{52}  & \ding{55}&  \ding{52} &  \ding{52} &  \ding{52} & \ding{52}&  \ding{52} &  \ding{52}  &  \ding{52} & \ding{52}& \ding{55} & \ding{55} &  \ding{52} \\
  & Pooling & \ding{52} &  \ding{52}  & \ding{55}&  \ding{52} &  \ding{52} &  \ding{52} & \ding{52}&  \ding{52} &  \ding{52}  &  \ding{52} & \ding{52}& \ding{55}  & \ding{55} &  \ding{52} \\
  \midrule
 % \multirow{3}{*}{\begin{tabular}[c]{@{}l@{}}Supported/\\Evaluated\\ Activation\\Functions\end{tabular}} & Sigmoid & \boldsymbol{-} &  \boldsymbol{-}  &  A &  A &  A & \ding{52}&  \ding{52} &  \ding{52}  &  \ding{52}&  \ding{52} &  \ding{70} \\
% & ReLU. & \boldsymbol{-} &  \ding{52}  &  \ding{52} &  \ding{52} &  \ding{52} & \ding{52}&  \ding{52} &  \ding{52}  &  \ding{52}&  \ding{55} &  \ding{52} \\
 % & Softmax & \ding{52} &  \boldsymbol{-}  &  \ding{55} & A &  A & \ding{52}&  \ding{52} &  \ding{52}  &  \ding{52}&  \ding{55} &  \ding{52} \\
 % & Square & \ding{52} &  \ding{52}  &  \ding{52} &  \ding{52} &  \ding{52} & \ding{52}&  \ding{52} &  \ding{52}  &  \ding{52}&  \ding{55} &  \ding{52} \\
\bottomrule
\end{tabular}
\captionsetup{width=\linewidth}
\caption{Qualitative comparison of private deep learning frameworks. Conf. stands for confidentiality. A and P stand for active and passive adversarial capabilities, respectively. GC, SS, HE denote garbled-circuits, secret sharing, and homomorphic encryption. Adversarial model* and collusion* take into account the servers responsible for the training/inference. $1$ P' denotes our interpretation as~\cite{CryptoNets},~\cite{Karthik2019}, and~\cite{cryptoDL} do not present an adversarial model. NA stands for not applicable.}
\label{table:comparison}
\end{table*}
%SH,MH,M stands for semi-honest, malicious with honest majority, and malicious.

\section{Approximated Activation Function Alternatives}\label{sec:approxAlternative}

For the piece-wise function ReLU, we propose two alternatives: (i) approximation of square-root for the evaluation of $\varphi(x) = 0.5(b + \sqrt{b^2})$ that is equivalent to ReLU, and  (ii) approximating the \textit{smooth} approximation of ReLU (SmoothReLU), or softplus, $\varphi(x)=\ln(1+e^x)$, both with least-squares. Our analysis shows that the latter achieves a better approximation for a degree $d_a=3$, whereas the former approximates better the exact ReLU if one increases the multiplicative depth by $1$ and uses $d_a=7$. In our evaluations, we use SmoothReLU for efficiency.

We note that the derivative of softplus is a sigmoid function, and we evaluate the approximated sigmoid as the derivative, as this achieves better accuracy. Finally, the Lattigo cryptographic library~\cite{lattigo} comes with a native way of approximating functions using Chebyshev interpolants and an efficient algorithm to evaluate polynomials in standard or Chebyshev basis. The least-squares is the optimal solution for minimizing the squared error over an interval, whereas Chebyshev asymptotically minimizes the maximum error. Hence, Chebyshev is more appropriate for keeping the error bounded throughout the whole interval, but requires a larger degree for a high accuracy approximation.
%Thus, when a high accuracy is needed for the activation function, we suggest using a large degree Chevshev interpolant. The advantage of the Chebyshev approximation, along with its asymptotic optimality \jpb{ref needed} and simple computation, is that it ensures that the polynomial interpolant has small coefficients and is numerically stable regardless of its degree, which is well suited for homomorphic evaluation.
\section{Approximation of the Max/Min Pooling and Its Derivative}\label{sec:maxpool}
For the sake of clarity, we describe the max-pooling operation. Given a vector $x = (x[0], \dots, x[n-1])$ the challenge is to compute $y$ with $y[0\leq i<n] = \max(x)$.
%Once $y$ is computed, we can also compute the vector $x_{\textsf{max}} = x \odot y$ which stores $\max(x)$ at the index of the maximum and zeros at all other indices. 
%\added{A possible solution is to compute the index of the maximum value in the vector $x$ and then use that vector to extract the maximum value of $x$.} 
To approximate the index of $\max(x)$, which can then be used to extract the max value of $x$, we follow an algorithm similar to that presented in~\cite{approxMax}, described below.

Given two real values $a, b$, with $0 \leq a,b \leq 1$, we observe the following: If $a > b$, then $a-b < a^{d} - b^{d}$ for $d > 1$, i.e., with increasing $d$, smaller values converge to zero faster and the ratio between the maximum value and other values increases. The process can be repeated to increase the ratio between $a$ and $b$ but, unless $a = 1$, both values will eventually converge to zero. To avoid this, we add a second step that consists in renormalizing $a$ and $b$ by computing $a = a/(a+b)$ and $b = b/(a+b)$. Thus, we ensure that after each iteration, $a + b = 1$ and since $b$ will eventually converge to zero, $a$ will tend towards $1$. If $a = b$, both values will converge to $0.5$. This algorithm can be easily generalized to vectors: Given a vector $x = (x[0], \dots, x[n-1])$, at each iteration it computes $x[i] = x[i]^{d} / \sum_{j=0}^{n-1} x[j]^{d}$, 
and multiplies the result with the original vector to extract the maximum value.

%Although theoretically possible, this iterative algorithm for max-pooling is a time-consuming procedure. Indeed, at each iteration, it requires computing an expensive inverse function, especially if a high accuracy is desired \added{or if the input values are very small}. 
This max-pooling algorithm is a time-consuming procedure as it requires computing an expensive inverse function, especially if a high accuracy is desired or if the input values are very small.
Instead, we employ a direct approach using $\max(a,b) = \tfrac{1}{2}(a+b+\sqrt{(a-b)^2})$, where the square-root can be approximated by a polynomial. 
To compute the maximum value for a kernel $f=k \times k$, we iterate $\log(f)$ times $\bm{c}_{i+1} = \max(\bm{c}_{i}, \textsf{RotL}_{2^{i}}(\bm{c}_{i}))$. 
%At the end of this procedure, the result of the max is stored in the first slot of the kernel. Using masking and replication we can copy it in the other slots of the kernel. 
%This procedure allows us to compute in parallel (per ciphertext) up to $(N/2) / (k + 2^{\ceil{\log(k)}})$ max-pooling
As each iteration consumes all levels, we use $\textsf{DBootstrap}(\cdot)$ $\log(f)$ times.
Hence, we suggest using the average-pooling instead, which is more efficient and precise, e.g., Dowlin et al.~\cite{CryptoNets} show that low-degree approximations of max-pooling will converge to a scalar multiple of the mean of $k$ values. 
%Hence, using average-pooling is much more efficient in the encrypted domain. 
We provide microbenchmarks of both max and average-pooling in Appendix~\ref{sec:microbenchmarks}.
\section{Technical details of Distributed Bootstrapping with Arbitrary Linear Transformations ($\textsf{DBootstrapALT}(\cdot)$)}\label{sec:DBootstrapAltDetails}

A linear transformation $\phi(\cdot)$ over a vector of $n$ elements can be described by a $n\times n$ matrix. 
The evaluation of a matrix-vector multiplication requires a number of rotations proportional to the square-root of its non-zero diagonals, thus, this operation becomes prohibitive when the number of non-zero diagonals is large.

Such a linear transformation can be, however, efficiently carried out \textit{locally} on a secret-shared plaintext, as $\phi(msg + M) = \phi(msg) + \phi(M)$ due to the linearity of $\phi(\cdot)$. Moreover, because of the magnitude of $msg + M$ (100 to 200 bits), arbitrary precision complex arithmetic with sufficient precision should be used for $\textsf{Encode}(\cdot)$, $\textsf{Decode}(\cdot)$, and $\phi(\cdot)$ to preserve the lower bits. The collective bootstrapping protocol in~\cite{mouchet2019distributedbfv} is performed through a conversion of an encryption to secret-shared values and a re-encryption in a refreshed ciphertext. We leverage this conversion to perform the aforementioned linear transformation in the secret-shared domain, before the refreshed ciphertext is reconstructed. This is our $\textsf{DBootstrapALT}(\cdot)$ protocol (Protocol~\ref{algorithm:cbootwithpermute}).

When the linear transformation is simple, i.e., it does not involve a complex permutation or requires a small number of rotations, the $\textsf{Encode}(\cdot)$ and $\textsf{Decode}(\cdot)$ operations in Line 8, Protocol~\ref{algorithm:cbootwithpermute} can be skipped. Indeed, those two operations are carried out using arbitrary precision complex arithmetic. In such cases, it is more efficient to perform the linear transformation directly on the encoded plaintext.\\
\descr{Security Analysis of $\textsf{DBootstrapALT}(\cdot)$.} This protocol is a modification of the $\textsf{DBootstrap}(\cdot)$ protocol  of Mouchet et al.~\cite{mouchet2019distributedbfv}, with the difference that it includes a product of a public matrix. Both $\textsf{DBootstrap}(\cdot)$ and $\textsf{DBootstrapALT}(\cdot)$ for CKKS differ from the BFV version proposed in~\cite{mouchet2019distributedbfv} in which the shares are not unconditionally hiding, but statistically or computationally hiding due to the incomplete support of the used masks.
Therefore, the proof follows analogously the passive adversary security proof of the BFV $\textsf{DBootstrap}(\cdot)$ protocol in~\cite{mouchet2019distributedbfv}, with the addition of Lemma~\ref{lemmaDBoot} which guarantees the statistical indistinguishablity of the shares in $\mathbb{C}$.
While the RLWE problem and Lemma~\ref{lemmaDBoot} do not rely on the same security assumptions, the first one being computational and the second one being statistical, given the same security parameter, they share the same security bounds. 
Hence, $\textsf{DBootstrap}(\cdot)$ and  $\textsf{DBootstrapALT}(\cdot)$ provide the same security as the original protocol of Mouchet et al.~\cite{mouchet2019distributedbfv}.

\begin{lemma}
Given the distribution $P_{0} = (a+b)$ and $P_{1} = c$ with $0 \leq a < 2^{\delta}$ and $0 \leq b, c < 2^{\lambda + \delta}$ and $b$, $c$ uniform, then the distributions $P_{0}$ and $P_{1}$ are $\lambda$-indistinguishable; i.e., a probabilistic polynomial adversary $\mathcal{A}$ cannot distinguish between them with probability greater than $2^{-\lambda}$: $|\text{Pr}[\mathcal{A}\rightarrow 1 | P = P_{1}] - \text{Pr}[\mathcal{A}\rightarrow1|P = P_{0}]|\leq 2^{-\lambda}$.
\label{lemmaDBoot}
\end{lemma}

 We refer to Algesheimer et. al~\cite[Section 3.2]{Algesheimer2002}, and Schoenmakers and Tuyls~\cite[Appendix A]{Schoenmakers2006}, for the proof of the statistical $\lambda$-indistinguishability.

%$\text{Pr}[(a + b) \geq 2^{\lambda + \delta}] \leq 2^{-\lambda}$ implies that the only way to distinguish the distribution of $a + b$ from the distribution of $c$ is to have $a+b \geq 2^{\lambda + \delta}$. 
%The distribution of $a$ is not known a priori but is bounded by $2^{\delta}-1$.
%In the worst-case, $a = 2^{\delta}-1$ so there will be an overflow whenever $b>2^{\delta + \lambda} - 2^{\delta}$ + 1, which, as $b$ is uniform, happens with probability $(2^{\delta}-1)/(2^{\delta + \lambda}-1) < 2^{-\lambda}$.
We recall that an encoded message $msg$ of $\mathcal{N}/2$ complex numbers with the CKKS scheme is an integer polynomial of $\mathbb{Z}[X]/(X^{\mathcal{N}}+1)$. Given that $||msg|| < 2^{\delta}$, and a second polynomial $M$ of $\mathcal{N}$ integer coefficients with each coefficient uniformly sampled and bounded by $2^{\lambda +\delta} -1$ for a security parameter $\lambda$, Lemma~\ref{lemmaDBoot} suggests that Pr$[||msg^{(i)} + M^{(i)}|| \geq 2^{\lambda + \delta}] \leq 2^{-\lambda}$, for $0 \leq i < \mathcal{N}$ and where $i$ denotes the $i^{th}$ coefficient of the polynomial.
That is, the probability of a coefficient of $msg + M$ to be distinguished from a uniformly sampled integer in $[0, 2^{\lambda + \delta})$ is bounded by $2^{-\lambda}$. 
Hence, during Protocol~\ref{algorithm:cbootwithpermute} each party samples its polynomial mask $M$ with uniform coefficients in $[0, 2^{\lambda + \delta})$. 
The parties, however, should have an estimate of the magnitude of $msg$ to derive $\delta$, and a probabilistic upper-bound for the magnitude can be computed by the circuit and the expected range of its inputs.

In Protocol~\ref{algorithm:cbootwithpermute}, the masks $M_{i}$ are added to the ciphertext of $R_{Q_{\ell}}$ during the decryption to the secret-shared domain.
To avoid a modular reduction of the masks in $R_{Q_{\ell}}$ and ensure a correct re-encryption in $R_{Q_{L}}$, the modulus $Q_{\ell}$ should be large enough for the additions of $N$ masks. 
Therefore, the ciphertext modulus size should be greater than $(N+1) \cdot ||M||$ when the bootstrapping is called. For example, for $N = 10$, a $Q_{L}$ composed of a 60 bits modulus, a message $msg$ with $||msg||<2^{55}$ (taking the scaling factor $\Delta$ into account) and $\lambda  = 128$, we should have $||M_{i}|| \geq 2^{183}$ and $Q_{\ell} > 11 \cdot 2^{183}$. Hence, the bootstrap should be called at $Q_{3}$ because $Q_{2} \approx 2^{180}$ and $Q_{3} \approx 2^{240}$. 
Although the aforementioned details suggest that $\textsf{DBootstrapALT}(\cdot)$ is equivalent to a depth 3 to 4 circuit, depending on the parameters, it is still compelling, as it enables us to refresh a ciphertext and apply an arbitrary complex linear transformation at the same time. Thus, its cost remains negligible compared to a centralized bootstrapping where any transformation is applied via rotations.

\section{Supplementary Experimental Results}
We provide further experimental results of \sys, that were left out of the main text due to space constraints. We provide the microbenchmarks and execution times of various NN architectures. 
\begin{table}[t]
\centering
\footnotesize
\setlength{\tabcolsep}{0.6pt}
\begin{tabular}{lccccc}
Topology & \textbf{MAP:}FF (s) & \textbf{MAP:}BP (s) & \textbf{REDUCE} (s)  & Comm. (s) & \textbf{Total (s)} \\
\toprule
(6, 1, 1, 2) & 0.40& 0.36& 0.05 &  0.47 & 1.28  \\

\midrule
(6, 2, 2, 2) & 0.44 & 0.43& 0.04  & 0.52& 1.43 \\
\midrule
(16, 2, 2, 8) & 0.48 & 0.42 & 0.03 & 0.54& 1.47  \\
\midrule
(16, 4, 4, 8) & 0.47 & 0.45 & 0.04 & 0.51 & 1.47 \\
\midrule
(32, 8, 8, 8) & 0.57 & 0.50 & 0.04  &0.45 & 1.56  \\
\midrule
(32, 16, 16, 8) & 0.55 & 0.52 & 0.03 & 0.47 & 1.57 \\
\midrule
(64, 8, 8, 8) & 0.55 & 0.50 & 0.04  & 0.45 & 1.54 \\
\midrule
 (64, 32, 32, 8) & 0.55 & 0.62 & 0.04  & 0.43 & 1.64  \\
\midrule
 (128, 32, 32, 8) & 0.60 & 0.63 & 0.04  & 0.38 & 1.65 \\
 \midrule
 (128, 64, 64, 8) & 0.78 & 0.80 & 0.05  & 0.56 & 2.19 \\
 \midrule
 (256, 64, 64, 8) & 1.04 & 1.36 & 0.06  & 0.38 & 2.84 \\
 \midrule
 (256, 128, 128, 8) & 2.01 & 2.62 & 0.11 & 0.61 & 5.35  \\
 \bottomrule
\end{tabular}
\captionsetup{width=\linewidth}
\caption{Execution times per-global-iteration of various NN architectures with batch size $B=120$, $N=10$ parties. \textbf{MAP:}FF, \textbf{MAP:}BP, Comm. stand for \textbf{MAP:}feed-forward, \textbf{MAP:} backpropagation, and communication respectively. } 
\label{table:supplementaryResults}
\end{table}

\subsection{Microbenchmarks}\label{sec:microbenchmarks}
We present microbenchmark timings for the various functionalities and sub-protocols of \sys in Table~\ref{table:microbenchmarks}. 
These are measured in an experimental setting with $N=10$ parties, a dimension of $d=32$ features, $h=64$ neurons in a layer or kernel size $k=3\times3$, and degree $d_a=3$ for the approximated activation functions for FC, CV, FC backpropagation, CV backpropagation, and average-pooling benchmarks. These benchmarks represent the processing of 1 sample per party, thus $b=1$. For max-pooling, we achieve a final precision of 7 bits with a square-root approximated by a Chebyshev interpolant of degree $d_a=31$. We observe that max-pooling is 6 times slower than average-pooling, has a lower precision, and needs more communication due to the large number of \textsf{DBootstrap}($\cdot$) operations. For 12-bits precision, max-pooling takes 4.72s. This supports our choice of using average-pooling instead of max-pooling in the encrypted domain.
The communication column shows the overall communication between the parties in MB. As several HE-based solutions
~\cite{CryptoNets,Gazelle,cryptoDL}, use square activation functions, we also benchmark them and compare them with the approximated activation functions with $d_a=3$.

We note that \textbf{PREPARE} stands for the offline phase and it incorporates the collective generation of the encryption, decryption, evaluation, and rotation keys based on the protocols presented in~\cite{mouchet2019distributedbfv}.
Most of the time and bandwidth are consumed by the generation of the rotation keys needed for the training protocol. We refer the reader to~\cite{mouchet2019distributedbfv,lattigo} for more information about the generation of these keys. Although we present the \textbf{PREPARE} microbenchmark to hint about the execution time and communication overhead of this offline phase, we note that it is a non-trivial task to extrapolate its costs for a generic neural network structure. \textbf{REDUCE} indicates the reducing step for 1 weight matrix (updating the weight matrix in root) and collectively refreshing it.
\begin{table}
\centering
\footnotesize
\begin{tabular}{llll}
Functionality  & Execution time (s) & Comm. (MB) \\
\toprule
%\multirow{2}{*}{\begin{tabular}[c]{@{}l@{}}ASigmoid/\\ ASmoothRelu\end{tabular}}& 0.050 & - \\
%\\
%\midrule
%\multirow{2}{*}{\begin{tabular}[c]{@{}l@{}}ASigmoidD/\\ ASmoothReluD\end{tabular}}& 0.022 & - \\
%\\
\textsf{ASigmoid/ASmoothRelu} & 0.050 &-  \\
\midrule
\textsf{ASigmoidD/ASmoothReluD} & 0.022 &-  \\
\midrule
\textsf{Square / SquareD} & 0.01 / 0.006 &-  \\
\midrule
\textsf{ASoftmax} & 0.07 &-  \\
\midrule
%\textsf{SquareD} &  0.006 &-  \\
%\midrule
$\textsf{DBootstrap}(\cdot)$ & 0.09 & 6.5   \\
\midrule
$\textsf{DBootstrapALT}(\cdot)$ ($\log_2{(h)}$ rots) & 0.18 & 6.5\\
\midrule
$\textsf{DBootstrapALT}(\cdot)$ with Average Pool & 0.33 & 6.5\\
\midrule
\textsf{MaxPooling}  & 2.08 & {19.5}\\
\midrule
\textsf{FC layer / FC layer-backprop} & 0.09 / 0.13 & -\\
\midrule
\textsf{CV layer / CV layer-backprop} & 0.03 / 0.046 & -\\
\midrule
%\added{\textsf{FC layer-backprop}} & \added{0.13} & -\\
%\midrule
%\added{\textsf{CV layer-backprop}} & \added{0.046} & -\\
%\midrule
\textsf{DKeySwitch} & 0.07 & 23.13  &  \\
\midrule
\textbf{PREPARE} (offline) & 18.19   & 3.8k \\
\midrule
\textbf{MAP} (only communication) & 0.03 &  18.35 \\
\midrule
\textbf{COMBINE} & 0.09 & 7.8  \\
\midrule
\textbf{REDUCE} & 0.1 & 6.5 &  \\
\bottomrule
\end{tabular}
\captionsetup{width=\linewidth}
\caption{Microbenchmarks of different functionalities for  $N = 10$ parties, $d=32$, $h=64$, $\mathcal{N}=2^{13}$, $d_a=3$, $k=3\times3$.} 
\label{table:microbenchmarks}
\vspace{-0.5em}
\end{table}

We show how to use these microbenchmarks to \textit{roughly} estimate the \textit{online} execution time and communication overhead of one global iteration for a chosen neural network structure. We combine the results of Table~\ref{table:microbenchmarks} for layers/kernels with specific size, fixed $\mathcal{N}$, $d_a$, and $N$, with those of Table~\ref{table:theoryAnalysis} that show \sys's linear scalability with $N$ for the operations requiring communication, linear scalability with $\mathcal{N}$, and logarithmic scalability with $d$. We scale the execution time of each functionality for the various parameters depending on the theoretical complexity. Here exemplify the time for computing one global iteration with $N=50$ parties, for a CNN with $32\times32$ input images, 1 CV layer with kernel size $k=6\times6$, 1 average-pooling layer with $k=3\times3$, and 1 FC layer with $h=128$ neurons. We observe that the number of parties $N$ is 5 times bigger than the setting of Table~\ref{table:microbenchmarks}, thus yields one round of communication of \textbf{MAP} and \textbf{COMBINE} as $0.03\times5=0.15$s and $0,09\times5=0.45$s, respectively. The \textbf{REDUCE} microbenchmark is calculated for 1 weight matrix, thus with 2 weight matrices, \textbf{REDUCE} will consume $0.2$s. For the LGD computation, we start with the CV layer with $k=6\times6$ kernel size. We remind that CV layers are represented by FC layers, thus the kernel size affects the run-time logarithmically; we multiply the CV layer execution time by 2 ($0.03\times2=0.06$) followed by an activation execution time of $0.05$s. For more than 1 filter per CV layer, this number should be multiplied by the number of filters (assuming no parallelization). Then, we use $\textsf{DBootstrapALT}(\cdot)$ with average pooling to refresh the ciphertext, compute the pooling together with the backpropagation values yielding an execution time of $0.33$s scaled to $50$ parties as $0.33\times5=1.65$s. Lastly, since its execution time scales logarithmically with the number of neurons, the FC layer will be executed in $0.09/\log_2(64)*\log_2(128)=0.105$s followed by another activation of $0.05$s. A similar approach is then used for the backward pass and with FC layer-backprop, CV layer-backprop, and using the derivatives of the activation functions. The microbenchmarks are calculated using 1 sample per-party; thus, to extrapolate the time for $b>1$ \textit{without any parallelization}, the total time for the forward and backward passes should be multiplied by $b$. Finally, as this example is a CNN, we already refresh the ciphertexts after each CV layer both in the forward and backward pass, to compute pooling or to re-arrange the slots. To extrapolate the times for MLPs, the number of bootstrappings are calculated as described in Section~\ref{sec:paramSelect} and this is multiplied by the $\textsf{DBootstrap}(\cdot)$ benchmark. Extrapolating the communication overhead for a global iteration is straightforward: As the communication scales linearly with the number of parties, we scale the overheads given in Table~\ref{table:microbenchmarks} with $N$. For example, with $N=50$ parties, \textbf{MAP} and \textbf{COMBINE}  consume $18.32\times5=91.6$MB and $7.8\times5=39$MB, respectively. Similarly, the total number of $\textsf{DBootstrap}(\cdot)$, and its variants, should be multiplied by $5$. Lastly, in this example, the weight or kernel matrices fit in one ciphertext ($\mathcal{N}/2=4,096$ slots); if more than 1 cipher per weight matrix is needed, the aforementioned numbers should be multiplied by the number of ciphertexts.

\subsection{Benchmarks on Various Neural Network Topologies}
\label{sec:supplementaryResults}
We provide execution times of different network topologies in Table~\ref{table:supplementaryResults}. ``Topology'' represents number of features ($d$), hidden neurons in each layer ($h_1,h_2$), and number of output labels ($h_3$) as $(d,h_1,h_2,h_3)$. We use local batch size $b=12$ and global batch size $B=120$ for $N=10$ parties. We use $\textsf{ASigmoid}$ with $d_a=3$ as an activation function. The execution times indicate the time required for one global iteration, i.e., a processing of the global batch, and we report forward pass, backpropagation and the number of communications in separate columns. The ``Communication'' column includes the communication required for the \textbf{COMBINE} phase and for $\textsf{DBootstrap}(\cdot)$ operations. We provide the feed-forward and backpropagation times in \textbf{MAP} separately. 
\subsection{Benchmarks on Various Convolutional Neural Network Topologies}
\label{sec:supplementaryResultsCNN}
We provide extrapolated execution times of different CNN topologies in Table~\ref{table:supplementaryResultsCNN}. As we introduce several operations (derivative of pooling) in the forward pass to bootstrapping function, we do not separate between forward pass and backpropagation times, and we introduce the overall execution times. ``Topology'' represents the padded (power-of-two) number of features ($d$),  kernel size for CV layer ($CV[n\times n]$), kernel size for average pooling layer ($P[n\times n]$), and $h$ number of neurons in the last FC layer connected to $h_{\ell}$ output layers ($FC[h:h_{\ell}]$) as $(d,CV[n\times n],P[n\times n],FC[h:h_{\ell}])$.
\begin{table}[t]
\centering
\footnotesize
\begin{tabular}{ll}
Topology &  Execution Time (s) \\
\toprule
$(256,CV[2\times 2],P[2\times 2],FC[16:2])$ &  1.42  \\
\midrule
$(512,CV[2\times 2],P[2\times 2],FC[16:2])$ & 1.52   \\
\midrule
$(512,CV[2\times 2],P[2\times 2],FC[32:2])$  & 1.88   \\
\midrule
$(784,CV[2\times 2],P[2\times 2],FC[32:2])$ &  2.12  \\
\midrule
$(784,CV[2\times 2],P[2\times 2],FC[32:10])$ &  2.56  \\
\midrule
$(784,CV[2\times 2],P[2\times 2],CV[2\times 2],P[2\times 2],FC[32:10])$&  3.88 \\
\bottomrule
\end{tabular}
\captionsetup{width=\linewidth}
\caption{Execution times per-global-iteration of various CNN architectures, with batch size $B=120$, $N=10$ parties.} 
\label{table:supplementaryResultsCNN}
\end{table}

\section{Extensions}\label{sec:extensions}
We introduce here several security, learning, and optimization extensions that can be integrated to \sys.

\subsection{Security Extensions}\label{securityExtensions}
We provide several security extensions that can be integrated to \sys as a future work.

\descr{Active Adversaries:} \sys preserves the privacy of the parties under a passive-adversary model with up to $N-1$ colluding parties, motivated by the cooperative federated learning scenario presented in Sections~\ref{intro} and~\ref{sec:system-threat}. If applied to other different scenarios, our work could be extended to an active-adversarial setting by using standard verifiable computation techniques, e.g., resorting to zero-knowledge proofs and redundant computation. This would, though, come at the cost of an increase in the computational complexity, that will be analyzed as future work.

\descr{Out-of-the-Scope Attacks:} We briefly discuss here out-of-the-scope attacks and countermeasures. By maintaining the intermediate values of the learning process and the final model weights under encryption, during the training process, we protect data and model confidentiality. As such, \sys protects against federated learning attacks~\cite{Nasr2019,Melis2019,Hitaj2017,NIPS2019_9617,Wang2019}. Nonetheless, there exist inference attacks that target the outputs of the model's predictions, e.g., membership inference~\cite{shokri2017membership}, model inversion~\cite{fredrikson2015model}, or model stealing~\cite{Florian2016}. Such attacks can be mitigated via complementary countermeasures that can be easily integrated to \sys: (i) limiting the number of prediction queries for the queriers, and (ii) adding noise to the prediction's output to achieve differential privacy guarantees. The choice of the differential privacy parameters in this setting remains an interesting open problem.

%TO-DO: Add paragraph on active adversaries

\subsection{Learning Extensions}\label{sec:learnextensions}

% \subsubsection{Early Stop}
% \label{sec:earlystop}

\descr{Early Stop.} There are several techniques proposed for the early stopping of the training of a neural network. They also prevent over-fitting as described and evaluated by Prechelt~\cite{Prechelt1998}. These approaches are: (i) $\text{GL}_{\alpha}$: stop when the generalization loss exceeds a certain threshold $\alpha$, (ii) $\text{PQ}_{\alpha}$: stop when the quotient of generalization loss and progress exceeds a certain threshold $\alpha$, and (iii) $\text{UP}_{s}$: stop when the generalization error increased in $s$ successive strips. The generalization error is estimated by the error on a validation set. We note that these methods can be seamlessly integrated into \sys by dividing each party's data into training and validation sets. Depending on the threshold and the method, the privacy-preserving implementation would require the homomorphic aggregation of the generalization error evaluated on each $P_i$'s validation set and a collective decryption of the error, after a number of global iterations $t$. As the error is the averaged scalar value. The leakage from the loss remains negligible when there are sufficient validation samples.

\descr{Availability, Data Distribution, and Asynchronous Distributed Neural Networks.}
In this work, we rely on a multiparty cryptographic scheme that assumes that the parties are always available. We here note that \sys can support asynchronous distributed neural network training~\cite{Downpour_SGD} without waiting for all parties to send the local gradients. As such, a time threshold could be used for updating the global model. However, we note that the collective cryptographic protocols (e.g., $\textsf{DBootstrap}(\cdot)$ and $\textsf{DBootstrapALT}(\cdot)$) require that all the parties be available. Changing \sys's distributed bootstrapping with a centralized one that achieves a practical security level would require increasing the size of the ciphertexts and result in higher computation and communication overhead.

For the evaluation of \sys, we evenly distribute the dataset across the parties; we consider the effects of uneven distributions or the asynchronous gradient descent to the model accuracy --- which are studied in the literature~\cite{Lian2018,Downpour_SGD,Wang2018} --- orthogonal to this work. However, a preliminary analysis with the MNIST dataset and the NN structure defined in our evaluation (see Section~\ref{sec:evaluation}) shows that asynchronous learning decreases the model accuracy between $1$ and $4\%$ when we assume that a server is down with a failure probability between $0.4$ and $0.8$, i.e., when there is between $40$ and $80\%$ chance of not receiving the local gradients from a server in a global iteration. Finally, we find that the uneven distribution of the MNIST dataset for $N=10$ parties with one party holding $90\%$ of the data results to a $6\%$ decrease in the model accuracy. Lastly, we note that the non-iid distribution of the data in federated learning settings causes weight/parameter divergence~\cite{li2020convergence,zhao2018federated}. The proposed mitigation techniques, however, do not change the working principle of POSEIDON and can be seamlessly integrated, e.g., by adjusting hyperparameters~\cite{li2020convergence} or creating and globally sharing a set of data with uniform distribution among the participants~\cite{zhao2018federated}.

% We here note that the the effects of uneven distribution of data or asynchronous SGD on the model accuracy in the context of distributed learning . 

\descr{Other Neural Networks.} In this work, we focus on the training of MLPs and CNNs and present our packing scheme and cryptographic operations for these neural networks. For other structures, e.g., long short-term memory (LSTM), recurrent neural networks (RNN), and residual neural networks (ResNet), \sys requires modifications of the LGD-computation phase according to their forward and backward pass operations and of the packing scheme. For example, ResNet has skip connections to jump over some layers, thus the shape of the packed ciphertext after a layer skip should be aligned according to the weight matrix that it is multiplied with. This can be ensured by using the $\textsf{DBootstrapALT}(\cdot)$ functionality (to re-arrange the slots of the ciphertext). We note that \sys's packing protocols are tailored to MLPs and CNNs and might require adaptation for other neural network structures.
% We here note that the the effects of uneven distribution of data or asynchronous SGD on the model accuracy in the context of distributed learning . 

\subsection{Optimization Extensions}\label{sec:optimizationExtensions}
% \label{sec:convOpt}
\descr{Optimizations for Convolutional Neural Networks.} We present a scheme for applying the convolutions on the slots, similar to FC layers, by representing them with a matrix multiplication. Convolution on a matrix, however, can be performed with a simple polynomial multiplication by using the coefficients of the polynomial. This operation requires a Fast-Fourier Transform (FFT) from slots (Number Theoretic Transform (NTT)) to coefficients domain, and vice versa (inverseFFT) for switching between CV to pooling or FC layers. Although it achieves better performance for CV layers, domain-switching is expensive. In the case of multiple CV layers before an FC layer, this operation could be embedded into the distributed bootstrapping ($\textsf{DBootstrapALT}(\cdot)$) for efficiency. The evaluation of the trade-off between the two solutions for larger matrix dimensions is an interesting direction for future work.

\descr{Graphics Processing Units (GPUs).} In this work, we evaluate our system on CPUs. 
Using GPUs to improve 
\sys's performance requires GPU-compatible cryptographic functions, i.e., extending the underlying cryptographic library Lattigo~\cite{lattigo}. 
In a recent work, Badawi et al.~\cite{Badawi2019} proposed the first GPU implementation of the full RNS-variant of the CKKS scheme, for which they report speedups of one to two orders of magnitude over a CPU implementation. Hence, GPU-accelerated FHE is an option that could greatly improve the practicality of \sys.

\end{document}